\documentclass[longauth]{aa}
\usepackage{graphicx}
\usepackage{txfonts}
\usepackage{natbib}
\usepackage[colorlinks=true,citecolor=blue]{hyperref}
\usepackage{xspace}
\usepackage{subfig}
\usepackage{lscape} 
\bibpunct{(}{)}{;}{a}{}{,}

\newcommand{\MJup}{\ensuremath{M_{\mathrm{Jup}}}\xspace}

\newcommand{\MSun}{\ensuremath{M_{\odot}}\xspace}

\def\hr8799{{HR\,8799}}

\def\gpi{GPI}

\def\megno{{MEGNO}}

\def\hst{HST}

\def\sphere{SPHERE}
\def\irdis{IRDIS}

\def\emcee{{\tt emcee}}

\newcommand{\msun}{{M}_{\odot}}

\newcommand{\mJ}{{M}_{\idm{Jup}}}

\newcommand{\Mmean}{\mathcal{M}}

\newcommand{\Ym}{\langle Y \rangle}

\def\idm#1{{\mbox{\scriptsize #1}}}
\newcommand{\mic}{$\mu$m\xspace}
\newcommand{\as}{\hbox{$^{\prime\prime}$}\xspace}

\newcommand{\epk}{1998.830}

\newcommand{\ifs}{IFS}

\newcommand{\Chinu}{\chi^2_{\nu}}
\newcommand{\Chi}{\chi^2}
\newcommand{\gm}{GM2020}
\newcommand{\DEC}{Dec}
\newcommand{\RA}{RA}
\def\drad{\,rad$\cdot$d$^{-1}$}

\begin{document}

\title{Orbital and dynamical analysis of the system around HR\,8799.}
\subtitle{New astrometric epochs from VLT/SPHERE and LBT/LUCI.}

   \author{A. Zurlo\inst{\ref{udp}, \ref{udp2}, \ref{lam}}, K. Go\'zdziewski\inst{\ref{pol}}, C. Lazzoni\inst{\ref{ex}, \ref{oapd}}, D. Mesa\inst{\ref{oapd}}, P. Nogueira\inst{\ref{udp}}, S. Desidera\inst{\ref{oapd}}, R. Gratton\inst{\ref{oapd}}, F. Marzari\inst{\ref{unipd}},\\ E. Pinna\inst{\ref{inaf-ar}}, G. Chauvin\inst{\ref{ipag}}, P. Delorme\inst{\ref{ipag}}, J.H. Girard\inst{\ref{stsci}}, J. Hagelberg\inst{\ref{geneva}}, Th. Henning\inst{\ref{mpia}}, M. Janson\inst{\ref{stuniv}}, E. Rickman\inst{\ref{esa}}, P. Kervella\inst{\ref{lesia}}, H. Avenhaus\inst{\ref{mpia}}, T. Bhowmik\inst{\ref{udp}}, B. Biller\inst{\ref{supa}}, A. Boccaletti\inst{\ref{lesia}}, M. Bonaglia\inst{\ref{inaf-ar}}, M. Bonavita\inst{\ref{mk7},\ref{oapd}}, M. Bonnefoy\inst{\ref{ipag}}, F. Cantalloube\inst{\ref{lam}}, A. Cheetham\inst{\ref{geneva}}, R. Claudi\inst{\ref{oapd}}, V. D'Orazi\inst{\ref{oapd}}, M. Feldt\inst{\ref{mpia}}, R. Galicher\inst{\ref{lesia}}, E. Ghose\inst{\ref{inaf-ar}}, A.-M. Lagrange\inst{\ref{ipag}}, M. Langlois\inst{\ref{cral},\ref{lam}}, H. le Coroller\inst{\ref{lam}}, R. Ligi\inst{\ref{lagrange}}, M. Kasper\inst{\ref{eso}}, A.-L. Maire, F. Medard\inst{\ref{ipag}}, M. Meyer\inst{\ref{eth}}, S. Peretti\inst{\ref{geneva}}, C. Perrot\inst{\ref{lesia}}, A.T. Puglisi\inst{\ref{inaf-ar}}, F. Rossi\inst{\ref{inaf-ar}}, B. Rothberg\inst{\ref{lbt},\ref{fairfax}}, T. Schmidt, E. Sissa\inst{\ref{oapd}}, A. Vigan\inst{\ref{lam}}, Z. Wahhaj\inst{\ref{esochile}} }
   \institute{N\'ucleo de Astronom\'ia, Facultad de Ingenier\'ia, Universidad Diego Portales, Av. Ejercito 441, Santiago, Chile \label{udp} \and Millennium Nucleus on Young Exoplanets and their Moons (YEMS), Universidad Diego Portales, Av. Ejercito 441, Santiago, Chile \label{udp2} \and
  Aix-Marseille Universit\'e, CNRS, LAM (Laboratoire d'Astrophysique de Marseille) UMR 7326, 13388, Marseille, France \label{lam} \and Institute of Astronomy, Faculty of Physics, Astronomy and Informatics, Nicolaus Copernicus University, Grudziadzka 5, 87-100 Toru\'n, Poland \label{pol} \and University of Exeter, Astrophysics Group, Physics Building, Stocker Road, Exeter, EX4 4QL, UK \label{ex} \and INAF - Osservatorio Astronomico di Padova, Vicolo dell'Osservatorio 5, 35122, Padova, Italy \label{oapd} \and Department of Physics and Astronomy, University of Padova, Via Marzolo 8, I-35131 Padova, Italy \label{unipd} \and INAF - Osservatorio Astrofisico di Arcetri, L.go E. Fermi 5, 50125, Firenze, Italy \label{inaf-ar} \and IPAG, Univ. Grenoble Alpes, CNRS, IPAG, 38000, Grenoble, France \label{ipag} \and Space Telescope Science Institute (STScI), 3700 San Martin Dr, Baltimore MD, 21218, USA \label{stsci} \and Geneva Observatory, University of Geneva, 51 ch. Pegasi, 1290, Versoix, Switzerland \label{geneva} \and Max Planck Institute for Astronomy, Königstuhl 17, 69117, Heidelberg, Germany \label{mpia} \and Department of Astronomy, Stockholm University, SE-106 91 Stockholm, Sweden \label{stuniv}  \and European Space Agency (ESA), ESA Office, Space Telescope Science Institute, 3700 San Martin Drive, Baltimore, MD 21218, USA \label{esa} \and LESIA, Observatoire de Paris, Universit\'e PSL, CNRS, Sorbonne Universit\'e, Universit\'e de Paris, 5 place Jules Janssen, 92195 Meudon, France \label{lesia} \and SUPA, Institute for Astronomy, University of Edinburgh, Blackford Hill, Edinburgh, EH9 3HJ, UK \label{supa} \and School of Physical Sciences, Faculty of Science, Technology, Engineering and Mathematics,
The Open University, Walton Hall, Milton Keynes, MK7 6AA \label{mk7} \and CRAL, UMR 5574, CNRS, Universit\'e de Lyon, ENS, 9 avenue Charles Andr\'e, 69561, Saint Genis Laval Cedex, France \label{cral} \and Institute for Particle Physics and Astrophysics, ETH Zurich, Wolfgang-Pauli-Strasse 27, 8093, Zurich, Switzerland \label{eth} \and Universit\'e C\^ote d'Azur, Observatoire de la C\^ote d'Azur, CNRS, Laboratoire Lagrange, France \label{lagrange} \and European Southern Observatory, Karl-Schwarzschild-Strasse 2, D-85748 Garching, Germany \label{eso} \and LBT Observatory, University of Arizona, 933 N.Cherry Ave,Tucson AZ 85721, USA \label{lbt} \and George Mason University, Department of Physics \& Astronomy, MS 3F3, 4400 University Drive, Fairfax, VA 22030, USA \label{fairfax} \and European Southern Observatory, Alonso de Cordova 3107, Casilla 19001 Vitacura, Santiago 19, Chile \label{esochile}}

%K.Go. ORCID ID 0000-0002-8705-1577
   \date{Submitted/Accepted}

% \abstract{}{}{}{}{}
% 5 {} token are mandatory

  \abstract
  {HR\,8799 is a young planetary system composed of four planets and a double debris belt. Being the first multi-planetary system discovered with the direct imaging technique, it { has been} observed extensively  since 1998. This wide baseline of astrometric measurements, counting over 50 observations in 20 years, permits a detailed orbital and dynamical analysis of the system. }
    {To explore the orbital parameters of the planets, their dynamical history, and the planet-to-disk interaction, we made follow-up observations of the system during the VLT/SPHERE guaranteed time observation program. We obtained 21 observations, most of them in favorable conditions. In addition, we observed HR\,8799 with the instrument LUCI at the Large Binocular Telescope (LBT). }
    {All the observations were reduced with state-of-the-art algorithms implemented to apply the spectral and angular differential imaging method. We re-reduced the SPHERE data obtained during the commissioning of the instrument and in three open-time programs to have homogeneous astrometry. The precise position of the four planets with respect to the host star was calculated by exploiting the fake negative companions method. We obtained an astrometric precision of the order of 6 mas in the worst case and 1 mas in the best case. To improve the orbital fitting, we also took into account all of the astrometric data available in the literature. From the photometric measurements obtained in different wavelengths, we estimated the masses of the planets following the evolutionary models.  }
    {We obtained updated parameters for the orbits with the assumption of coplanarity, relatively small eccentricities, and periods very close to the 2:1 resonance. We also refined the dynamical mass of each planet and the parallax of the system (24.49 $\pm$ 0.07 mas), which overlap with the recent GAIA { eDR3/DR3} estimate. Hydrodynamical simulations suggest that inward migration of the planets caused by the interaction
with the disk might be responsible for the planets being locked in resonance. We also conducted detailed $N$-body simulations indicating possible positions of a~putative fifth innermost planet with a mass below the present detection limits of $\simeq 3$~\MJup.}
   {}


\titlerunning{Dynamical analysis of HR\,8799 from SPHERE and LUCI astrometry.}
\authorrunning{Zurlo et al.}
\maketitle
%
%________________________________________________________________
\section{Introduction}

Among the exoplanets discovered with the high-contrast imaging technique \citep[e.g.,][]{2004A&A...425L..29C,2010Sci...329...57L,2013ApJ...772L..15R,2018A&A...617A..44K,2021A&A...648A..73B}, the system around HR\,8799 is undoubtedly one of the most interesting. This is mainly because HR\,8799 hosts a greater number of planets  detected with high-contrast imaging than any other system, only three systems that host two planets were detected: PDS\,70 \citep{2018A&A...617A..44K,2019NatAs...3..749H}, TYC\,8998-760-1 \citep{2020ApJ...898L..16B}, and $\beta$\,Pic \citep{2010Sci...329...57L, 2020A&A...642L...2N}. HR\,8799 is a perfect laboratory with which to study dynamical interaction in young planetary systems, with its four planets \citep[HR\,8799\,bcde;][]{2008Sci...322.1348M,2010Natur.468.1080M} and a double debris belt \citep[see, e.g.,][]{2009ApJ...705..314S,2011ApJ...740...38H,2014ApJ...780...97M, Booth,2021AJ....161..271F}. The host star HR\,8799 is a young \citep[$\sim$ 42 Myr,][]{2010ApJ...716..417H,2011ApJ...732...61Z,2015MNRAS.454..593B}, $\gamma$ Dor-type variable star \citep{1999AJ....118.2993G} with $\lambda$ Boo-like abundance patterns. The mass of the star is 1.47$^{+0.12}_{-0.17}$ \MSun \citep{2022AJ....163...52S} and its distance is 40.88 $\pm$ 0.08 pc from GAIA measurements \citep{2020yCat.1350....0G}.

This system is an optimal target for high-contrast imaging observations, as its four planets are easily detectable with state-of-the-art imagers; their contrasts are about 2--8 $\times$ 10$^{-6}$; and the separation of the closest planet HR\,8799\,e is $\sim$ 390 mas, which is further than the inner working angle of most current instruments designed for direct imaging. For these reasons, the system has been observed dozens of times starting from 1998, with different instruments, wavelengths, and configurations. Thanks to this rich pool of archival data, HR\,8799 can be studied in detail from an astrometric point of view, being the only multi-planetary system for which there are tens of astrometric data points on a baseline of more than 20 years. These measurements began with uncertainties of $\sim$ 20 mas, but have since reached unprecedented precision of below a milliarcsecond (0.1--0.2 mas) with optical interferometry \citep{2019AA...623L..11G}. 

Works regarding the analysis of the dynamical interaction of the four planets have been presented; most of these agree on the near coplanarity of the planets, and that they are locked in a 1:2:4:8 resonance, \citep[see, e.g.,][]{2010ApJ...710.1408F,2013AA...549A..52E, Konopacky2016, 2016A&A...587A..57Z, 2018AJ....156..192W,  2019AA...623L..11G, GM2020}. From the age of the system and the luminosity of the planets, evolutionary hot-start models derive masses of around 5--7 \MJup \citep{2010Natur.468.1080M,2011ApJ...729..128C,2012ApJ...755...38S}. These masses are compatible with the dynamical studies because small masses help the orbits to remain stable \citep[e.g.,][]{2018AJ....156..192W}. On the other hand, \cite{2021ApJ...915L..16B} found a dynamical mass for the innermost planet HR\,8799 e of 9.6$^{+1.9}_{-1.8}$ \MJup assuming that planets c, d, and e share the same mass within $\sim$ 20\%. This result is 2 \MJup higher than the prediction from the evolutionary models.

In this paper, we present all the astrometric measurements obtained during the whole guaranteed time observation (GTO) of VLT/SPHERE for HR\,8799\,bcde. {Ad hoc} observations were designed for the orbital follow-up of the four planets; HR\,8799 was observed 21 times in total with SPHERE, and only four observations were rejected based on quality criteria. To complement the SPHERE measurement, we obtained one astrometric epoch with the instrument LUCI, installed at the LBT. This instrument observed HR\,8799 as part of the commissioning phase of the new adaptive optics (AO) system. With a total of 18 epochs and the addition of all the astrometric measurements available in the literature (69 astrometric epochs in total), we performed the dynamical analysis of the system and the planet-to-disk interaction.

The outline of the paper is as follows: in Sect.~\ref{sec:obs}, we present SPHERE and LUCI observations; in Sect.~\ref{sec:red}, we describe the reduction methods applied and the astrometric results that we obtained. In Sect.~\ref{Sec:astro} we present the astrometric fitting for the four planets of HR\,8799 and in Sect.~\ref{Sec:his} a possible interpretation of the history of the system. We also explored the possibility of the presence of a fifth planet, studying the planets--disk interaction (Sect.~\ref{s:planetdisk}).  We provide our conclusions in Sect.~\ref{sec:con}.

%__________________________________________________________________

\section{Observations}
\label{sec:obs}
\subsection{VLT/SPHERE}
HR\,8799 was observed several times during the SpHere INfrared survey for Exoplanets \citep[SHINE, papers I, II, III;][]{ 2021A&A...651A..70D, 2021A&A...651A..72V, 2021A&A...651A..71L} during the GTO of VLT/SPHERE. The instrument SPHERE \citep{2019A&A...631A.155B} is a planet finder equipped with an extreme AO system \citep[SAXO;][]{Fusco:06, 2014SPIE.9148E..0OP} to characterize substellar companions with high-contrast imaging. The near-infrared (NIR) arm includes the IR dual-band imager and spectrograph \citep[IRDIS;][]{Do08} and an integral field spectrograph  \citep[IFS;][]{Cl08}. During the observations, these two subsystems observed the target in parallel. HR\,8799 was periodically observed for astrometric monitoring, the setup of the observations included three different filter pairs: IRDIS in $H2H3$ bands ($\lambda_{H2}$=1.593 \mic, $\lambda_{H3}$=1.667 \mic), in BB\_H ($\lambda_{H}$=1.625 \mic),  and $K1K2$ bands ($\lambda_{K1}$=2.102 \mic, $\lambda_{K2}$=2.255 \mic). The coronagraph used for the shortest wavelengths was an apodized Lyot with a mask diameter of 185 mas and an inner working angle (IWA) of 0\farcs09 \citep[see][]{Boc08}, while for $K1K2$ band the IWA is $\sim$ 0\farcs12. For a detailed description of the observing sequence, we refer the reader to \citet{2014A&A...572A..85Z, 2016A&A...587A..57Z}. In general, the working sequence includes an image of the off-axis star point spread function (PSF) for the flux calibration, a long coronagraphic sequence with the satellite spots \citep{2013aoel.confE..63L} mode, a second image of the stellar PSF, and the sky images. The waffle mode was used on purpose to assure maximum astrometric precision. The long waffle sequence was taken in pupil stabilized mode in order to apply the angular differential imaging \citep[ADI;][]{2006ApJ...641..556M} method. While IRDIS has a field of view of $\sim$11$\times$11 arcseconds, IFS is smaller (1.7\as$\times$1.7\as), and only the two interior planets are visible. The SHINE observations are summarized in Table~\ref{t:obs}. For almost all the epochs, the observation was with IRDIS and IFS working in parallel. On a few occasions, IRDIS was used alone; these observations are marked in the table. We discarded data for time periods where the conditions did not permit a clear detection of the planets or there was a very poor signal-to-noise ratio (S/N). In particular, we rejected data from 2014 August 13 ( only IFS data were considered), 2014 December 08, 2015 July 29, and 2017 October 07. All the other observations are taken into account in this analysis.

\begin{table}
\begin{minipage}{0.5\textwidth}
\caption{Summary of the observations of HR\,8799 with IRDIS and IFS during SHINE.} 
\label{t:obs}
\renewcommand{\footnoterule}{}  % to avoid a line before footnotes
\centering
\begin{tabular}{lcccc}
\hline
\hline
           UT date           &  Filter &$\Delta_{FoV}$ ($^{\circ}$) & S/N$_{Min}$ & S/N$_{Max}$ \\
           \hline

2014-07-12      &       DB\_H23 &       17      &       9       &       44      \\
2014-08-13\footnote{Discarded for bad weather}  &       BB\_J   &       23      &       0       &       36      \\
2014-12-04\footnote{IRDIS alone}        &       BB\_H   &       9       &       14      &       24      \\
2014-12-05$\color{red}^b$       &       BB\_H   &       8       &       11      &       25      \\
2014-12-06$\color{red}^b$       &       BB\_H   &       8       &       15      &       25      \\
2014-12-08$\color{red}^{ab}$    &       BB\_H   &       8       &       6       &       25      \\
2015-07-03      &       DB\_K12 &       18      &       12      &       75      \\
2015-07-29$\color{red}^{ab}$    &       DB\_J23 &       82      &       4       &       51      \\
2015-07-30$\color{red}^b$       &       DB\_K12 &       81      &       21      &       123     \\
2015-09-27      &       DB\_K12 &       24      &       10      &       71      \\
2016-11-17      &       DB\_H23 &       17      &       15      &       52      \\
2017-06-14      &       DB\_H23 &       19      &       14      &       67      \\
2017-10-07$\color{red}^a$       &       BB\_H   &       76      &       0       &       0       \\
2017-10-11      &       BB\_H   &       74      &       45      &       90      \\
2017-10-12      &       BB\_H   &       78      &       24      &       100     \\
2018-06-18      &       DB\_H23 &       34      &       20      &       78      \\
2018-08-17      &       BB\_H   &       73      &       29      &       57      \\
2018-08-18      &       BB\_H   &       73      &       42      &       106     \\
2019-10-31      &       DB\_K12 &       24      &       22      &       62      \\
2019-11-01      &       DB\_K12 &       54      &       38      &       114     \\
2021-08-20      &       DB\_H23 &       20      &       21      &       59      \\
\hline
\end{tabular}
\end{minipage}
\end{table}

\subsection{Large Binocular Telescope/SOUL-LUCI}

 HR\,8799 was observed with LBT/SOUL-LUCI1  during its commissioning on night 2020 September 29. LUCI1 \citep{2010SPIE.7735E..7WS} is a NIR spectro-imager that can work in a diffraction-limited regime with a sampling of 15~mas/pix. The AO correction is provided by SOUL \citep{Pinna2019}, the upgrade of the FLAO system \citep{Esposito2010}. During this observation, the SOUL system was correcting 500 modes with a rate of 1.7~kHz.
 No coronagraphic masks are available on LUCI, and therefore to avoid any saturation, HR\,8799 was observed adopting a sub-windowing of 256$\times$256 pixels (corresponding to a field of view (FoV) of 3.84$\times$3.84\as). In this way, we were able to set a minimum exposure time of 0.34~s. We observed this target in pupil stabilization mode with two narrow-band filters: FeII ($\lambda_{FeII}$=1.646~\mic; 
$\Delta\lambda$=0.018~\mic) and H2 ($\lambda_{H2}$=2.124~\mic; $\Delta\lambda$=0.023~\mic). The
observations with the two filters were alternated during the whole sequence.

\section{Data reduction}
\label{sec:red}
\subsection{SPHERE}
The reduction of the IRDIS data was carried out entirely with the Data Center \citep[DC;][]{2017sf2a.conf..347D} which uses the standard SPHERE pipeline SpeCal \citep{2018A&A...615A..92G}. For the astrometric calibration, we used a true north orientation of -1.75 deg and a pixel scale value of 12.25 mas as reported in \citet{2016SPIE.9908E..34M}. The reduction algorithm used by the DC for HR\,8799 is the T-LOCI \citep{marois14}. To provide a homogeneous reduction of all the IRDIS data, we processed the observations presented in \citet{2016A&A...587A..57Z}  again, which were previously reduced with custom routines, which included seven epochs in 2014 (we excluded the sequence of 2014 August 13 for poor quality). The four epochs taken for variability monitoring ---once per night (2014 December 4, 2014 December 5, 2014 December 6, 2014 December 8)--- and presented in \citet{2016ApJ...820...40A} were reprocessed with the DC. As in \citet{2016ApJ...820...40A}, we excluded the sequence of December 8 for poor conditions. In the same way, we used the DC to reduce the data from the variability monitoring of \citet{2021MNRAS.503..743B}, with observation dates: 2015 July 29, 2015 July 30, 2017 October 07, 2017 October 11, and 2017 October 12 (two epochs were excluded for poor quality), and 2018 August, 17-18. Finally, we re-processed the IRDIS data presented by \citet{2021A&A...648A..26W} and taken on October 31, 2019, and November 1, 2019. In the DC, a ``fake negative planets'' \citep[e.g.,][and references therein]{2014A&A...572A..85Z} algorithm is implemented to calculate the precise position of each planet with respect to the central star. A summary of all the SPHERE astrometric points, both from the SHINE survey and published observations presented in \citet{2016A&A...587A..57Z}, \citet{2016ApJ...820...40A}, \citet{2021MNRAS.503..743B}, and \citet{2021A&A...648A..26W} is listed in Table~\ref{t:astroirdis} (IRDIS) and \ref{t:astroifs} (IFS). We refer the reader to these latter publications for further information about the observing conditions. The results of the updated astrometry are also listed in Table~\ref{t:lite}, together with all the other data from different instruments presented in the literature.

From the IRDIS photometry, we also estimated the mass of each planet using evolutionary models. In particular, we applied evolutionary models from \cite{2003A&A...402..701B,2015A&A...577A..42B} to the IRDIS filters photometry. The age of the system used in this estimation is 42 Myr and the parallax is 24.46 $\pm$ 0.05 mas, as in GAIA EDR3. The error reported is the standard deviation between different epochs taken with the same filter. Results are listed in Table~\ref{t:mass}.

%\begin{landscape}
\begin{table*}
\caption{List of astrometric points from IRDIS observations.}
\label{t:astroirdis}
\centering
\begin{tabular}{lcccc}
\hline
\hline
%Date &  \multicolumn{2}{c}{Planet b} & \multicolumn{2}{c}{Planet c} & \multicolumn{2}{c}{Planet d} & \multicolumn{2}{c}{Planet e}\\
Date &  Planet b & Planet c & Planet d &Planet e\\
 & $\Delta$RA ; $\Delta$Dec (mas)  & $\Delta$RA ; $\Delta$Dec (mas)  & $\Delta$RA  ; $\Delta$Dec (mas) &  $\Delta$RA  ; $\Delta$Dec (mas)  \\
\hline
2014-07-12              &       1570    $\pm$   3       ;       704     $\pm$   3       &       -521    $\pm$   3       ;       790     $\pm$   9       &       -391    $\pm$   2       ;       -530    $\pm$   2       &       -387    $\pm$   2       ;       -10     $\pm$   3       \\
2014-12-04              &       1574    $\pm$   3       ;       701     $\pm$   2       &       -514    $\pm$   3       ;       798     $\pm$   4       &       -399    $\pm$   4       ;       -525    $\pm$   4       &       -389    $\pm$   8       ;       11      $\pm$   4       \\
2014-12-05              &       1574    $\pm$   4       ;       701     $\pm$   3       &       -512    $\pm$   3       ;       798     $\pm$   4       &       -400    $\pm$   4       ;       -523    $\pm$   4       &       -390    $\pm$   7       ;       12      $\pm$   4       \\
2014-12-06              &       1573    $\pm$   3       ;       701     $\pm$   3       &       -512    $\pm$   3       ;       797     $\pm$   4       &       -403    $\pm$   4       ;       -524    $\pm$   4       &       -383    $\pm$   8       ;       11      $\pm$   4       \\
2015-07-03              &       1579    $\pm$   1       ;       694     $\pm$   1       &       -498    $\pm$   1       ;       806     $\pm$   1       &       -417    $\pm$   1       ;       -517    $\pm$   1       &       -383    $\pm$   9       ;       33      $\pm$   5       \\
2015-07-30              &       1580    $\pm$   5       ;       689     $\pm$   3       &       -495    $\pm$   2       ;       806     $\pm$   2       &       -419    $\pm$   2       ;       -516    $\pm$   1       &       -386    $\pm$   1       ;       36      $\pm$   1       \\
2015-09-27              &       1580    $\pm$   1       ;       688     $\pm$   1       &       -494    $\pm$   1       ;       811     $\pm$   1       &       -426    $\pm$   1       ;       -512    $\pm$   1       &       -382    $\pm$   9       ;       50      $\pm$   5       \\
2016-11-17              &       1589    $\pm$   2       ;       666     $\pm$   1       &       -464    $\pm$   2       ;       824     $\pm$   2       &       -454    $\pm$   2       ;       -490    $\pm$   2       &       -378    $\pm$   4       ;       90      $\pm$   2       \\
2017-06-14              &       1591    $\pm$   1       ;       653     $\pm$   1       &       -449    $\pm$   1       ;       835     $\pm$   1       &       -472    $\pm$   2       ;       -482    $\pm$   2       &       -373    $\pm$   3       ;       118     $\pm$   2       \\
2017-10-11              &       1595    $\pm$   1       ;       647     $\pm$   1       &       -441    $\pm$   1       ;       839     $\pm$   1       &       -480    $\pm$   1       ;       -477    $\pm$   1       &       -369    $\pm$   1       ;       128     $\pm$   1       \\
2017-10-12              &       1595    $\pm$   1       ;       647     $\pm$   1       &       -441    $\pm$   1       ;       839     $\pm$   1       &       -480    $\pm$   1       ;       -477    $\pm$   1       &       -371    $\pm$   2       ;       128     $\pm$   2       \\
2018-06-18              &       1601    $\pm$   1       ;       635     $\pm$   1       &       -424    $\pm$   1       ;       848     $\pm$   1       &       -497    $\pm$   2       ;       -463    $\pm$   2       &       -358    $\pm$   2       ;       156     $\pm$   2       \\
2018-08-17              &       1601    $\pm$   2       ;       632     $\pm$   3       &       -421    $\pm$   1       ;       850     $\pm$   1       &       -502    $\pm$   1       ;       -461    $\pm$   1       &       -357    $\pm$   1       ;       162     $\pm$   2       \\
2018-08-18              &       1600    $\pm$   1       ;       632     $\pm$   1       &       -421    $\pm$   1       ;       851     $\pm$   1       &       -502    $\pm$   2       ;       -458    $\pm$   1       &       -358    $\pm$   2       ;       163     $\pm$   1       \\
2019-10-31              &       1606    $\pm$   2       ;       615     $\pm$   2       &       -392    $\pm$   2       ;       875     $\pm$   2       &       -532    $\pm$   3       ;       -425    $\pm$   2       &       -338    $\pm$   2       ;       215     $\pm$   2       \\
2019-11-01              &       1611    $\pm$   1       ;       611     $\pm$   1       &       -388    $\pm$   2       ;       870     $\pm$   2       &       -530    $\pm$   2       ;       -430    $\pm$   1       &       -335    $\pm$   1       ;       210     $\pm$   1       \\
2021-08-20              &       1626    $\pm$   1       ;       578     $\pm$   2       &       -339    $\pm$   2       ;       890     $\pm$   2       &       -563    $\pm$   2       ;       -391    $\pm$   2       &       -287    $\pm$   4       ;       272     $\pm$   2       \\

\hline
\end{tabular}
\end{table*}
%\end{landscape}

\begin{table*}
\caption{List of astrometric points from IFS observations.}
\label{t:astroifs}
\centering
\begin{tabular}{lcc}
\hline
\hline
%Date &  \multicolumn{2}{c}{Planet b} & \multicolumn{2}{c}{Planet c} & \multicolumn{2}{c}{Planet d} & \multicolumn{2}{c}{Planet e}\\
Date &   Planet d &Planet e\\
 & $\Delta$RA ; $\Delta$Dec (mas)  & $\Delta$RA ; $\Delta$Dec (mas)   \\
\hline

2014-07-12 &    -400    $\pm$   4       ;       -512    $\pm$   4       &       -389    $\pm$   1       ;       -22     $\pm$   2       \\
2014-08-13 &    -396    $\pm$   1       ;       -524    $\pm$   1       &       -389    $\pm$   1       ;       -17     $\pm$   2       \\
2015-07-03 &    -424    $\pm$   4       ;       -509    $\pm$   3       &       -391    $\pm$   1       ;       33      $\pm$   2       \\
2015-09-27 &    -420    $\pm$   4       ;       -513    $\pm$   4       &       -392    $\pm$   1       ;       39      $\pm$   3       \\
2016-11-17 &    -464    $\pm$   1       ;       -486    $\pm$   2       &       -382    $\pm$   2       ;       94      $\pm$   6       \\
2017-06-14 &    -473    $\pm$   2       ;       -476    $\pm$   2       &       -377    $\pm$   1       ;       115     $\pm$   3       \\
2017-10-11 &    -480    $\pm$   2       ;       -478    $\pm$   2       &       -371    $\pm$   1       ;       129     $\pm$   2       \\
2017-10-12 &    -492    $\pm$   5       ;       -463    $\pm$   6       &       -370    $\pm$   2       ;       135     $\pm$   3       \\
2018-06-18 &    -495    $\pm$   1       ;       -460    $\pm$   2       &       -360    $\pm$   2       ;       162     $\pm$   2       \\
2018-08-17 &    -509    $\pm$   2       ;       -452    $\pm$   3       &       -361    $\pm$   2       ;       166     $\pm$   1       \\
2018-08-18 &    -503    $\pm$   1       ;       -456    $\pm$   2       &       -359    $\pm$   1       ;       167     $\pm$   2       \\
2019-10-31 &    -527    $\pm$   3       ;       -432    $\pm$   3       &       -337    $\pm$   2       ;       210     $\pm$   3       \\
2019-11-01 &    -528    $\pm$   1       ;       -435    $\pm$   2       &       -337    $\pm$   2       ;       208     $\pm$   1       \\
2021-08-20 &    -569    $\pm$   1       ;       -390    $\pm$   2       &       -292    $\pm$   2       ;       276     $\pm$   2       \\

\hline
\end{tabular}
\end{table*}

\begin{table*}
\caption{Values for the mass of each planet calculated using the evolutionary models in the different IRDIS filters. The age of the system is assumed to be 42 Myr. The error is the standard deviation between different observations with the same filter. }
\label{t:mass}
\centering
\begin{tabular}{lcccc}
\hline
\hline
%Date &  \multicolumn{2}{c}{Planet b} & \multicolumn{2}{c}{Planet c} & \multicolumn{2}{c}{Planet d} & \multicolumn{2}{c}{Planet e}\\
Filter &   Planet e (\MJup) &Planet d (\MJup) & Planet c (\MJup) & Planet b (\MJup)\\
\hline
BB\_H   &               8.9 $\pm$       1.1 &   8.4$\pm$        0.9&    8.5$\pm$        0.9&    6.4$\pm$        0.6 \\
D\_H2 &7.0$\pm$ 0.3&    6.6$\pm$        0.3&    6.7$\pm$        0.3&    4.6$\pm$        0.3  \\
D\_H3& 8.5$\pm$ 0.3&    8.1$\pm$        0.2&    8.3$\pm$        0.3&    6.8$\pm$        0.3 \\
D\_K1 & 9.6$\pm$        0.3&    9.3$\pm$        0.3&    9.3$\pm$        0.2&    6.7$\pm$        0.3 \\
D\_K2 & 10.5$\pm$       0.1&    10.5$\pm$       0.1&    10.4$\pm$       0.1&    9.0$\pm$        0.3 \\
\hline
\end{tabular}
\end{table*}

\subsection{SOUL+LUCI}

The data were taken with a randomized offset of the position of the PSF on the detector. This was
used to create a background obtained by calculating the median of all the frames. This background was then
subtracted from each science frame. The final dataset was composed of 60 and 61 files for the $FeII$
and $H2$ observations, respectively. In both cases, each file was composed of 112 frames for a total
exposure time of 2284.80~s and 2322.88~s, respectively. During the observations, the derotator was switched off to allow the rotation of the FOV and be able to use the ADI method. Before performing the high-contrast imaging method, we obtained the position of the stellar PSF for each frame using the FIND procedure as
implemented for IDL. Exploiting these positions, we then precisely registered the whole dataset, positioning the stellar PSFs at the center of each frame. The ADI was then implemented by exploiting the principal component analysis \citep[PCA; ][]{2012ApJ...755L..28S} technique. For the reduction, we tested a
different number of principal components, but found that the best solution was to use ten of them.
We were able to recover all the known companions with S/N ranging between 7 and 10. \par
We obtained the precise position of each planet, introducing fake negative companions
and changing its position to minimize the standard deviation in a small region 
around the planet itself. The results of this procedure are listed in Table~\ref{t:luci}.

%\begin{landscape}
\begin{table*}
\caption{List of astrometric points from the LBT/LUCI observation.}
\label{t:luci}
\centering
\begin{tabular}{lcccc}
\hline
\hline
%Date &  \multicolumn{2}{c}{Planet b} & \multicolumn{2}{c}{Planet c} & \multicolumn{2}{c}{Planet d} & \multicolumn{2}{c}{Planet e}\\
Date &  Planet b & Planet c & Planet d &Planet e\\
 & $\Delta$RA ; $\Delta$Dec (mas)  & $\Delta$RA ; $\Delta$Dec (mas)  & $\Delta$RA  ; $\Delta$Dec (mas) &  $\Delta$RA  ; $\Delta$Dec (mas)  \\
\hline
2020-09-29 & 1620       $\pm$ 1  ; 591  $\pm$ 3  &         -364         $\pm$   2 ; 883   $\pm$   1  &    -551    $\pm$   1 ;     -415    $\pm$   4 &     -315            $\pm$ 3 ; 242         $\pm$ 5 \\

\hline
\end{tabular}
\end{table*}
%\end{landscape}

\section{Astrometric fit of the orbits}
\label{Sec:astro}
\subsection{Dynamical constraints on the astrometry}

Given the literature regarding astrometric fits of the \hr8799{} system \cite[e.g.,][and references therein]{2018AJ....156..192W,GM2020}, it is now widely recognized that the present observation time-window does not make it possible to determine long-term stable orbital solutions of HR \hr8799 without invoking particular dynamical constraints.  Such constraints may arise from the coplanarity of the planets' orbits and their relatively small eccentricities, which implies a ratio of orbital periods close to 2:1 for successive pairs of planets, that is, 2:1 mean-motion resonances (MMRs). This assumption can be further supported by the likely origin of the system from planetary migration  \cite[e.g.,][]{2018AJ....156..192W}; see also Sect.~\ref{Sec:his} in this paper.  Also, the recent analysis of high-contrast images and resulting orbital solutions in \cite{2021A&A...648A..26W} indicate that planets \hr8799{}e and \hr8799{}c are most consistent with co-planar and resonant orbits. The dynamical multiple MMR scenario therefore seems to be well justified and documented in the literature.  Recently, \cite{GM2020} (\gm\  hereafter) reported an exact-resonance configuration (or ``periodic orbits
model'' (PO) hereafter). This model is designed to explain the astrometric measurements through constraints on geometry, planetary masses, and astrometric parallax of the system. Direct determination of planetary masses is crucial for calibrating astrophysical cooling models and for determining the origin and long-term orbital evolution of the entire system, including its debris disk components.

Here we consider a less stringent condition on the stability of the system, assuming that it may be close to exact resonance but not necessarily fully periodic. This assumption may be natural in the sense that migration may result in a system that is not exactly in a configuration of PO \cite[e.g.][]{2017A&A...602A.101R}. We want to determine whether such near-resonance models can yield statistically different or perhaps even better best-fitting solutions than those that appear in the PO scenario.  
We define{ the following merit functions as in} 
%the posterior the same, as in
\cite{GM2018,GM2020}:
\begin{eqnarray}
%\label{eq:L}
 \ln {\cal L}(\vec{x}) & = &
-\frac{1}{2} \chi^2(\vec{x}) 
-\frac{1}{2}\sum_{i=1}^{N_{\idm{obs}}}\left[ \ln \theta^2_{i,\alpha} + \ln \theta^2_{i,\delta} \right]- N_{\idm{obs}} \ln (2\pi),  
\nonumber
\\ 
\label{eq:L}
%\label{eq:chi2}
\chi^2(\vec{x}) &=& \sum_{i=1}^{N_{\idm{obs}}}
\left[ \frac{[\alpha_i-\alpha(t_i,\vec{x})]^2}{\theta^2_{i,\alpha}} +
       \frac{[\delta_i-\delta(t_i,\vec{x})]^2}{\theta^2_{i,\delta}} \right],
\end{eqnarray}
where $(\alpha_i,\delta_i)$ are the right ascension (\RA) and declination (\DEC) \ measurements at time $t_i$; $\alpha(t_i,\vec{x}),\delta(t_i,\vec{x})$ are for their ephemeris (model) values   at the same moments as implied by the adopted $\vec{x}$ parameters; and  $\theta^2_{i,\alpha}$ and $\theta^2_{i,\delta}$ are the nominal measurement uncertainties in \RA{} and \DEC{} scaled in quadrature with the so-called error floor,  $\theta^2_{i,\alpha}=(\sigma^2_{i,\alpha} + \sigma^2_{\alpha,\delta} )$ and  $\theta^2_{i,\delta}=(\sigma^2_{i,\delta} + \sigma^2_{\alpha,\delta} )$,  for each datum, respectively. Also, if $N_{\idm{obs}}$ is the number of observations then $N=2 N_\idm{obs}$, because \RA{} and \DEC{} are measured in a single detection.  The error floor can be introduced for unmodeled measurement uncertainties,  such that the resulting best-fitting model should yield the reduced $\chi_\nu^2 \simeq 1$.  However, in this work,  as we rely on MCMC sampling and the error floor introduces little qualitative variability into the solutions, we omit this parameter, thereby also simplifying the astrometric model.

\subsection{Mass and parallax priors}
\label{sec:priors}

We conducted MCMC sampling from the posterior defined through the general merit function in Eq.~\ref{eq:L} and astrophysical and geometrical priors. The most important astrophysical priors are the masses of the star and the planets, and the geometrical prior is the parallax.

Regarding the stellar mass, we considered two fixed values of $m_{\star}=1.52\,\msun$ determined by \cite{2012ApJ...761...57B} as 
$m_\star=1.516^{+0.038}_{-0.024}\,\msun$ for the age of $\simeq 33$~Myr, and  $m_{\star}=1.47^{+0.12}_{-0.17}\,\msun$, following the most recent dynamical estimate by \citet{2022AJ....163...52S}. The same value was used by \cite{2018AJ....156..192W} in their earlier work. \cite{2022A&A...657A...7K} determined $m_\star=1.50 \pm 0.08\,\msun$, which overlaps with the two estimates. We note here that the stellar mass is fixed in our astrometric $N$-body model because it also constrains the parallax $\Pi$ of the whole system. These two parameters are strongly correlated through the near-III Keplerian law. Moreover, the parallax tends to be systematically tightly bounded in subsequent GAIA catalogs. The most recent GAIA eDR3 estimate of $\Pi = 24.460 \pm 0.045$~mas is accurate to $0.1\%$, and it is a meaningful prior for the MCMC sampling. {We note that the parallax did not change in the final Gaia DR3 catalog; now it is listed as $\Pi = 24.462 \pm 0.046$~mas.}

We also implied Gaussian priors ${\cal N}_m=[ 8.7 \pm 1.7, 8.7 \pm 1.0, 8.7 \pm 1.0, 6 \pm 0.7]~$\MJup for the masses of the innermost to the outermost planet, respectively, based on the hot-start cooling models by \cite{2003A&A...402..701B},  following discussion in \cite{2018AJ....156..192W} and confirmed by our estimates collected in Table~\ref{t:mass}. The mass ranges for BB\_H are consistent with the priors in \cite{2018AJ....156..192W}, and D\_H3, D\_K1 overlap with the determinations from the earlier astrometric model in \gm{}. As also demonstrated by \gm{} and confirmed below, there are many discrepancies regarding the mass hierarchy; according to the resonant model, \hr8799d{} is the most massive one, and the masses of \hr8799{}e and \hr8799{}c appear strongly anti-correlated. Masses for D\_K2 seem to be too far into the high-end range regarding the dynamical stability, especially that of planet \hr8799{}b.

It is common in the literature to assume smaller planet masses, that is, of $\simeq 7\,$\MJup, given the results of dynamical $N$-body simulations. In particular, \cite{2018AJ....156..192W} report difficulty in finding dynamically stable solutions for planet masses larger than $\simeq 7\,$\MJup. Simultaneously, \cite{GM2018,GM2020} found rigorously stable systems locked in the Laplace resonance for higher mass { ranges as well} that is, of, $\simeq 8$--$9\,$\MJup. Given systematic observationally constrained{ shifts} to higher masses, as predicted by the exact Laplace MMR model, we decided to apply the mass priors in the $\simeq 9~$\MJup range. \cite{2021ApJ...915L..16B} 
%{ proposed a mass of $9.6^{+1.9}_{-1.8}~$\MJup for \hr8799{}e} 
estimated the mass of { \hr8799{}e as} $9.6^{+1.9}_{-1.8}~$\MJup 
based on the secular variation of the proper motion of the parent star in GAIA and Hipparcos catalogs. 
Moreover, in a very recent work, 
\cite{2022A&A...657A...7K} similarly determined  the dynamical mass of $12 \pm 3.5~$\MJup for \hr8799{}e. This latter  is less precise than in the prior work of \cite{2021ApJ...915L..16B}, because Kervella et al. (2022) did not account for the actual orbital geometry. These high mass ranges for \hr8799{}e are consistent with the larger mass priors. Therefore, in some experiments we  also used the mass priors ${\cal N}_m$ with the innermost mass changed to the value computed in \cite{2021ApJ...915L..16B}.

As the parallax prior, we defined the most %the most
recent GAIA {eDR3/DR3} estimate of $\Pi = 24.46 \pm 0.05$~mas. This value is significantly shifted from $\Pi \simeq 24.22 \pm 0.08$~mas in the GAIA DR2 and $24.8 \pm 0.7$~{ mas} in GAIA DR1 { catalog}, respectively. The parallax determinations in the GAIA catalogs appear subtly biased, depending on the luminosity and spectral type \citep{Lindegren2021}. However, compared to the uncertainties of the dynamical estimates here, the predicted parallax correction for GAIA { eDR3/DR3} of $<0.1$~mas (a~few tens of $\mu$as) would be insignificant. Indeed, the parallax correction predicted by \cite{Lindegren2021} yields $24.50 \pm 0.05$~mas \citep{2022A&A...657A...7K}. This apparently very small shift of the order of $1\sigma$ uncertainty may still
be meaningful when compared to the dynamical estimates, as is found below.

\subsection{The exact Laplace resonance revisited}
\label{sec:PO}

In the first step, we verified whether or not the resonant model in \gm{} based on measurements in \cite{Konopacky2016} fits the recent measurements listed in Tables~\ref{t:astroirdis},~\ref{t:astroifs}, and~\ref{t:luci}. We found that this model visually and significantly deviates from the updated data set, especially for planet \hr8799{}b. Therefore, we refined the PO model using the same 
parametrization as for the merit function; see Eq.~\ref{eq:L} in \gm{}. In the formulation presented by these latter authors, the primary parameters of the astrometric model consist of all planet masses $m_i$, $i=1,\ldots,4$, the osculating period ratio $\kappa$ for the two innermost planets ---which selects the particular four-body Laplace resonance chain---, as well as the reference epoch $\tau$, three Euler angles $(I,\omega_{\rm rot},\Omega)$ rotating the coplanar resonant configuration to the sky (observer) plane, and the system parallax~$\Pi$. This means 11~free parameters. We note that all remaining orbital parameters are constrained through the $N$-body dynamics confined to the manifold of strictly periodic solutions. 

We performed the MCMC sampling from the posterior defined with the merit function in Eq.~\ref{eq:L}. This merit function is based on a numerical procedure for computing periodic solutions in a co-planar $N$-planet system, as described in \gm{} and implemented by Cezary Migaszewski (private communication). To conduct the MCMC sampling, we used the affine sampler in {\tt emcee} package \citep{ForemanMackey2013}. We initiated 1536 walkers in a small hyper-ball in the parameter space around the initial condition in Table~\ref{tab:tab1} determined with a search with the Powell non-gradient minimization algorithm. Since the auto-correlation time appears relatively short, of $\sim 100$ iterations only, we ended up with $1000$ iterations that make it possible to derive the posterior distribution, given the large number of walkers.

\begin{figure*}
\centerline{ 
\hbox{
\includegraphics[width=1\textwidth]{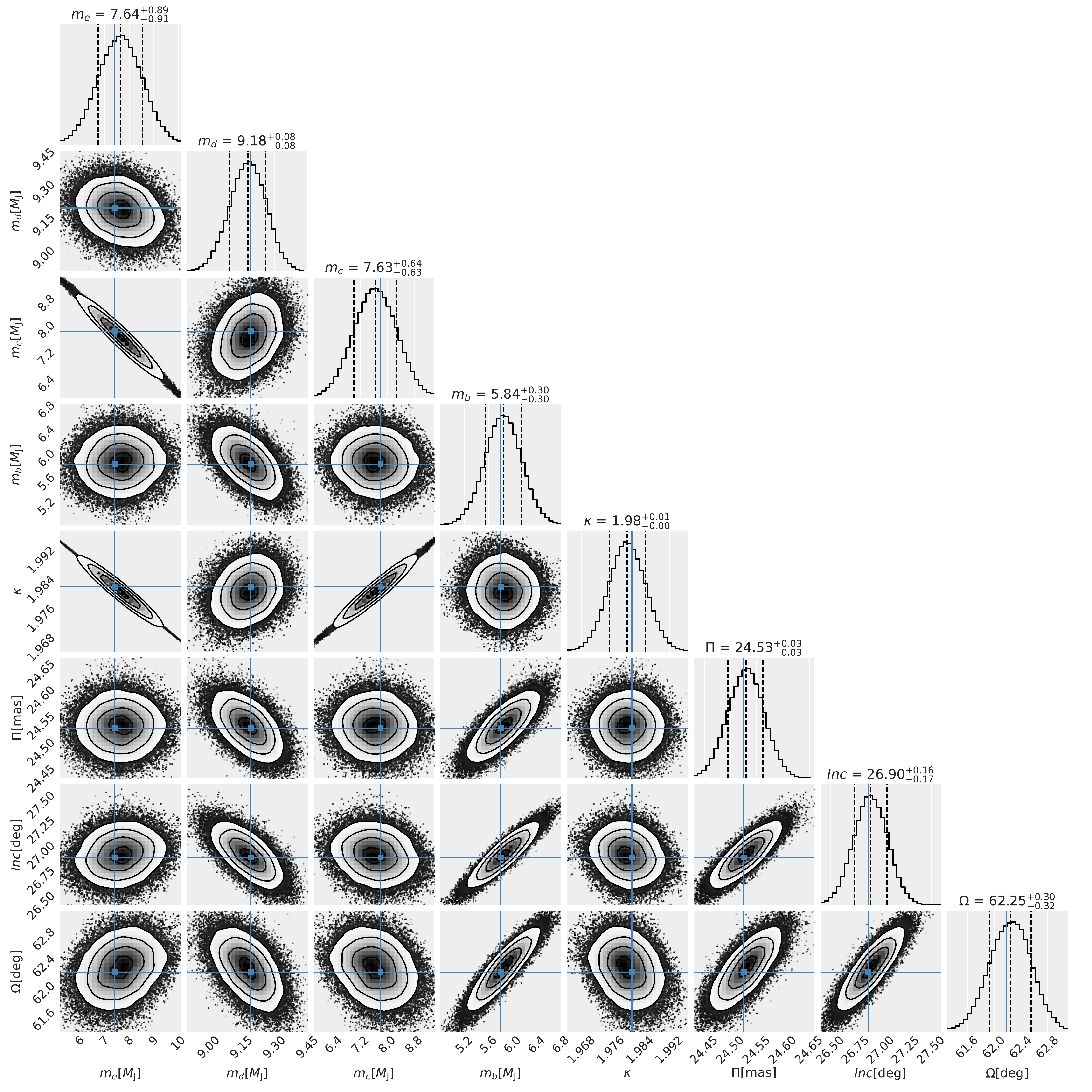}
}
}
\caption{
MCMC posterior for the best-fitting, strictly resonant model derived for $m_{\star}=1.47\,$\MJup. Parameters included in the diagram are for the planet masses and orbital period ratio $\kappa = P_\idm{d}/P_\idm{e} = \kappa$, and parallax $\Pi$, inclination $I,$ and the nodal angle $\Omega$ (the orbital scale $\rho$, PO epoch shift $t_0$ and the common rotation angle $\omega_\idm{rot}$ in the orbital planet are skipped). The crossed lines mark the best-fitting PO solution in terms of the smallest $\Chi$ that was used as a starting point to initiate the {\tt emcee} walkers; see Tables~\ref{tab:tab1} and~\ref{tab:tab3} (Model~1) for parameters of this solution. 
{ Uncertainties for the parameters are determined as the 16th and 86th percentile of the samples around the median values.}
}
\label{fig:fig1}
\end{figure*}

The derived posterior is illustrated in 2-dim projections and 1-dim histograms of the MCMC samples for the primary parameters (Fig.~\ref{fig:fig1}). This experiment reveals significant correlations between different parameters already reported in \gm{} that still cannot be reduced with the new data. Besides a~strong $m_\idm{e}-m_\idm{c}$ anti-correlation, there is also  $m_\idm{e}-\kappa$ correlation and $m_\idm{d}-\kappa$ anti-correlation. The remaining two masses are free from significant mutual correlations, however, the mass of \hr8799{}b appears strongly correlated with geometric parameters of the system, i.e., the Euler angles and the parallax. These parameters show mutually weaker but still significant correlations.

Parameters of the best-fitting solutions with their formal uncertainties and particular solutions from the posterior that may be useful for future numerical studies are collected in Table~\ref{tab:tab1}, referred to as Model IV$_{\rm PO}$ for the two stellar masses of $m_{\star}=1.52\msun$ and  $m_{\star}=1.47\msun$, respectively.
 
For a stellar mass of $m_{\star}=1.47\msun$, the \hr8799{}d mass seems to be constrained to $\simeq 9.2\,$\MJup within a relatively very small uncertainty of $\simeq 0.1\,$\MJup. This mass estimate is slightly larger than that found by \gm{}. The mass of the outermost planet \hr8799{}b may be determined as $\simeq 5.8$\MJup which is similar to  $\simeq 5.5\,$\MJup found by these latter authors, yet with also smaller uncertainty of $\simeq 0.3\,$\MJup. Moreover, for the larger stellar mass of $1.52\msun$, we derived slightly larger masses of \hr8799{}d  $\simeq 9.6 \pm 0.1 $\MJup and  \hr8799{}b $\simeq 6.0 \pm 0.4 $\MJup. 

The astrometric model for the updated data window yields $\Pi \simeq 24.26 \pm 0.05$~mas, which is consistent with the earlier estimate by \gm{}. The parallax estimate of $\Pi \simeq 24.53 \pm 0.04$~mas derived for $1.47\msun$ overlaps to $1\sigma$ with $\Pi \simeq 24.460 \pm 0.045$~mas in GAIA { eDR3/DR3}. Moreover, as noted above, when applying the parallax correction of the \cite{Lindegren2021} { eDR3 catalog}, \cite{2022A&A...657A...7K} obtained $\Pi = 24.50 \pm 0.05$ mas which is even closer to the value derived from the PO model. This relation confirms the predicted strong stellar mass--parallax correlation and indirectly favors $m_{\star}=1.47\,\msun$. The dynamical parallax determinations appear to be very accurate and meaningful, overlapping closely with the independent geometric GAIA measurements. 

Regarding the \hr8799{}e--\hr8799{}d mass anti-correlation, we conducted additional
MCMC sampling for the primary parameters in thePO model for fake data series. We prepared two or three  synthetic observations per year, extending the real observations window by $\simeq 20$~years around the best-fitting periodic model with deviations and uncertainties of $\sim 1$~mas. It appears that all correlations except those between the masses and $\kappa$ would be almost eliminated. A~probable cause of that effect may be the almost perfectly aligned gravitational tugs of these planets in the present particular time window of the observations. If the PO model is correct, then likely only highly precise data similar to the most accurate GRAVITY measurement in \cite{2019AA...623L..11G} could break this degeneracy.  This was indicated by \gm{}. We note that the GRAVITY measurement is depicted with a star in all figures illustrating orbital solutions.

\subsection{Near 8:4:2:1 MMR model}
To characterize the system near the 2:1~MMR chain, we rely on the notion of the proper mean motions as the fundamental frequencies $f_i$ in the framework of conservative $N$-body dynamics. We resolve these frequencies with the refined Fourier frequency analysis \cite[e.g.,][Numerical Analysis of Fundamental Frequencies, NAFF]{Laskar1993,Nesvorny1997}. To perform the NAFF, we consider the time series
\[
{\cal S}_{\lambda,i} = \{\, a_i(t_j) \, \exp \mbox{ i} \lambda_i(t_j) \,\}, \quad
{\cal S}_{{\cal M},i} = \{\, a_i(t_j) \, \exp \mbox{ i} {\cal M}_i(t_j) \,\}, 
\]
where $j=0,1,\ldots$, $t_j = j\Delta t$, and $a_i(t)$, $\lambda_i(t)$ and ${\cal M}_i$ are the canonical osculating semi-major axis, the mean longitude, and the mean anomaly for planet~$i$, $i=1,2,3,4$, respectively, and $\Delta{}t$ is the sampling interval ($\mbox{i}$ is the imaginary unit). These canonical elements inferred in the Jacobi or Poincar\'e reference frame make it possible to account for mutual interactions between the planets to the first order in the masses.

\begin{figure*}
\centerline{ 
%\vbox{
\hbox{
%\includegraphics[width=0.36\textwidth]{figs/hr8799eAE.pdf}
%\qquad
\includegraphics[width=0.33\textwidth]{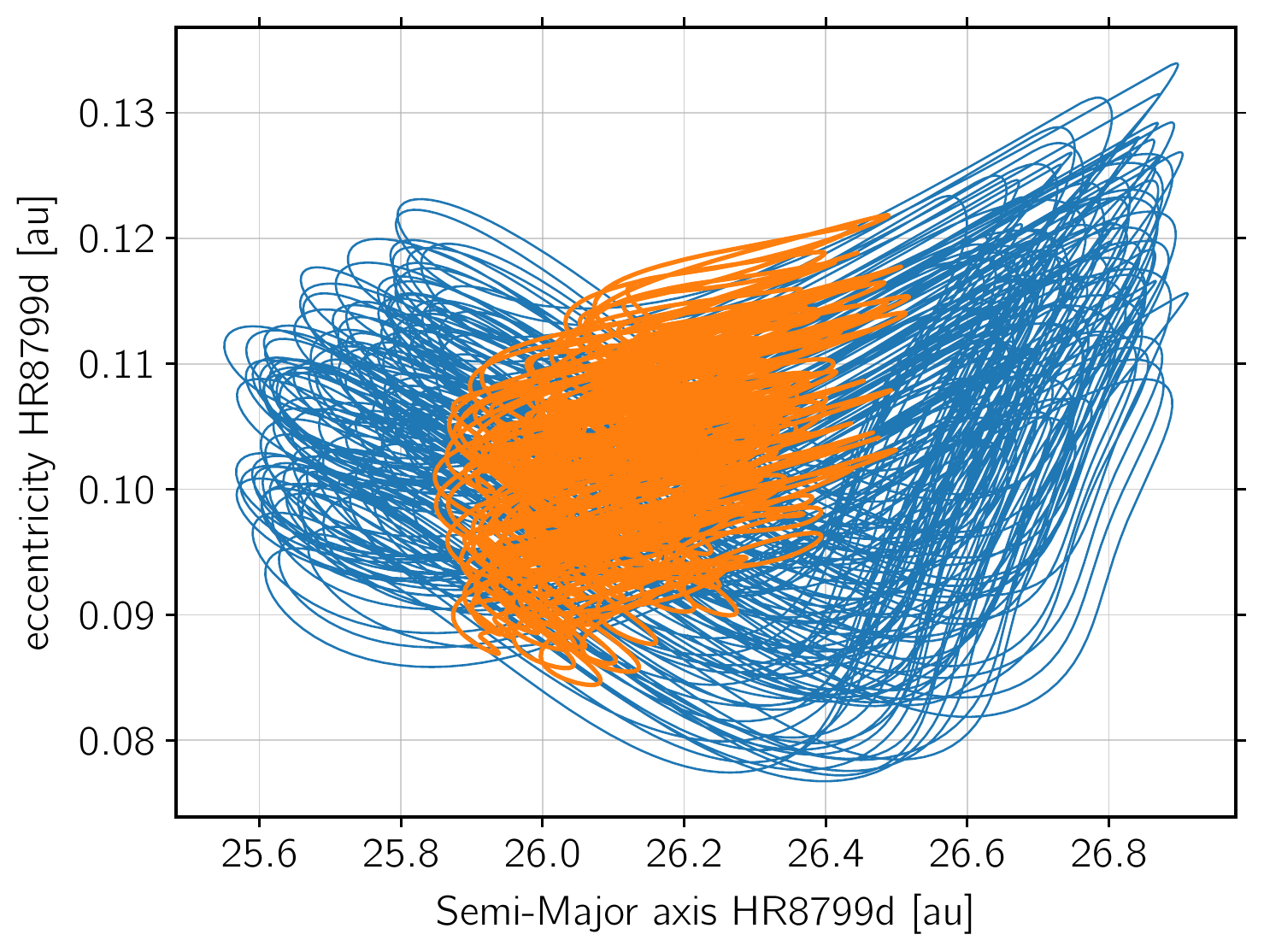}%hr8799dAE.pdf}
%\hbox{
\includegraphics[width=0.33\textwidth]{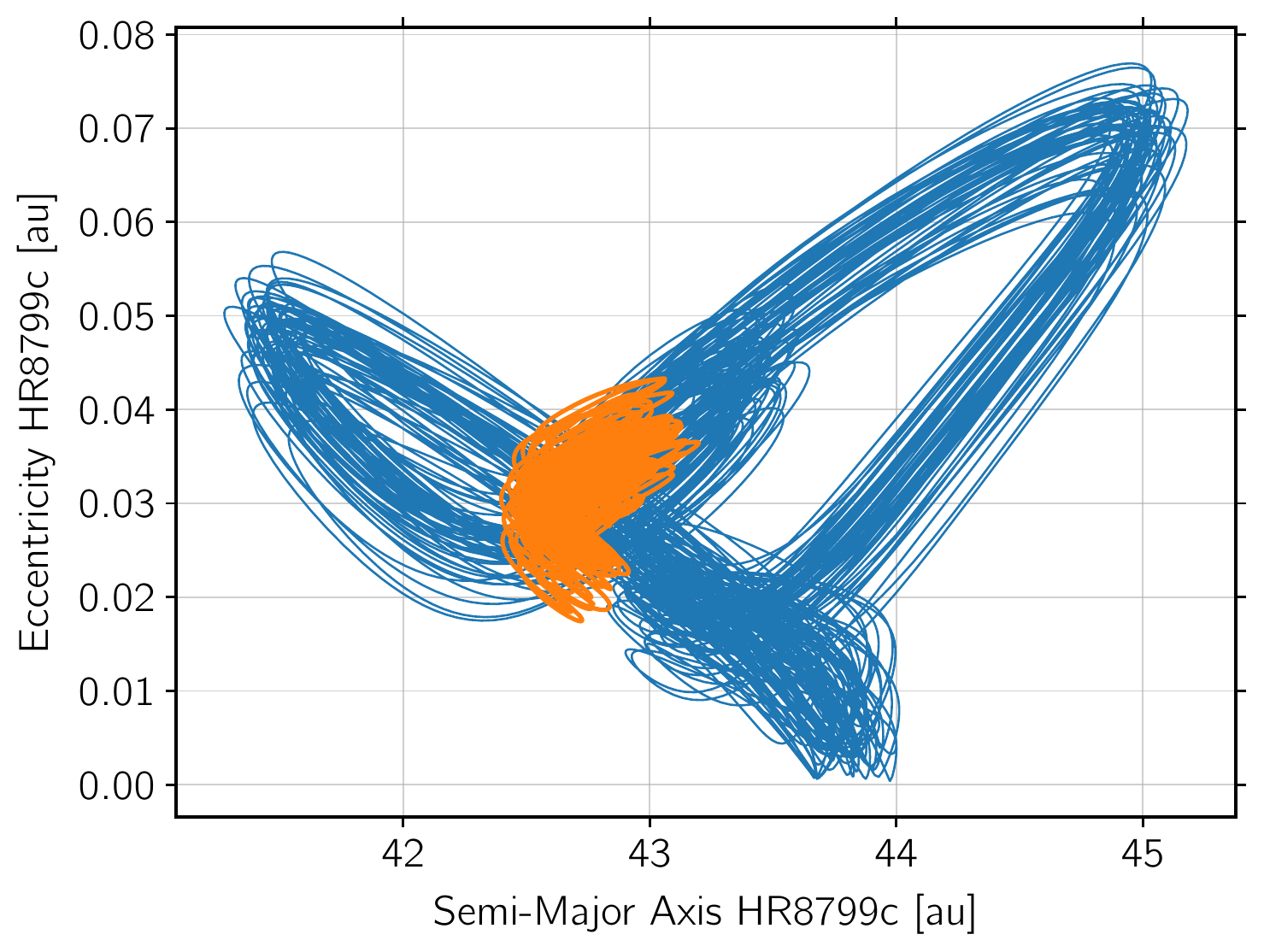}%hr8799cAE.pdf}
%\quad
\includegraphics[width=0.33\textwidth]{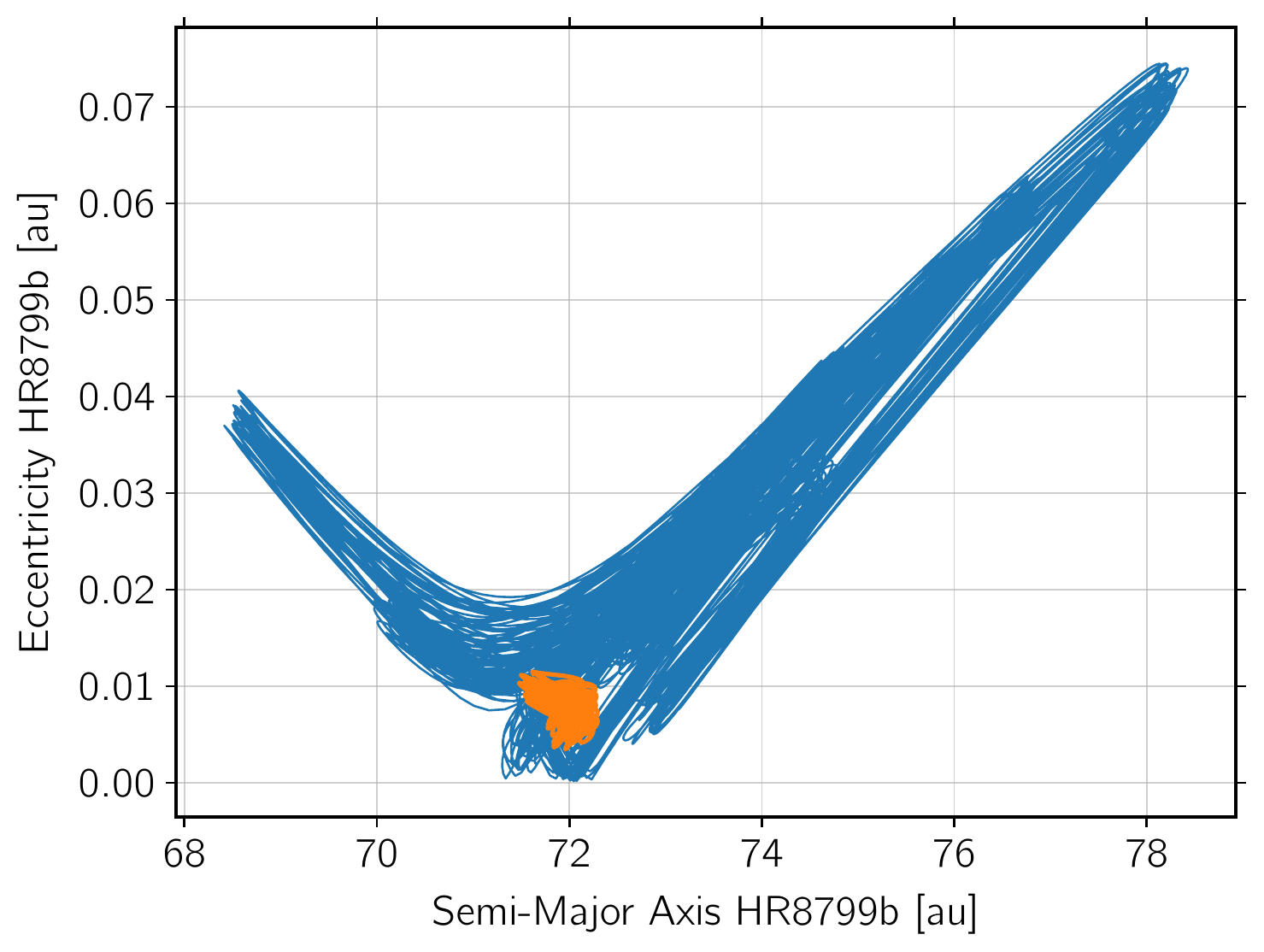}%hr8799bAE.pdf}
%}
}
}
\caption{
Temporal evolution of the osculating orbital elements in the astrocentric (blue curves) and Jacobian (yellow curves) frames, respectively, for the same $N$-body initial condition implying a stable, near 8:4:2:1~MMR system. We note that the elements overlap for the first planet according to the construction of the Jacobian reference frame.
}
\label{fig:fig2}
\end{figure*}

We note that the canonical orbital elements evolve differently from the common astrocentric elements. This is illustrated in Fig.~\ref{fig:fig2}. Subsequent panels are for the semi-major axis and eccentricity of  \hr8799{}d,c,b, respectively,  computed for the interval of a~few tens of outermost orbits, and marked with different colors. It turns out that these elements span much wider ranges in the common astrocentric Keplerian frame. This effect is particularly noticeable for the two outermost planets. A small change in the $N$-body initial condition may result in substantial displacement of the orbit in the Keplerian $(a,e)$--plane. Therefore, the geometric elements should be understood as a formal representation of the Cartesian $N$-body initial conditions, and the canonical elements are more suitable for the qualitative characterization of the orbits.

%In numerical experiments, in order to compute the proper mean-motions for all planets, we usually integrated the $N$-body equations of motion for a few multiples of 8192 intervals of 2048 or 3072~days that cover $\simeq 100$--$200$ outermost orbits.

Considering a particular four-body MMR chain that can explain astrometric observations of the \hr8799{} system, the zeroth-order mean-motion resonance implies one of the possible linear combinations of the fundamental frequencies (the proper mean-motions),
\begin{equation}
     \Delta f (\vec{k})\equiv \sum_{i=1}^N k_i f_i,
\end{equation}
where $\vec{k} \equiv [k_1,\ldots,k_N]$ is a vector of non-zero integers, and $\sum_{i=1}^N |k_i| \neq 0$, which yields small $|\Delta f|$.  Simultaneously, we may determine the critical angle corresponding to $\vec{\eta} = \mbox{arg} \min |\Delta f (\vec{n})|$:

\begin{equation}
 \theta_{\vec{\eta}}(t) \equiv \sum_{i=1}^N \eta_i \lambda_i,
\end{equation}

if $\Delta{}f(\vec{k})$ accounts only for the proper mean motions according to the { d'Alembert} rule. The resonant configuration may therefore be called the generalized Laplace resonance \citep{Papaloizou2015}. This type of MMR, which  possibly drives the orbital evolution of \hr8799,{} is characterized with the proper mean motions fulfilling the following relation (e.g., \gm{}): 
\begin{equation}
\label{eq:cond0}
  \Delta f (\vec{k}) =  n_1 - 2 n_2 +  n_3 - 2 n_4.
\end{equation}
Similarly to the classic Laplace resonance in the Galilean moons system, we consider the generalized MMR as a chain of two-body MMRs,
\[
  2 n_{i+1} - n_i + g_{j} 
  \simeq  2 n_{i+1} - n_i \simeq 0, \qquad i=1,2,3, \quad j=i,i+1, 
\]
for subsequent pairs of planets. Here,  $g_j$ are for the proper frequencies (mode) associated with the pericenter rotation. In the exact resonance, which we consider as the PO in the reference frame rotating with a selected planet \citep{Hadjidemetriou1976}, the apsidal lines of all planets rotate synchronously, with the same angular velocity relative to the inertial frame. The condition in Eq.~\ref{eq:cond0} for the exact resonance (periodic motion in the rotating frame) may be expressed through the proper mean motions of the planets \citep[e.g.,][]{Delisle2017}:
\begin{equation}
\label{eq:cond1}
 \Delta n \equiv \sum_{i=1}^3 \left( \frac{n_{i}}{n_{i+1}}-2 \right) = 0,
\end{equation}
given the proper mean-motions resolved from the time series $S_{{\cal M},i}$. Numerically, we find that the exact resonance yields $\Delta{}n \simeq 10^{-13}$\drad{} and $\Delta{}f \simeq 10^{-15}$\drad{} for $3\times 8192$ samples of $\Delta{}t =2048$~days.   

We used the resonance constraint in Eq.~\ref{eq:cond1} as one of the priors for the MCMC sampling of the posterior defined through the maximum likelihood function in Eq.~\ref{eq:L}.  This indirect frequency prior is Gaussian with a~small variance $\sigma_{\Delta n}$ to be determined later.
%of the order of $10^{-6}$--$10^{-8}$\drad{}.

We determined the frequency variance $\sigma_{\Delta n}$ based on extensive numerical experiments that relied on calibrating the $\Delta n$ range with the Lagrange (geometric) stability interval, which should not be shorter than the stellar age of $40$--$50$~Myr, as predicted in the most recent work \citep{2021ApJ...915L..16B}. These authors also rule out ages for \hr8799{} of greater than $\approx 300$~Myr.  The proper choice of the $\Delta{}n$ prior is crucial to bound (resolve) a long-term stable system in a very short integration time that spans merely a few hundred of the outermost orbits, $\simeq 2\times 10^{5}$~yr, which is also the instability timescale outside the Laplace resonance. We find, as described below in Sect.~\ref{sec:dyn}, that the astrometric \hr8799{} solutions are long-term stable if $|\Delta n|$ is roughly less than $10^{-5}$--$10^{-6}$\drad{}, as described above.

\subsection{Astrometric fits to near-resonance configurations}
\label{sec:nearfits}
To sample the posterior defined with Eq.~\ref{eq:L}, we adopted priors
described in Sect.~\ref{sec:priors}. We did not imply any other particular limits on the anticipated near-resonant $N$-body solutions, besides uniform priors for orbital parameters in sufficiently wide ranges. In this sense, the assumed prior set is minimal. Moreover, the $N$-body model is parameterized with $(m_i,a_i, x_i\equiv e_i \cos \varpi_i,  y_i \equiv e_i \sin \varpi_i, {\cal M}_i$), $i=1,\ldots,4$, that is, mass, semi-major axis, { Poincar\'e} elements composed of eccentricity and argument of pericenter, and the mean anomaly at the osculating epoch for each orbit, as well as two angles $(I,\Omega)$ determining the orbital plane of the system, and the astrometric parallax $\Pi$. This means 23~free parameters.

It should be noted that the implicitly defined frequency prior (Eq.~\ref{eq:cond1}) makes it difficult to derive the posterior distribution. In many experiments performed with a different number of MCMC walkers of up to 256 and 
up to 256 000 iterations, we could obtain an acceptance ratio as low as 0.1. Therefore, we understand the MCMC sampling with the $\Delta{}n$ prior as dynamically constrained optimization (rejection sampling) which helps to determine parameter ranges for the best-fitting, stable solutions. Also, due to implicit frequency prior definition, which has to be calibrated consistently with the expected dynamical stability of the system, the method has a heuristic character. Nevertheless, we may gain much insight into the parameter ranges of stable near-resonance models consistent with the astrometric observations.

\begin{figure*}
\centerline{ 
\vbox{
\hbox{
\includegraphics[width=1\textwidth]{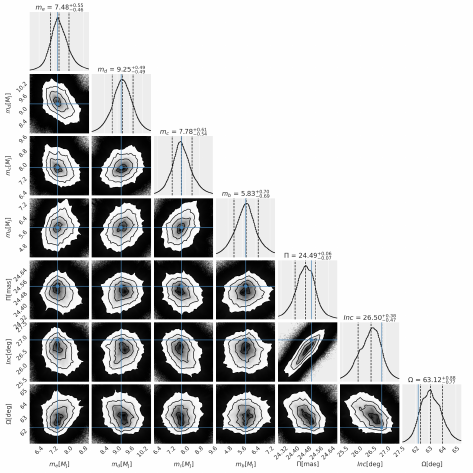}%TriangleFMA147F.pdf}
}
}
}
\caption{
Posterior samples for stable solutions characterized with $|\Delta{}n|<10^{-5}$~rad d$^{-1}$, and the semi-amplitude of the libration of the Laplace resonance argument $|\Delta\theta|<60^{\circ}$. The starting solution that initiated nearby MCMC walkers in the {\emcee{}} samplers is the periodic configuration in~Tables~\ref{tab:tab1} and~\ref{tab:tab3} (Model~1).
{ Uncertainties for the parameters are determined as the 16th and 86th percentiles of the samples around the median values.}
}
\label{fig:fig3}
\end{figure*}

We performed many MCMC experiments varying $\sigma_{\Delta{}n}$ within the range of $10^{-7}$--$10^{-9}$\drad{} and parameters of the {\emcee} samplers. The example posterior for $\sigma_{\Delta{}n}=10^{-8}$\drad{} is illustrated in Fig.~\ref{fig:fig3} for parameters selected consistently with the posterior for PO shown in Fig.~\ref{fig:fig1}. During the sampling, we recorded all elements of the tested models, and the semi-amplitude of the critical angles of the Laplace resonance for the integration time spanning $\simeq 200$ outermost periods. Fig.~\ref{fig:fig3} shows a collection of samples with $|\Delta{}n|<10^{-5}$\drad{}, and the semi-amplitude of Laplace resonance argument $|\Delta\theta|<60^{\circ}$. This choice is explained in the following Section~\ref{sec:dyn}, in which we analyze the orbital evolution of particular selected solutions.

\begin{figure*}
\centerline{ 
\vbox{
\hbox{
\includegraphics[height=0.22\textheight]{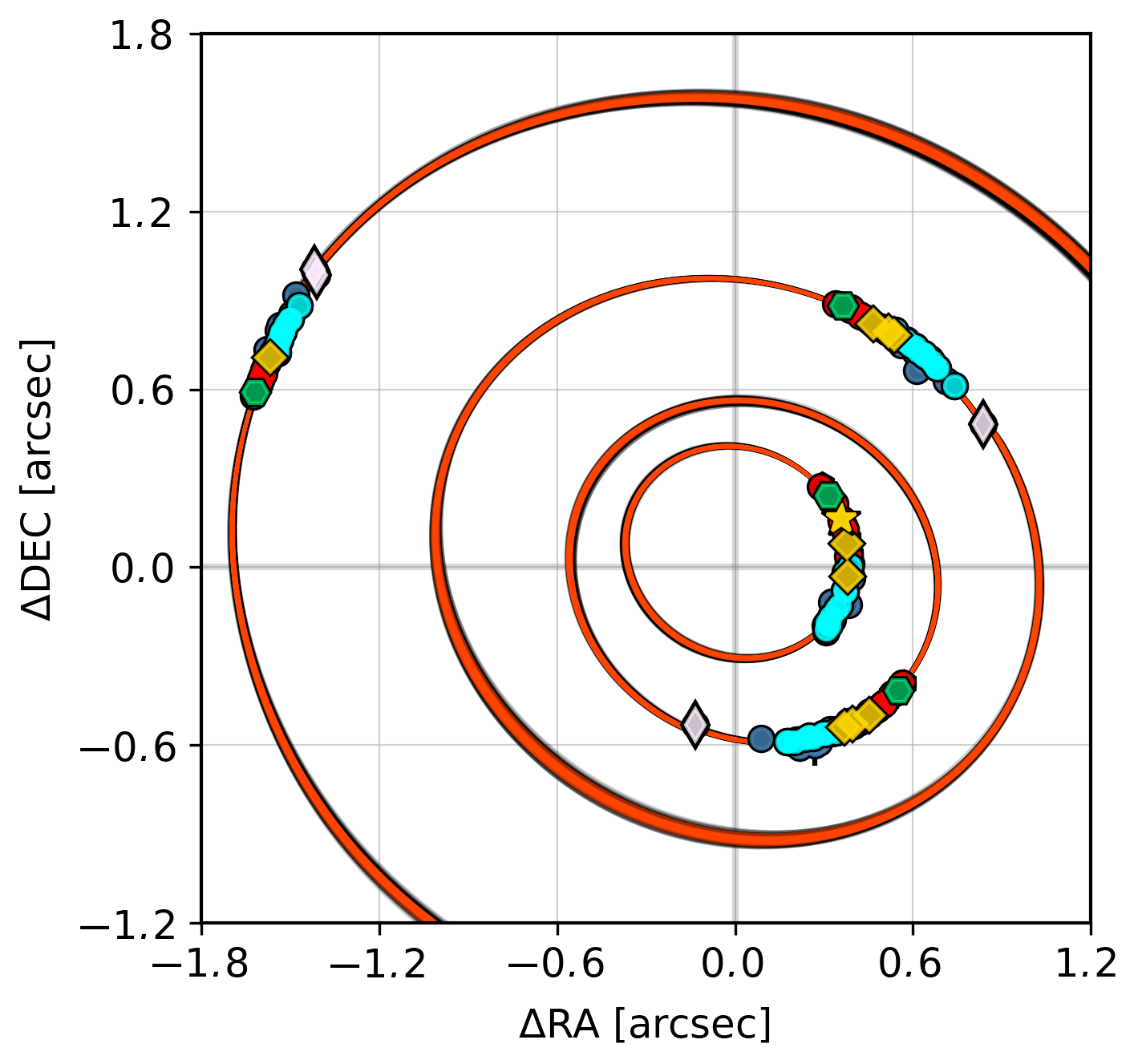} %7g.pdf}
%\quad
\includegraphics[height=0.22\textheight]{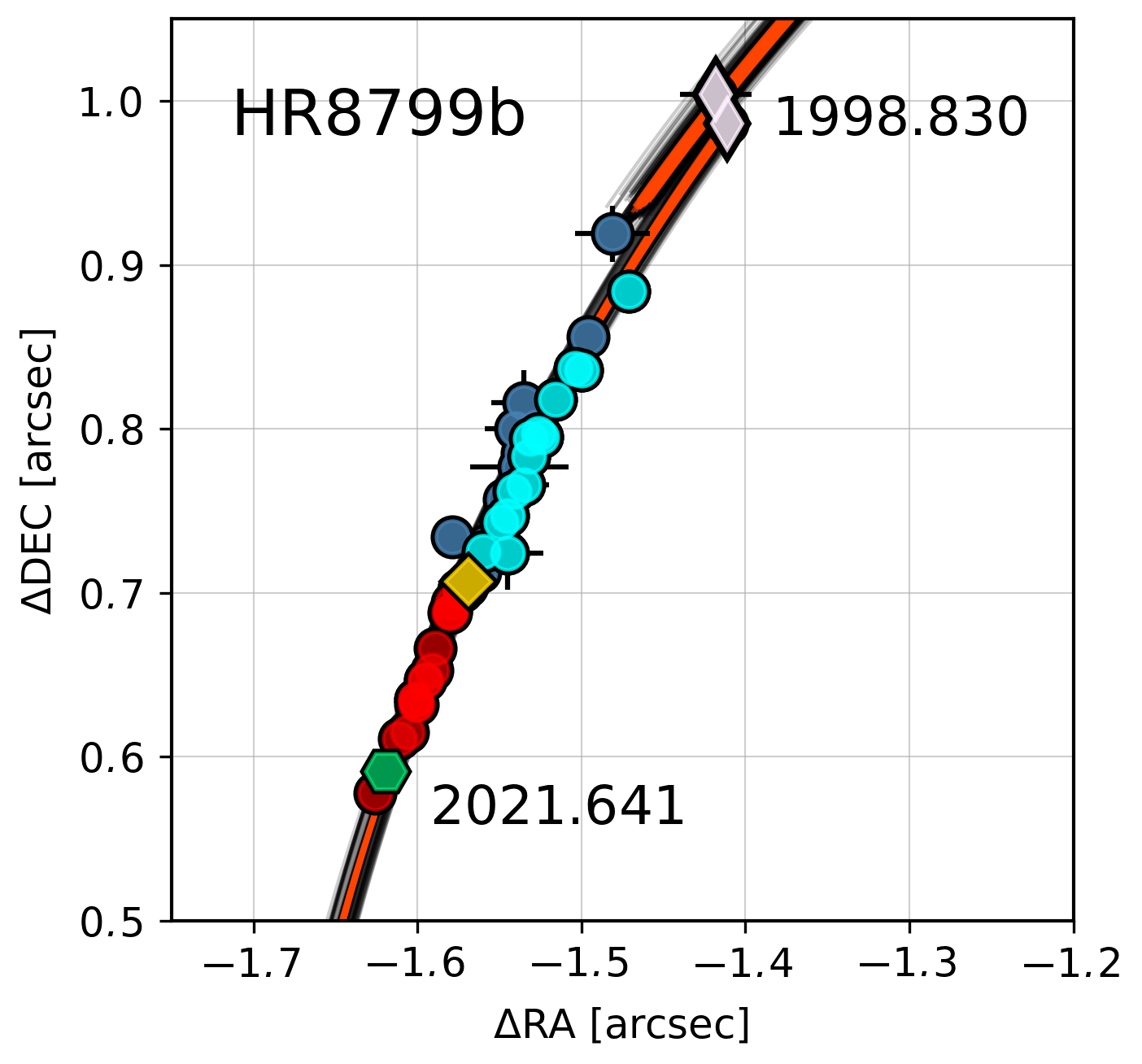} %7a.pdf}
\quad
\includegraphics[height=0.22\textheight]{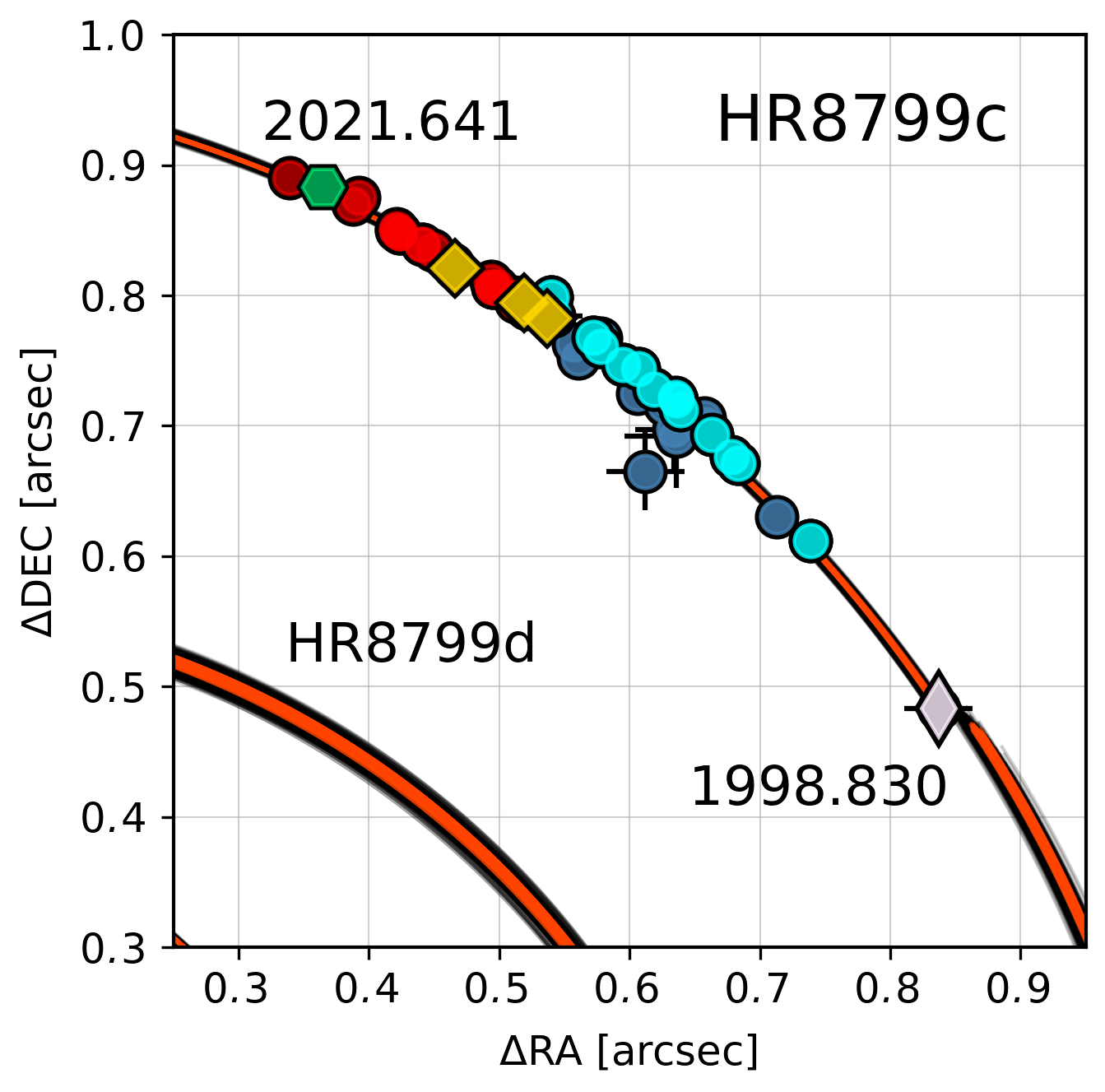} %7b.pdf}
}
\medskip 
\hbox{
\includegraphics[height=0.22\textheight]{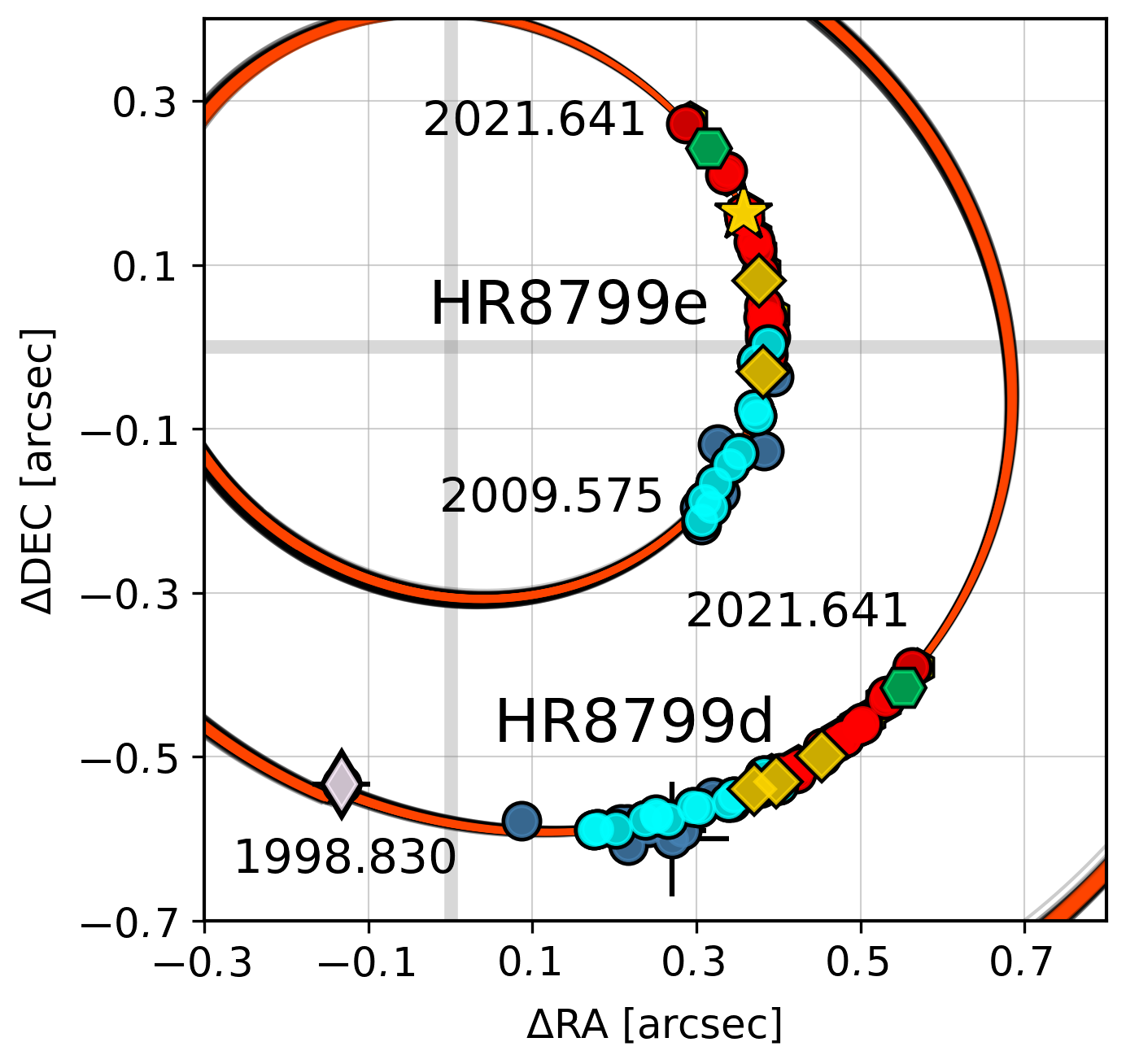} %7c.pdf}
%\quad
\includegraphics[height=0.22\textheight]{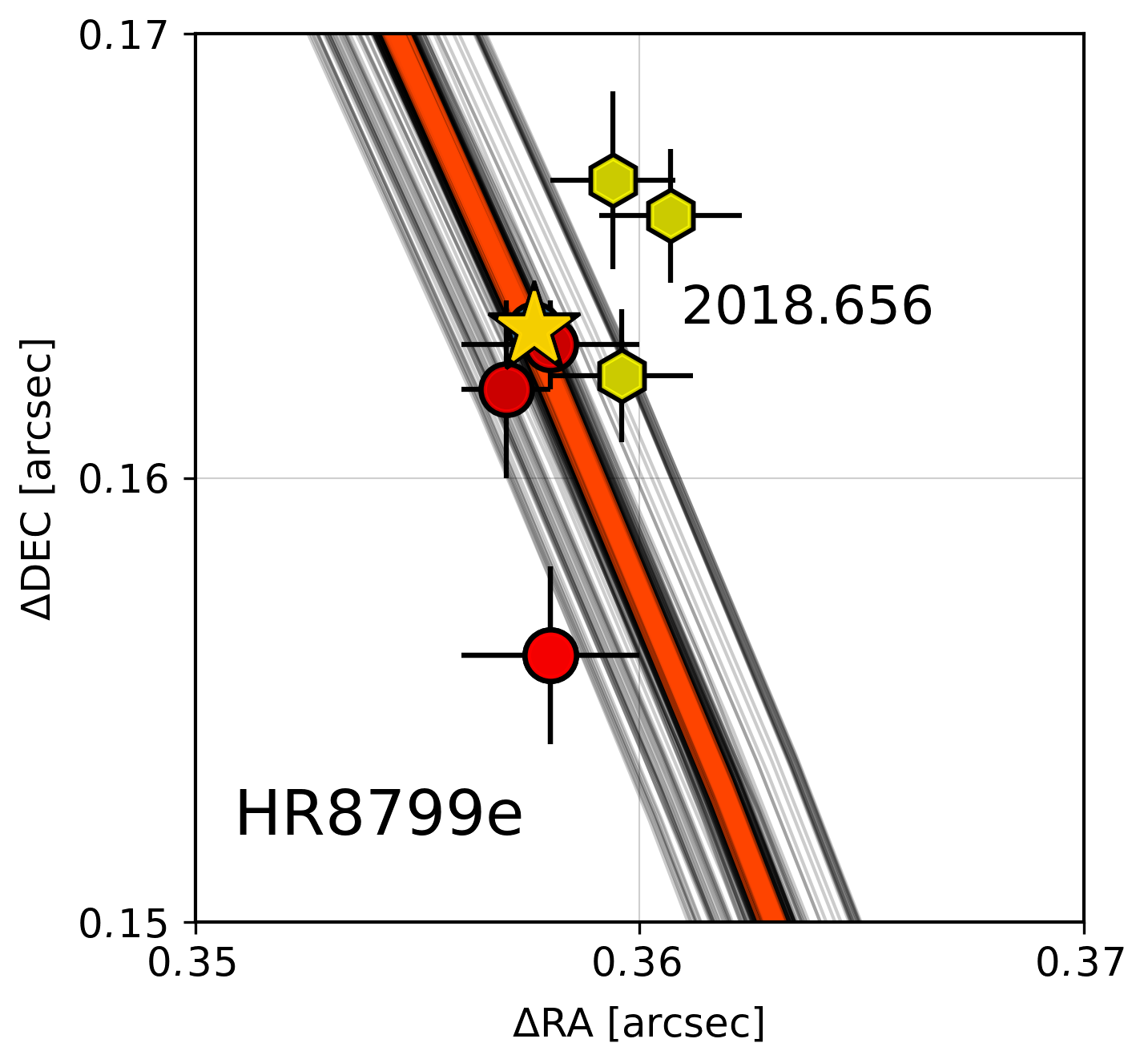} %7f.pdf}
%\quad
\includegraphics[height=0.22\textheight]{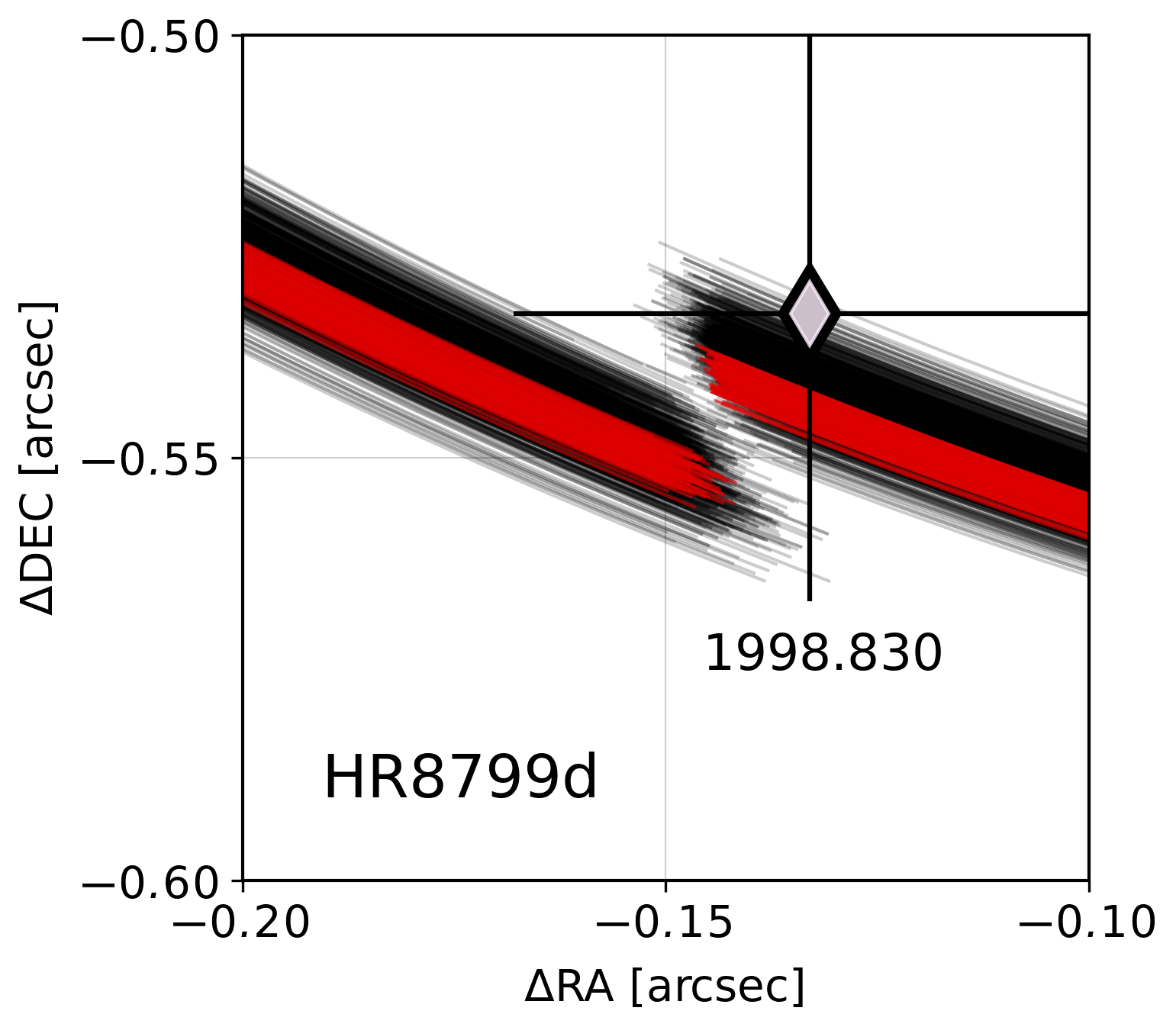}
}}}
\caption{Best-fitting solutions to the four-planet model in Table~\ref{tab:tab1} for $m_{\star}=1.47$\,\MSun, illustrated on the sky plane as randomly selected MCMC samples from the $\Delta{}n$ posterior. The $y$-axis corresponds to the north ($N$), and the $x$-axis corresponds to the east ($E$) direction, respectively (we note that the numerical values of $\Delta\alpha$ are opposite in sign with respect to the formal left-hand direction of the RA $\alpha${}). Red-filled circles are for the new or re-reduced IRDIS measurements in this paper, yellow hexagons are for the \ifs{} measurements, green hexagons are for the LUCI points, light-blue filled circles are for measurements in \cite{Konopacky2016}, and dark-blue circles are for data in other references collected in Tables~\ref{t:astroirdis}, \ref{t:astroifs}, and \ref{t:luci} and summarised in Table~\ref{t:lite}. Diamonds are for the \gpi{} data, and a star symbol is for the most accurate GRAVITY point. Red curves mark stable solutions with the lowest RMS $\simeq 7.6$--$8$ mas, randomly selected from the MCMC samples, and darker grey curves are for other orbital arcs derived for stable models up to RMS $\simeq 9$--$10$~mas. All model orbits have been derived for all measurements available to date, and cover roughly one osculating orbital period for every planet, respectively. The osculating epoch is \epk{}. {We also labeled the observational epochs encompassing the orbital arcs with data.} The upper left panel is for the global view of the system on the sky plane, and subsequent panels are for its close-ups and interesting regions.}
%{\color{red} I still need to add labels for planets and mark LUCI points with a different symbol (K.Goz.).}
\label{fig:fig4}
\end{figure*}

Overall, when comparing the posterior distributions in Fig.~\ref{fig:fig1} derived for the PO model, and in Fig.~\ref{fig:fig4} for the quasi-periodic model, we may notice similar ranges for the parameters. A basic distinction relies on the different character of the solutions: while the PO posterior covers strictly stable, resonant solutions, the $\Delta{}n$ posterior also involves weakly chaotic solutions around the Laplace resonance center, which is selected as the starting PO model and that may explain missing mass correlations found for the PO model. While the fit quality in terms of $\ln {\cal L}$ or $\Chi$ may be somewhat improved with the quasi-periodic model, the RMS of the best-fitting models remains almost the same in the two approaches. It may be concluded that the quasi-periodic model does not lead to any qualitative change in the dynamical status of the system, apart from making it possible to
%of it makes if possible to
systematically explore long-term stable solutions, which are not necessarily regular. We discuss this further in~Sect.~\ref{sec:dyn}.

Figure~\ref{fig:fig4} illustrates the best-fitting orbital solutions (red curves) over-plotted for models within $1\sigma$ (grey curves) randomly selected from the MCMC samples such that their RMS varies between $7$ and $9$~mas. The best-fitting models with RMS$\simeq 7.6$~mas are marked as red curves. All of those solutions are depicted for roughly one osculating period for each planet. Clearly, the orbits do not close for all companions. This effect is best visible for the outermost ({ top-middle} panel) and { \hr8799{}d} ({ bottom-right} panel) planet, respectively. In the latter case, the range of the orbit splitting can be compared to a relatively large uncertainty on the \hst{} measurements; for more accurate measurements, the splitting would be even more significant. This effect illustrates strong, short-term mutual interactions. Here, we follow \cite{GM2018}, who indicated that Keplerian orbits are already inadequate for representing the proper astrometric solutions despite limited observational orbital arcs. 

\begin{figure*}
\centerline{ 
\hbox{
\includegraphics[width=0.4\textwidth]{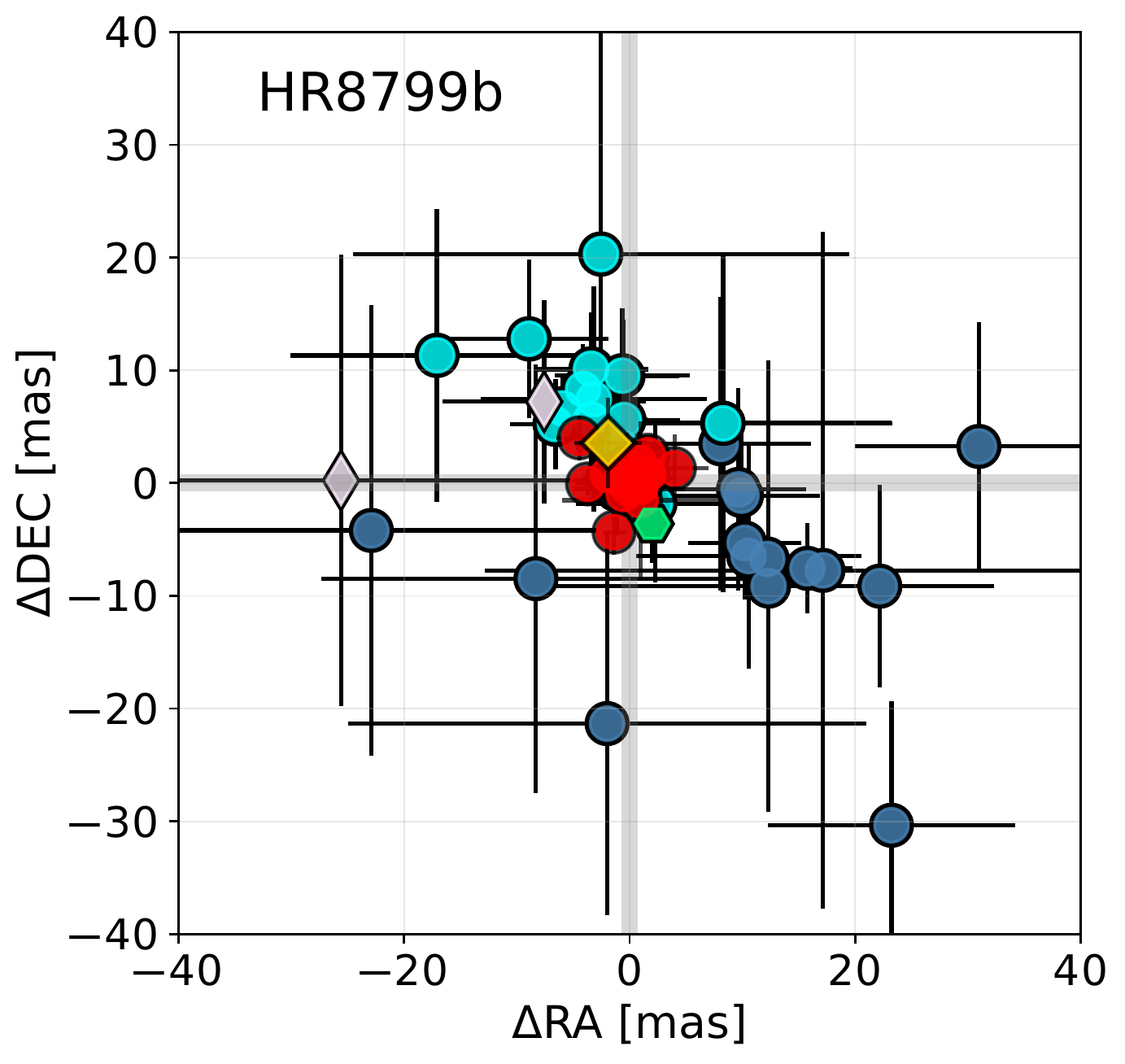}        %POresXY147HR8799b.pdf}
\quad
\vbox{
\hbox{\includegraphics[width=0.38\textwidth]{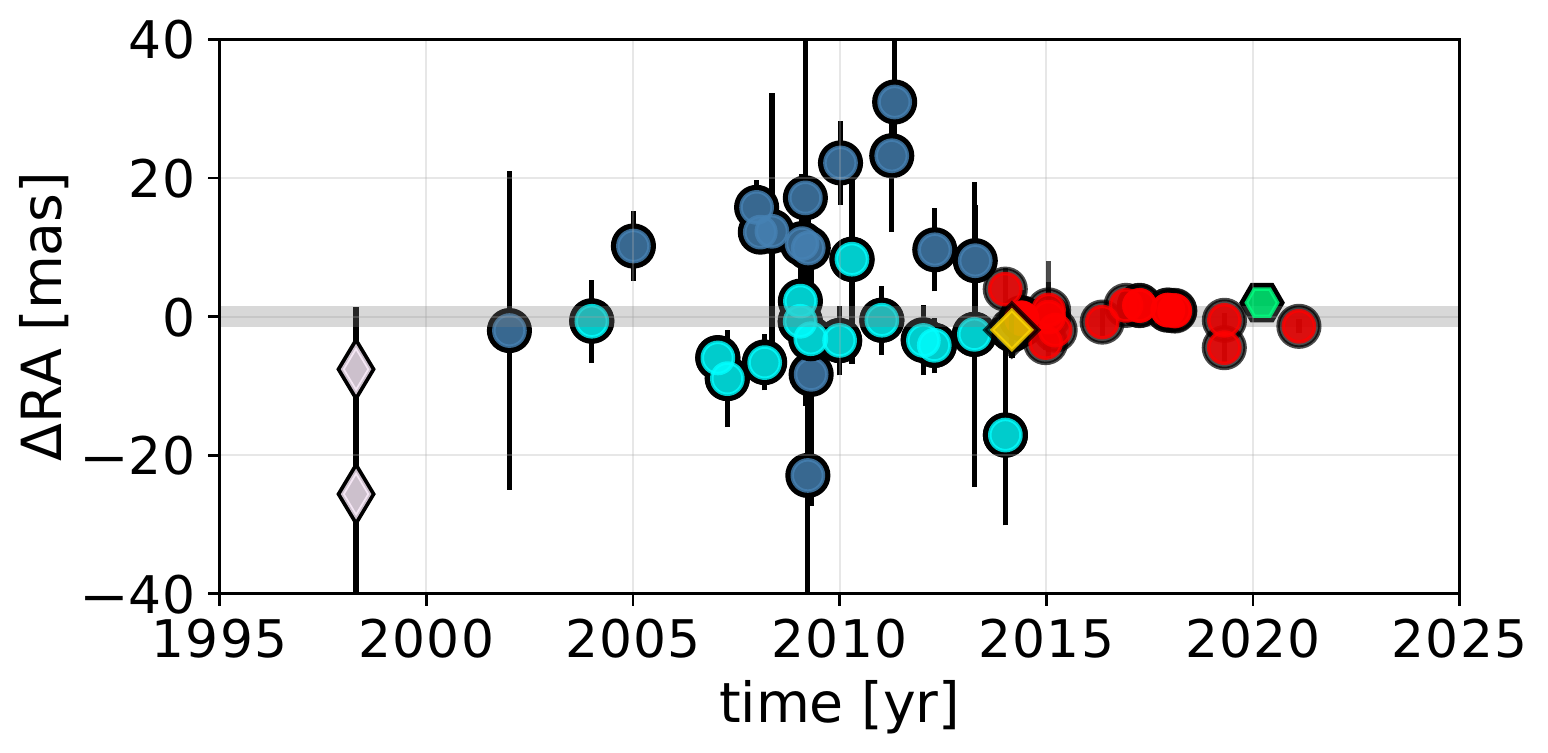}} %RA147HR8799b.pdf}}
\hbox{\includegraphics[width=0.38\textwidth]{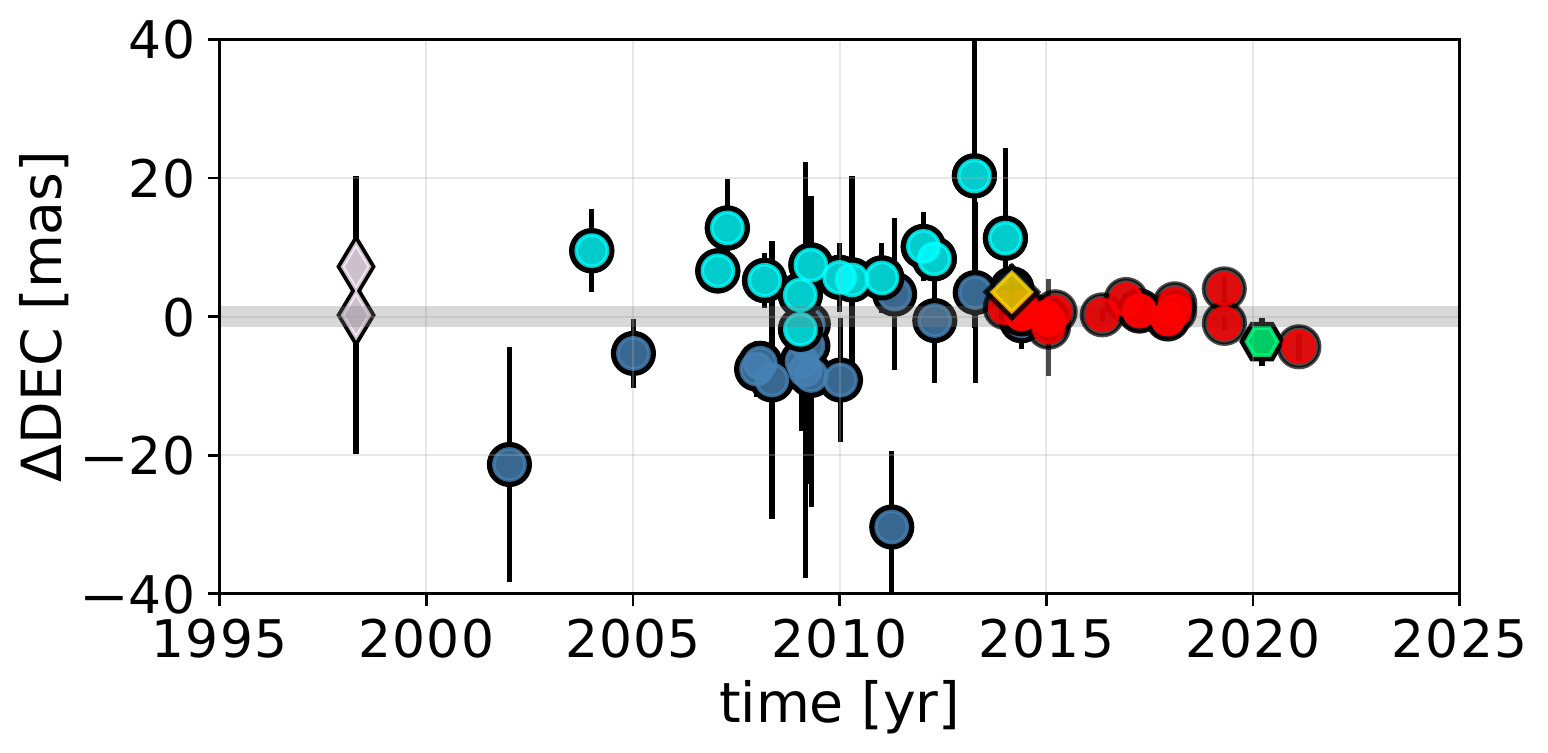}} %DEC147HR8799b.pdf}}
}}}
\centerline{ 
\hbox{
\includegraphics[width=0.4\textwidth]{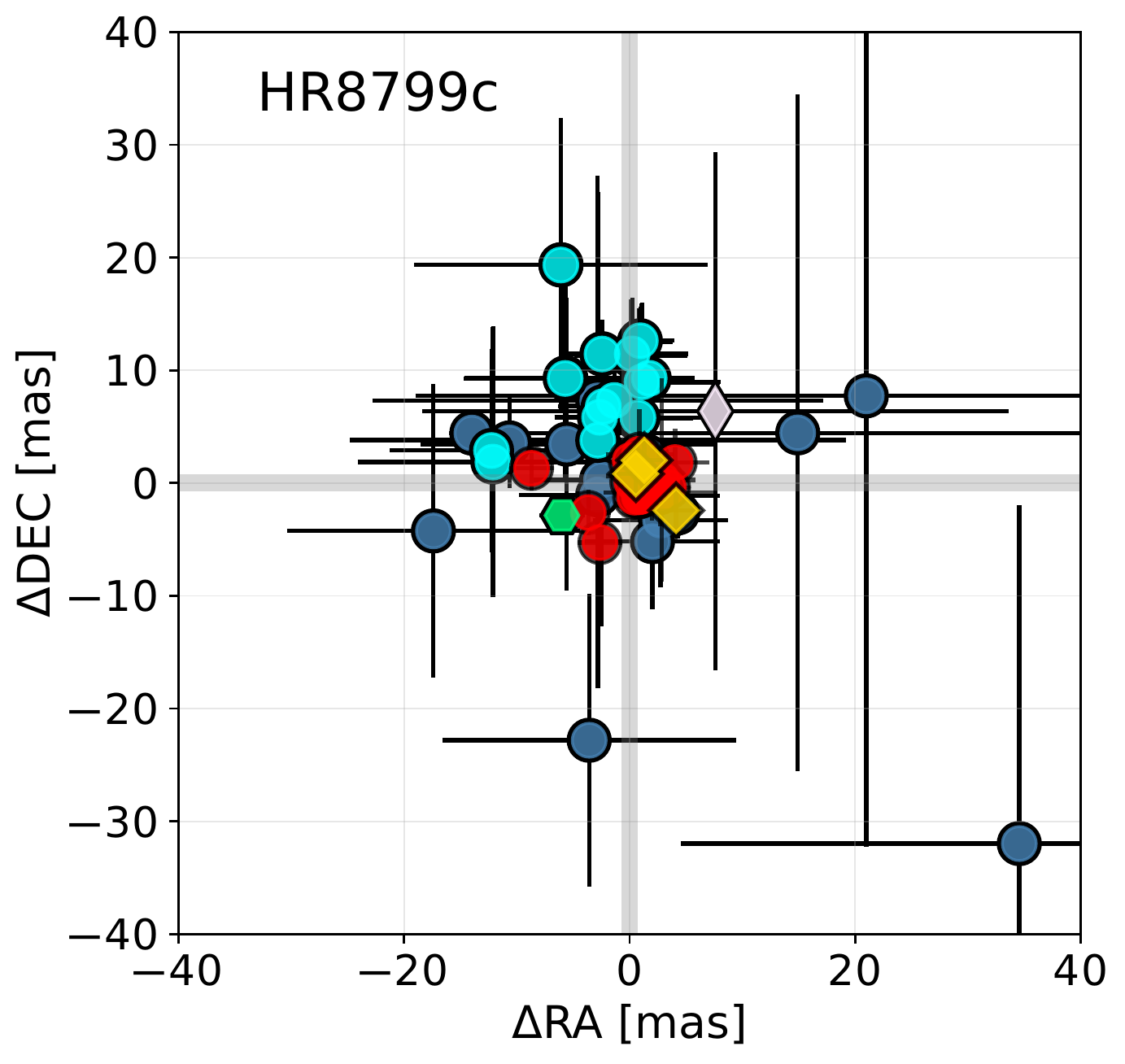}        %POresXY147HR8799c.pdf}
\quad
\vbox{
\hbox{\includegraphics[width=0.38\textwidth]{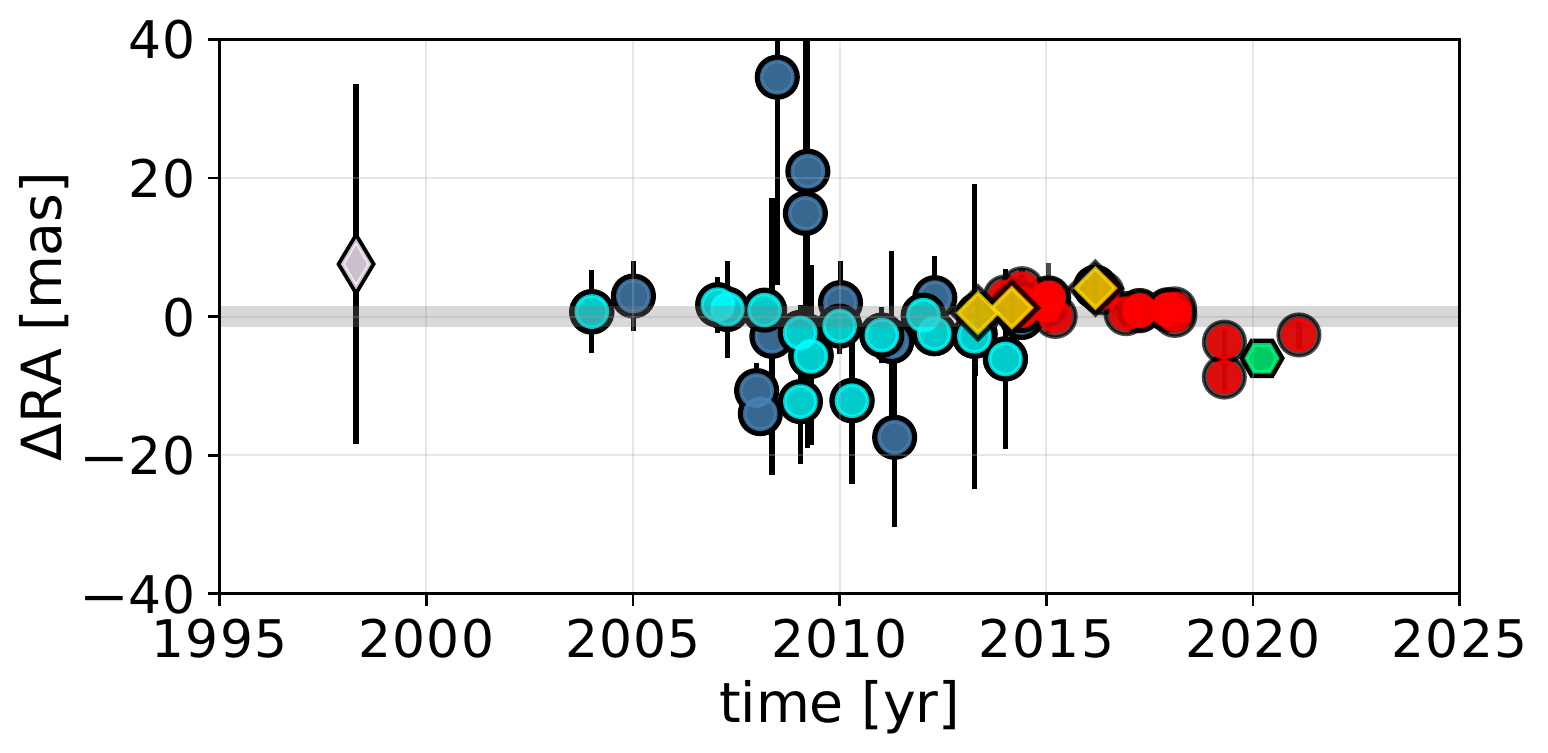}} %RA147HR8799c.pdf}}
\hbox{\includegraphics[width=0.38\textwidth]{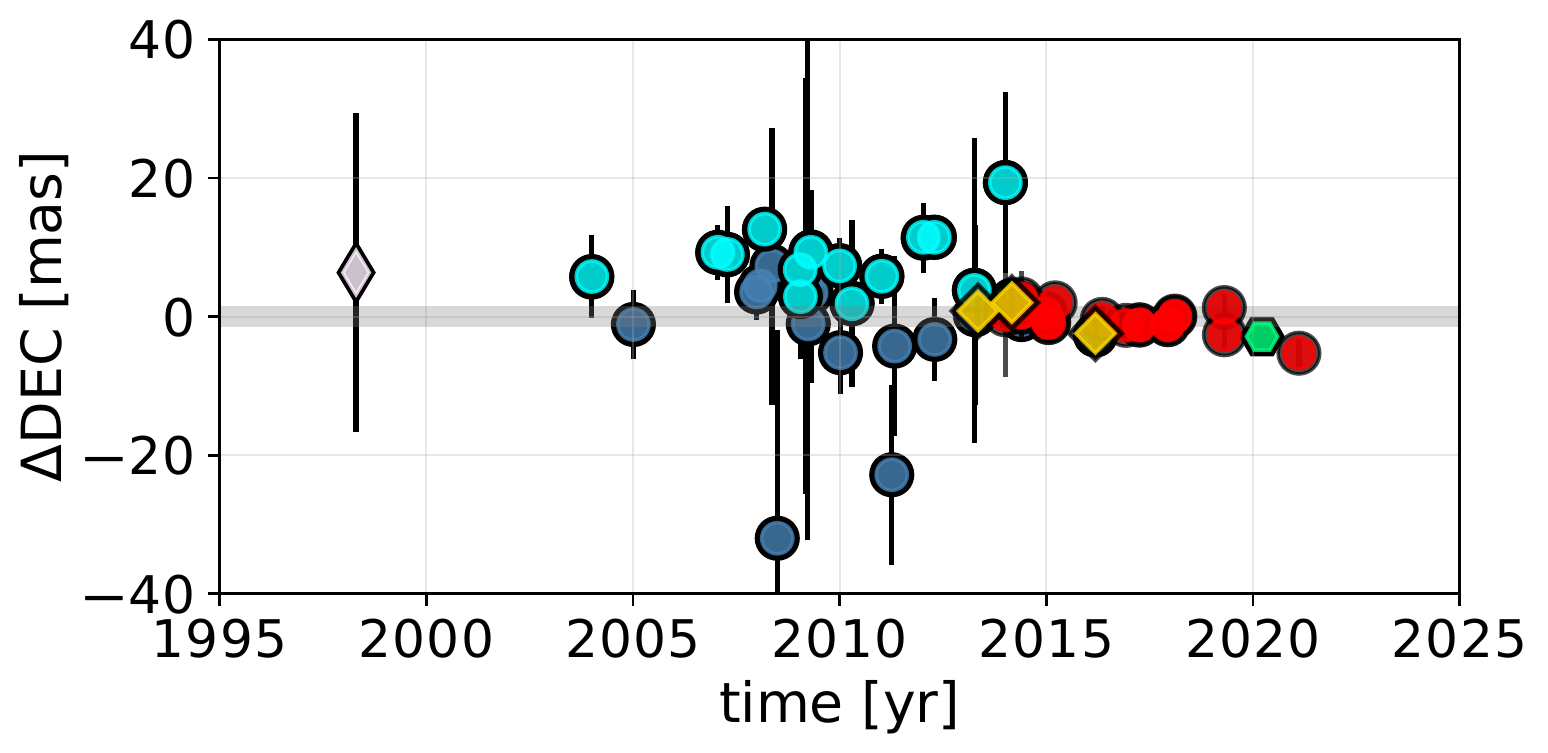}} %DEC147HR8799c.pdf}}
}}}
\caption{
Residuals to the selected best-fitting, near-resonant four-planet Model~2 in Table~\ref{tab:tab3} for planets \hr8799{}b and \hr8799{}c; stellar mass is $m_{\star}=1.47\msun$. The $y$-axis corresponds to the north ($N$), and the $x$-axis corresponds to the east ($E$) direction, respectively (we note that the numerical values of $\Delta$\RA{} are sign-opposite to regarding the formal left-hand direction of the \RA{} axis). Red filled circles, and {yellow and green hexagons are for the IRDIS, \ifs{}, and LUCI measurements} reported here,
respectively, and dark-blue and light-blue (light-grey) filled circles are for measurements in previous papers and in \cite{Konopacky2016}, respectively. 
% The yellow star marks the GRAVITY measurement.  
Yellow diamonds are for \gpi{}, the grey diamonds are for the early HST data.
}
\label{fig:fig5}
\end{figure*}

\begin{figure*}
\centerline{ 
\hbox{
\includegraphics[width=0.4\textwidth]{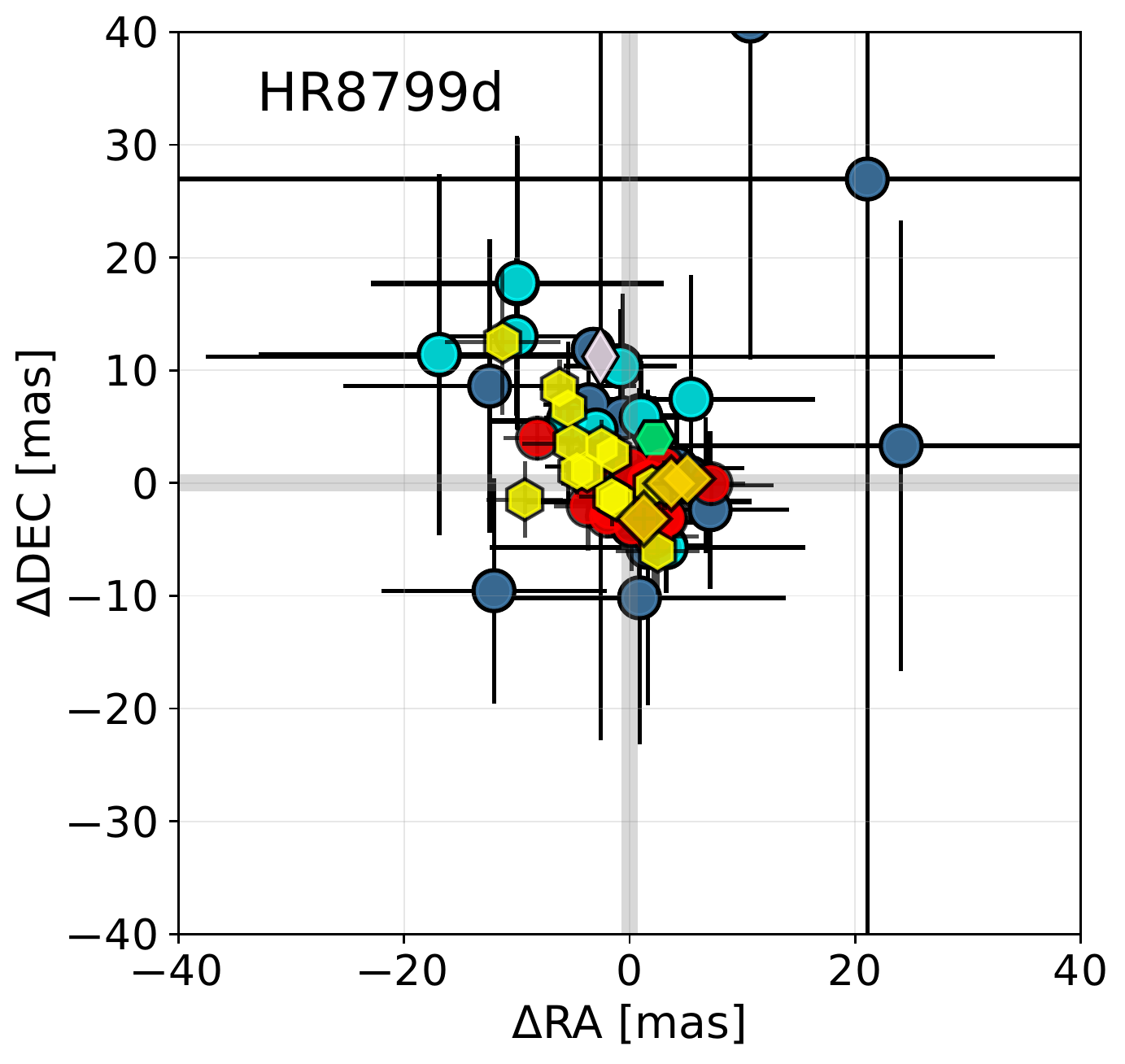}        %POresXY147HR8799d.pdf}
\quad
\vbox{
\hbox{\includegraphics[width=0.38\textwidth]{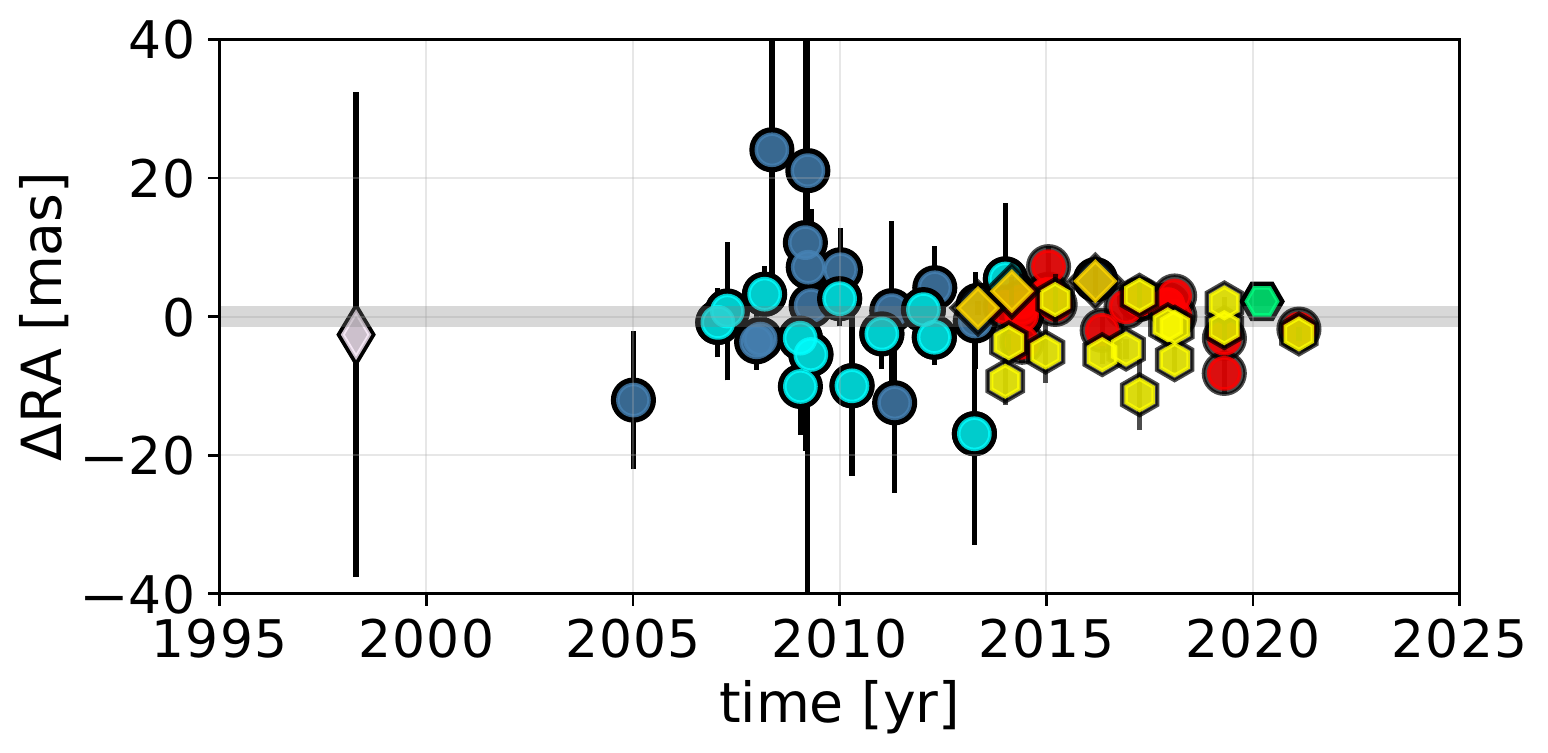}} %RA147HR8799d.pdf}}
\hbox{\includegraphics[width=0.38\textwidth]{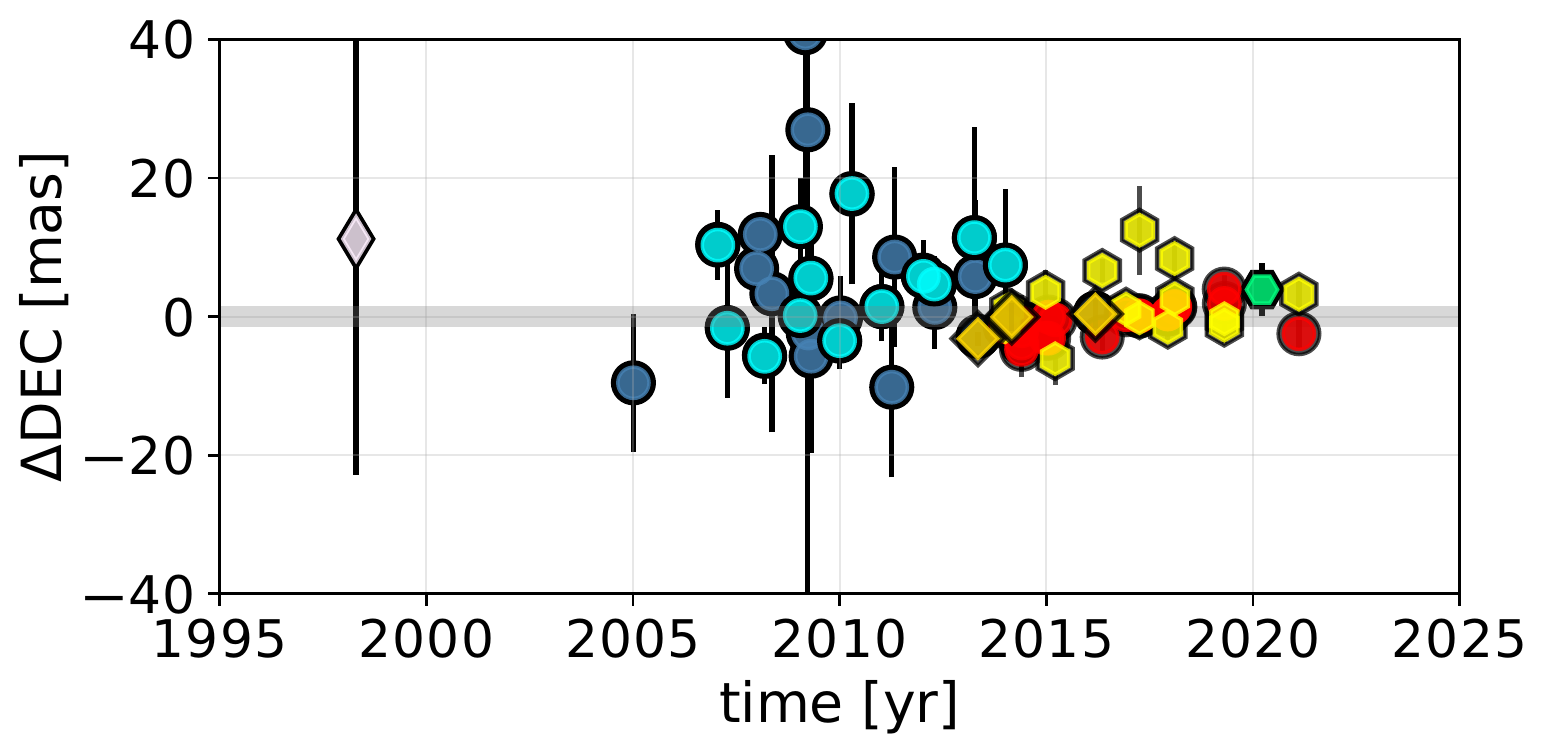}} %DEC147HR8799d.pdf}}
}}}
\centerline{ 
\hbox{
\includegraphics[width=0.4\textwidth]{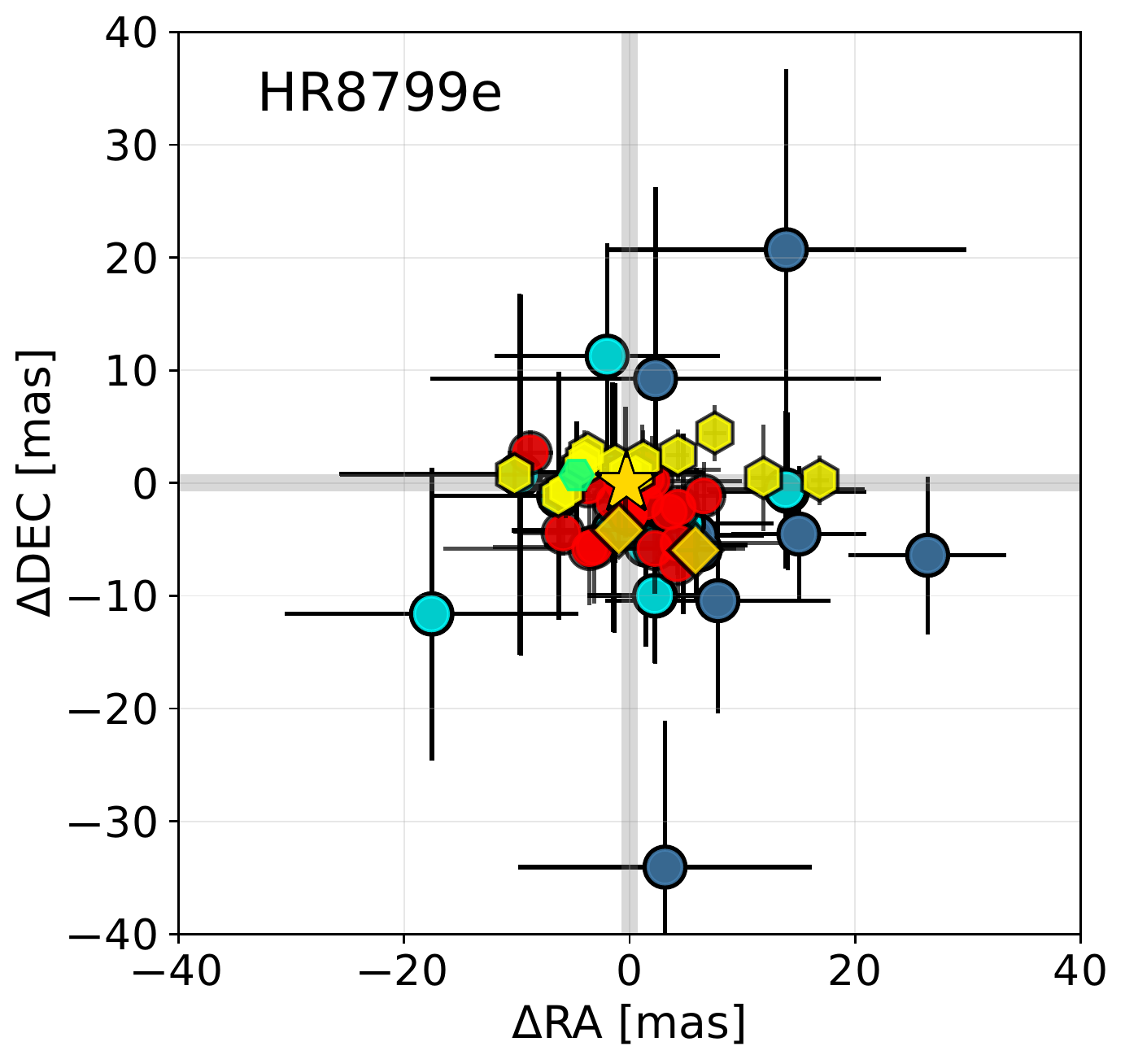}        %POresXY147HR8799e.pdf}
\quad
\vbox{
\hbox{\includegraphics[width=0.38\textwidth]{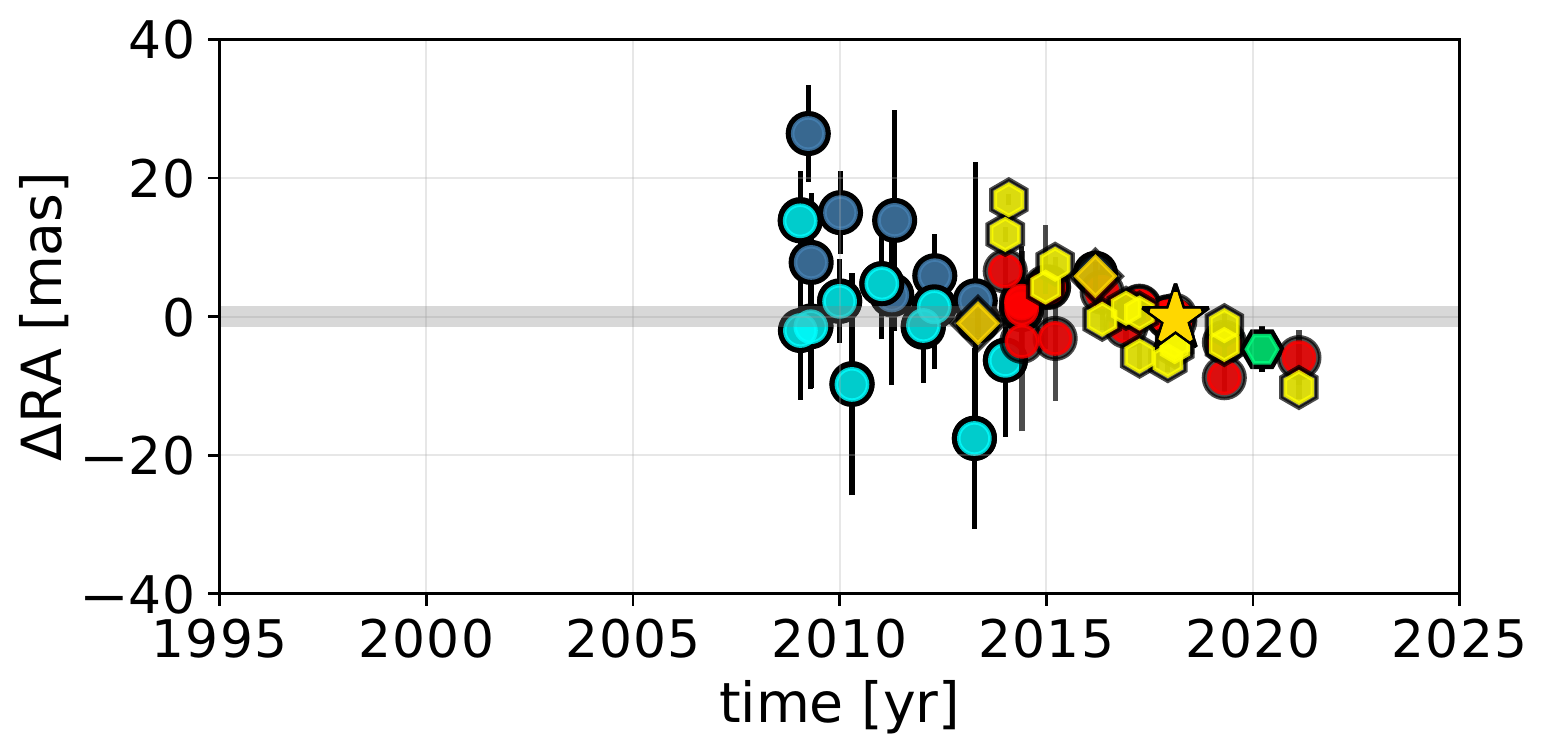}} %RA147HR8799e.pdf}}
\hbox{\includegraphics[width=0.38\textwidth]{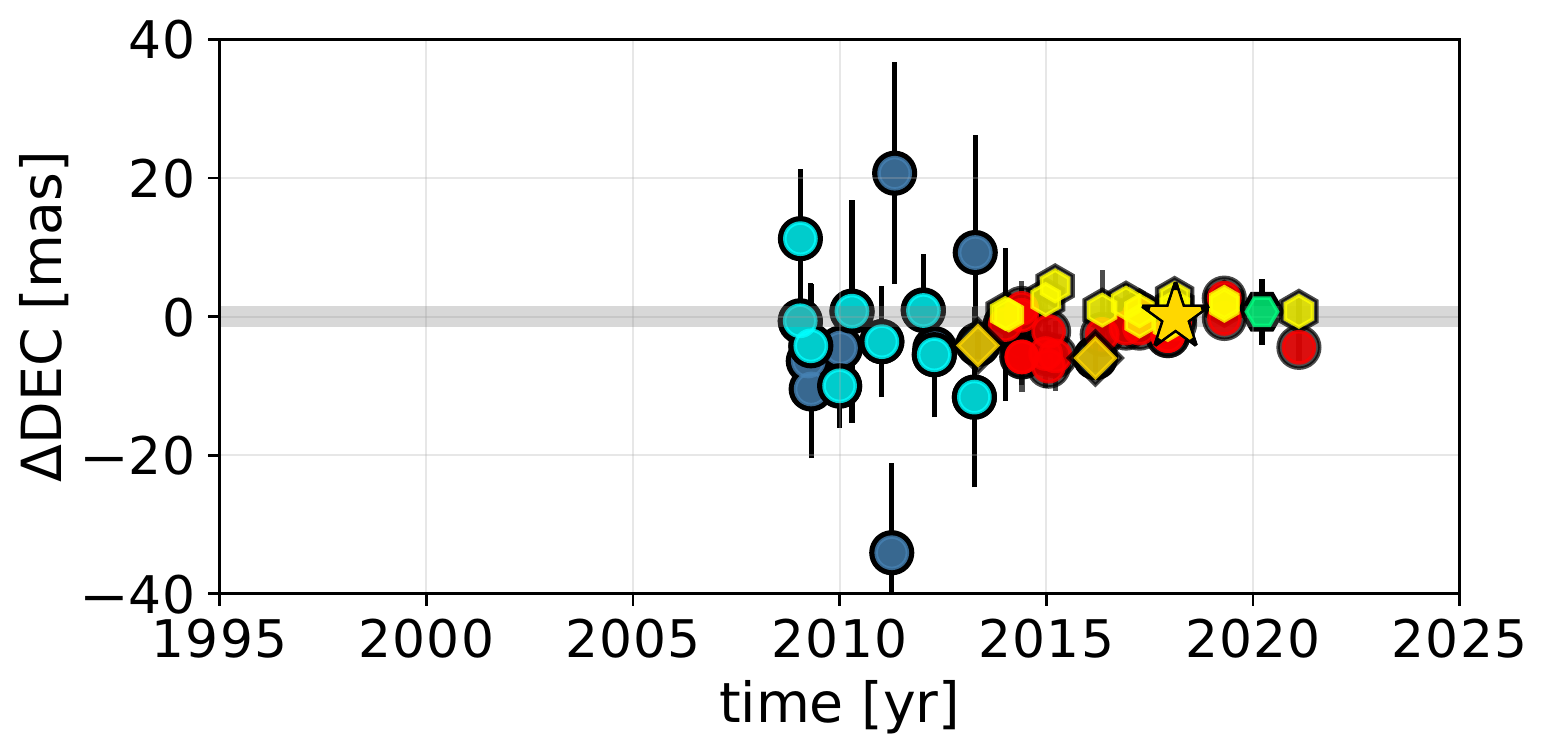}} %DEC147HR8799e.pdf}}
}}}
\caption{Residuals to the best-fitting, near-resonant four-planet Model~2 in Table~\ref{tab:tab3} for $m_{\star}=1.47\msun$, a continuation of Fig.~\ref{fig:fig8} for planets \hr8799{}d and \hr8799{}e. See the caption of Fig.~\ref{fig:fig5} for labels. The yellow star marks the GRAVITY measurement for \hr8799{}e}.
\label{fig:fig6}
\end{figure*}

In the top-right panel Fig.~\ref{fig:fig4}, for \hr8799{}c, blue circles representing measurements in \cite{Konopacky2016} and red circles and yellow hexagons for \sphere{} measurements are slightly shifted with respect to each other, and relative to the orbital model arcs. This is further visible in Figs.~\ref{fig:fig5} and~\ref{fig:fig6}, which illustrate residuals of the best-fitting model in Table~\ref{tab:tab1} in the $\Delta{}$\RA{}--$\Delta{}$\DEC{}-plane as well as individual $\Delta{}$\RA{}$(t)$ and $\Delta{}$\DEC{}$(t)$ panels. While the \sphere{} data are roughly uniformly clustered around the origin $(0,0)$, the blue circles exhibit systematic patterns, especially for planets \hr8799{}b and \hr8799{}c, besides a much larger spread. This effect is curious once we recall the two data sets consisting of uniformly reduced, essentially homogeneous measurements in \cite{Konopacky2016} and this work, respectively. It is likely that the instruments and/or reduction pipelines are not fully compatible. The effect is quite subtle but still noticeable. 

Another trend of residuals is revealed in Fig.~\ref{fig:fig6} for planet \hr8799{}e. The \sphere{} data seem to be clearly arranged along a curve in the \RA{} coordinate, revealing a systematic deviation in time with the ephemeris. This effect is especially clear for \irdis{} data and may be noted in the lower right $\Delta{}$\RA{}-$\Delta{}$\DEC{} panel and the $\Delta{}$RA$(t)$ panel, as the yellow hexagons spread along the $\Delta$\RA{}=0 axis. This may indicate a systematic effect in \irdis{} data reduction because it does not seem to appear for IFS data. However, the { time}--$\Delta$\RA{}$(t) $ panel is suggestive of this trend for both sets of measurements. If it has no instrumental origin, then it may indicate an unmodeled component of the astrometric model, such as the presence of a yet unknown object (e.g., a distant moon) or a different curvature of the orbit due to different parameters and geometry (e.g., noncoplanarity). We note here the perfect position of the GRAVITY measurement, but this one point is insufficient to dismiss the trend.

\subsection{Frequency priors versus the dynamical stability}
\label{sec:dyn}

\begin{figure*}
\centerline{ 
\vbox{
\hbox{
\includegraphics[width=0.33\textwidth]{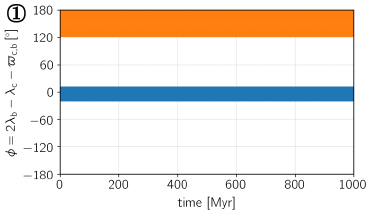}
\includegraphics[width=0.33\textwidth]{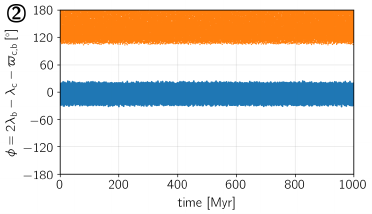}
\includegraphics[width=0.33\textwidth]{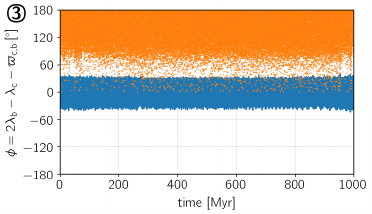}
}
\hbox{
\includegraphics[width=0.33\textwidth]{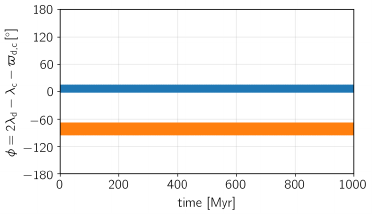}
\includegraphics[width=0.33\textwidth]{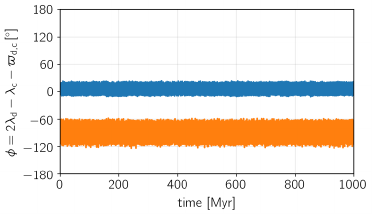}
\includegraphics[width=0.33\textwidth]{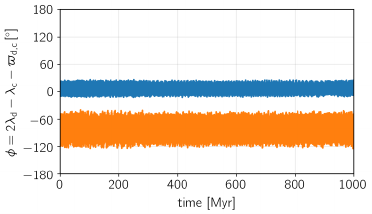}
}
\hbox{
\includegraphics[width=0.33\textwidth]{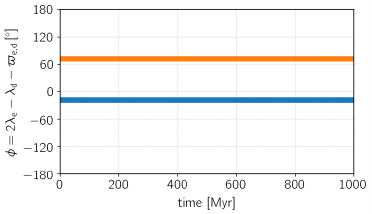}
\includegraphics[width=0.33\textwidth]{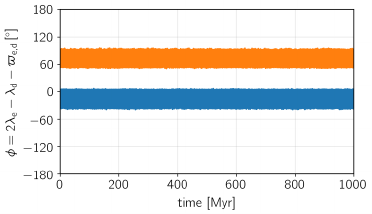}
\includegraphics[width=0.33\textwidth]{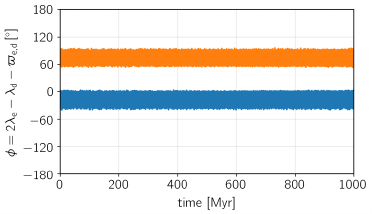}
}
\hbox{
\includegraphics[width=0.33\textwidth]{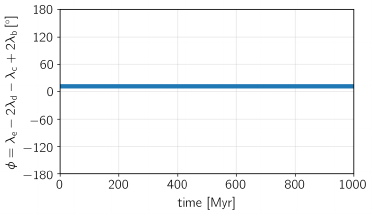}
\includegraphics[width=0.33\textwidth]{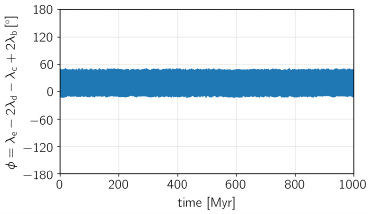}
\includegraphics[width=0.33\textwidth]{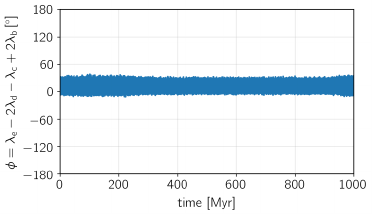}
}
\hbox{
\includegraphics[width=0.33\textwidth]{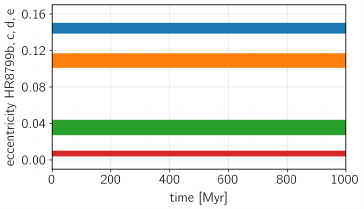}
\includegraphics[width=0.33\textwidth]{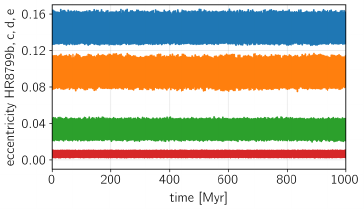}
\includegraphics[width=0.33\textwidth]{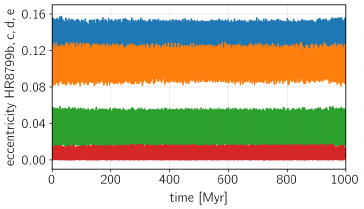}
}
}
}
\caption{
Orbital evolution of selected long-term-stable solutions with similar astrometric fit quality, but exhibiting qualitatively different stability signatures. The left column is for the exact Laplace resonance, the middle column is for a model displaced from the PO but rigorously stable, and the right column is for a $>1$~Gyr stable solution characterized by rotation of one critical angle of the 2:1~MMR in the outermost pair. The rows from the top to bottom are for all critical angles of the 2:1~MMR, the critical angle of the generalized Laplace MMR, and eccentricities for subsequent pairs of planets. See Table~\ref{tab:tab3} for the initial conditions labeled from 1 to 3 at the top-left corner of each column.
}
\label{fig:fig7}
\end{figure*}

\begin{figure*}
\centerline{ 
\vbox{
\hbox{
\includegraphics[width=0.33\textwidth]{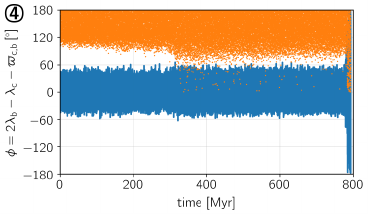}
\includegraphics[width=0.33\textwidth]{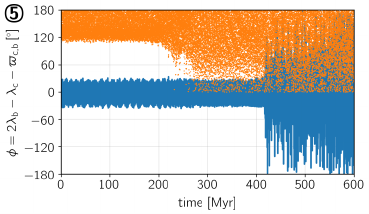}
\includegraphics[width=0.33\textwidth]{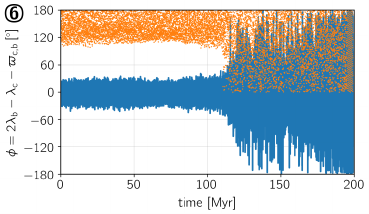}
}
\hbox{
\includegraphics[width=0.33\textwidth]{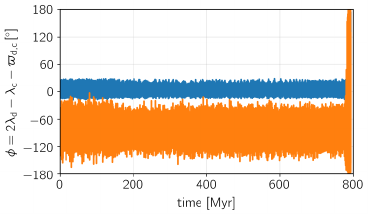}
\includegraphics[width=0.33\textwidth]{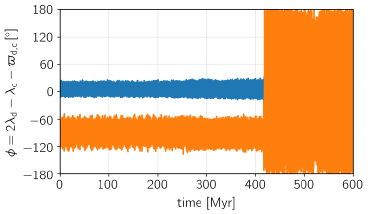}
\includegraphics[width=0.33\textwidth]{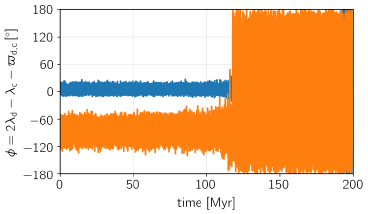}
}
\hbox{
\includegraphics[width=0.33\textwidth]{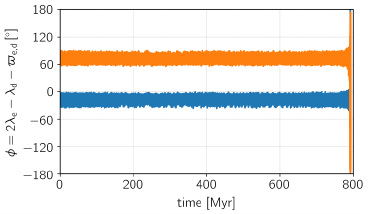}
\includegraphics[width=0.33\textwidth]{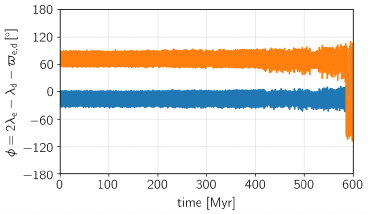}
\includegraphics[width=0.33\textwidth]{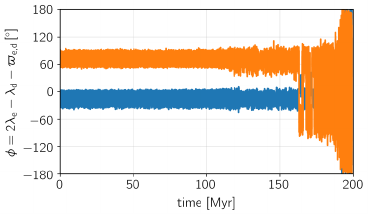}
}
\hbox{
\includegraphics[width=0.33\textwidth]{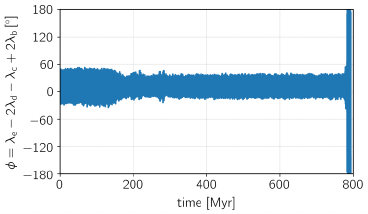}
\includegraphics[width=0.33\textwidth]{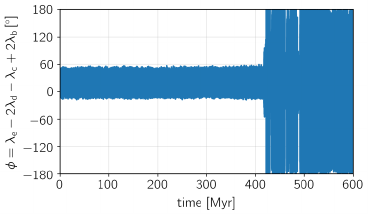}
\includegraphics[width=0.33\textwidth]{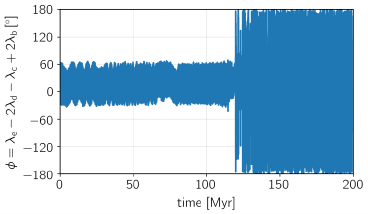}
}
\hbox{
\includegraphics[width=0.33\textwidth]{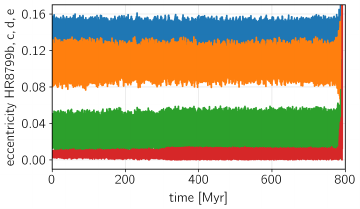}
\includegraphics[width=0.33\textwidth]{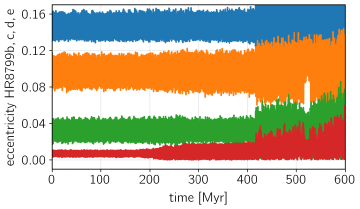}
\includegraphics[width=0.33\textwidth]{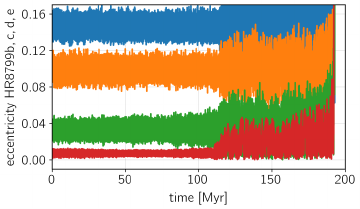}
}
}
}
\caption{
Orbital evolution of selected long-term-stable chaotic solutions with similar astrometric fit quality. The rows from the top to bottom are for all critical angles of the 2:1~MMR, the critical angle of the generalized Laplace MMR, and eccentricities for subsequent pairs of planets. See Table~\ref{tab:tab3} for the initial conditions labeled from 5 to 8 at the top of each column.
}
\label{fig:fig8}
\end{figure*}

Regarding the dynamical character of solutions derived in Sect.~\ref{sec:nearfits}, we analyzed representative examples of the best-fitting models marked in Fig.~\ref{fig:fig4}. We selected MCMC samples in Fig.~\ref{fig:fig3} for $m_\star=1.47\msun$ and from other MCMC sampling for $m_\star=1.52\msun$. These initial conditions were integrated with the IAS15 integrator from the REBOUND package \citep{reboundias15} for 1~Gyr or up to disruption of the system when a collision or ejection of a planet occurs. The results are shown in a sequence of plots in Fig.~\ref{fig:fig7} and Fig.~\ref{fig:fig8}.

Systems illustrated in Fig.~\ref{fig:fig7} are for long-term stable solutions that survived for at least 1~Gyr. The left column  is for exactly resonant  Model~1 and its elements displays  Table~\ref{tab:tab3} and also~Table~\ref{tab:tab1}. This solution is stable forever ``by design'' in the framework of the $N$ body dynamics. All but one critical angle of the two-body MMRs between subsequent pairs of planets librate around centers in the $[-90^{\circ},90^{\circ}]$--range. The exceptional critical argument is for the outermost pair of planets and librates around the center at $\simeq 180^{\circ}$. As it comes to the critical argument of the Laplace resonance, it librates with a very small amplitude of just $\simeq 5^{\circ}$ around the resonance center at $\simeq 15^{\circ}$. The periodic, perfectly regular character of this solution manifests as time evolution of the eccentricities.

\begin{figure*}
\centerline{ 
\vbox{
\hbox{
\includegraphics[width=0.4\textwidth]{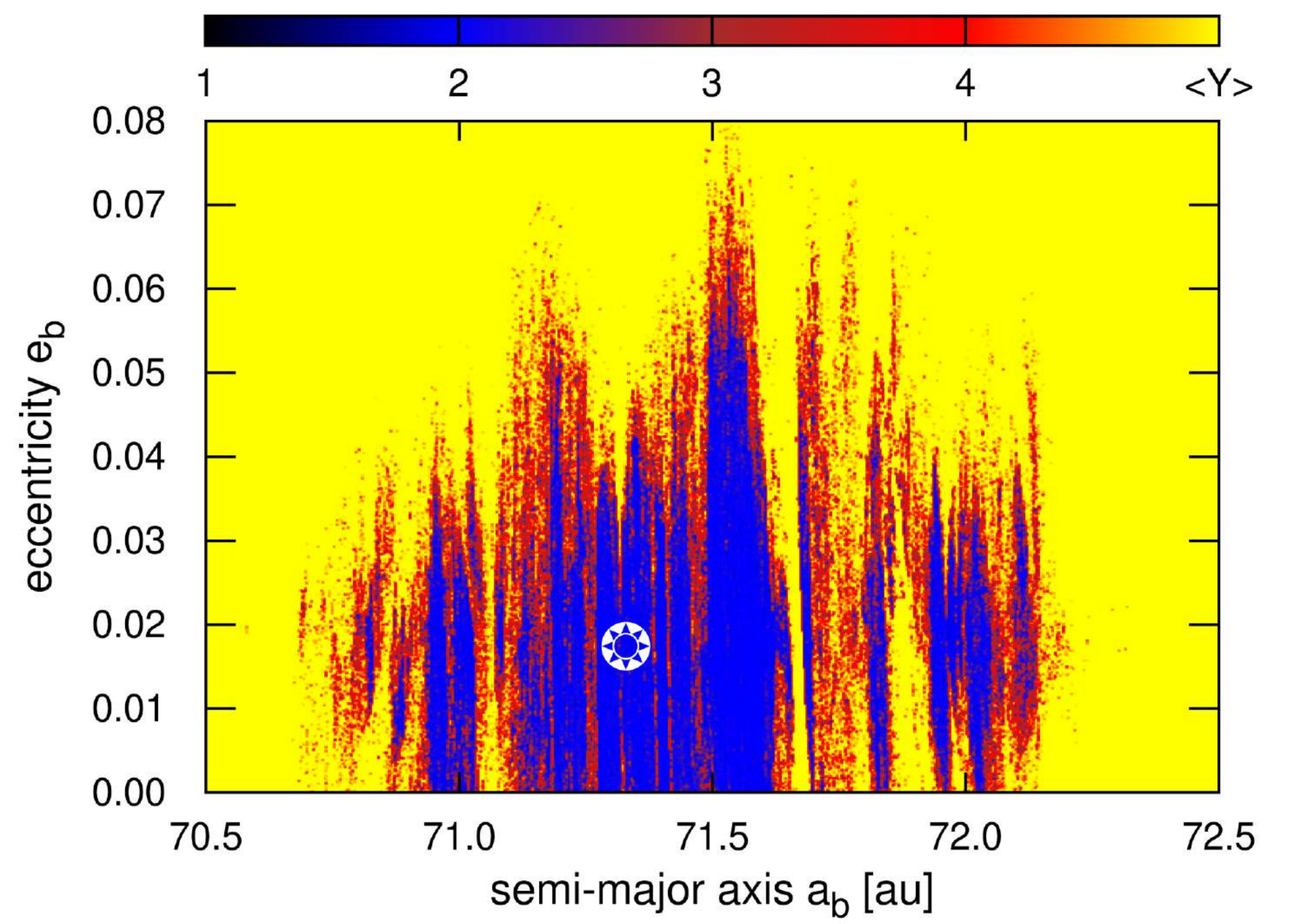} %hr8799bmapY.pdf}
\qquad
\includegraphics[width=0.4\textwidth]{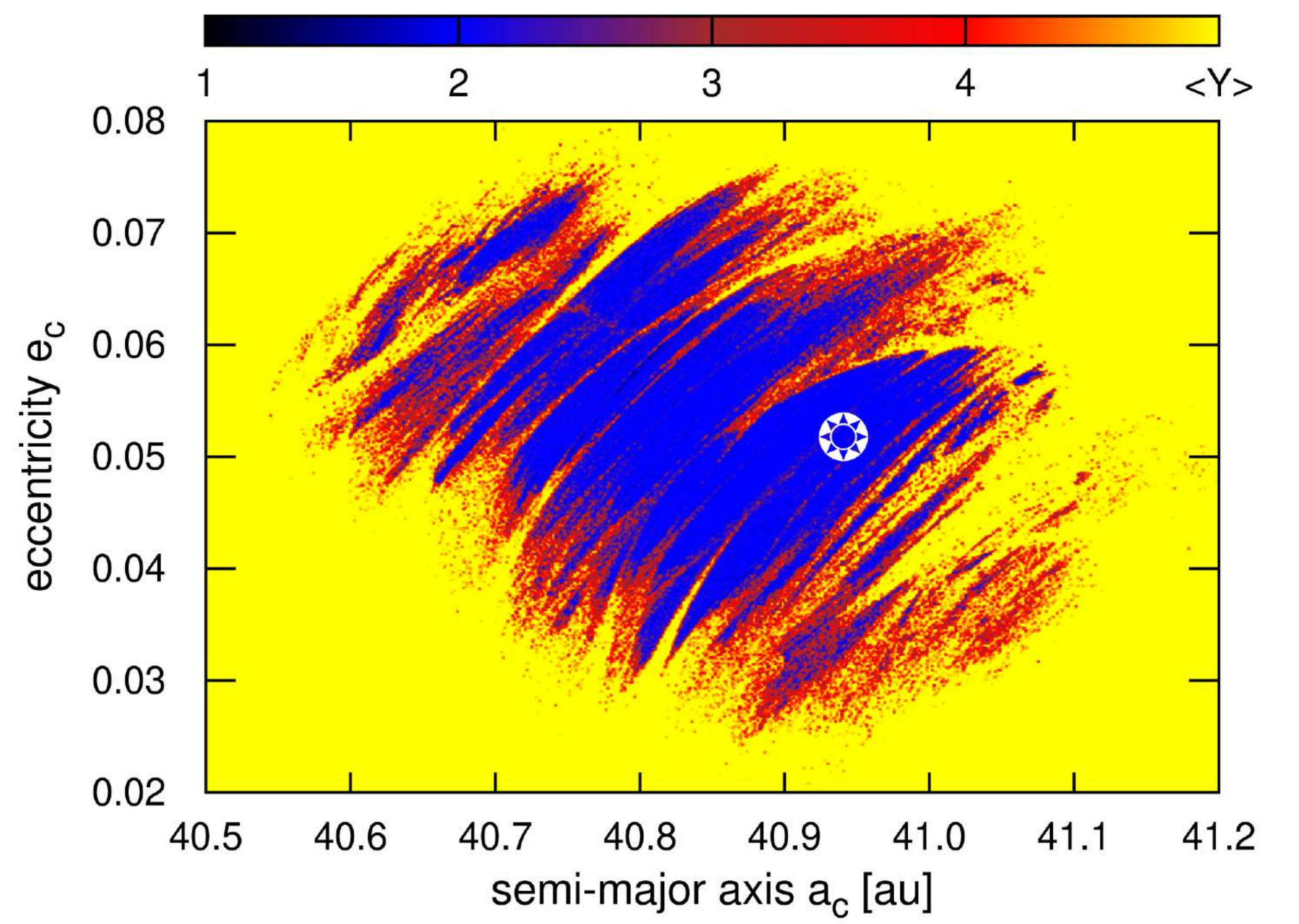} %hr8799cmapY.pdf}
}
%\qquad 
\hbox{
\includegraphics[width=0.4\textwidth]{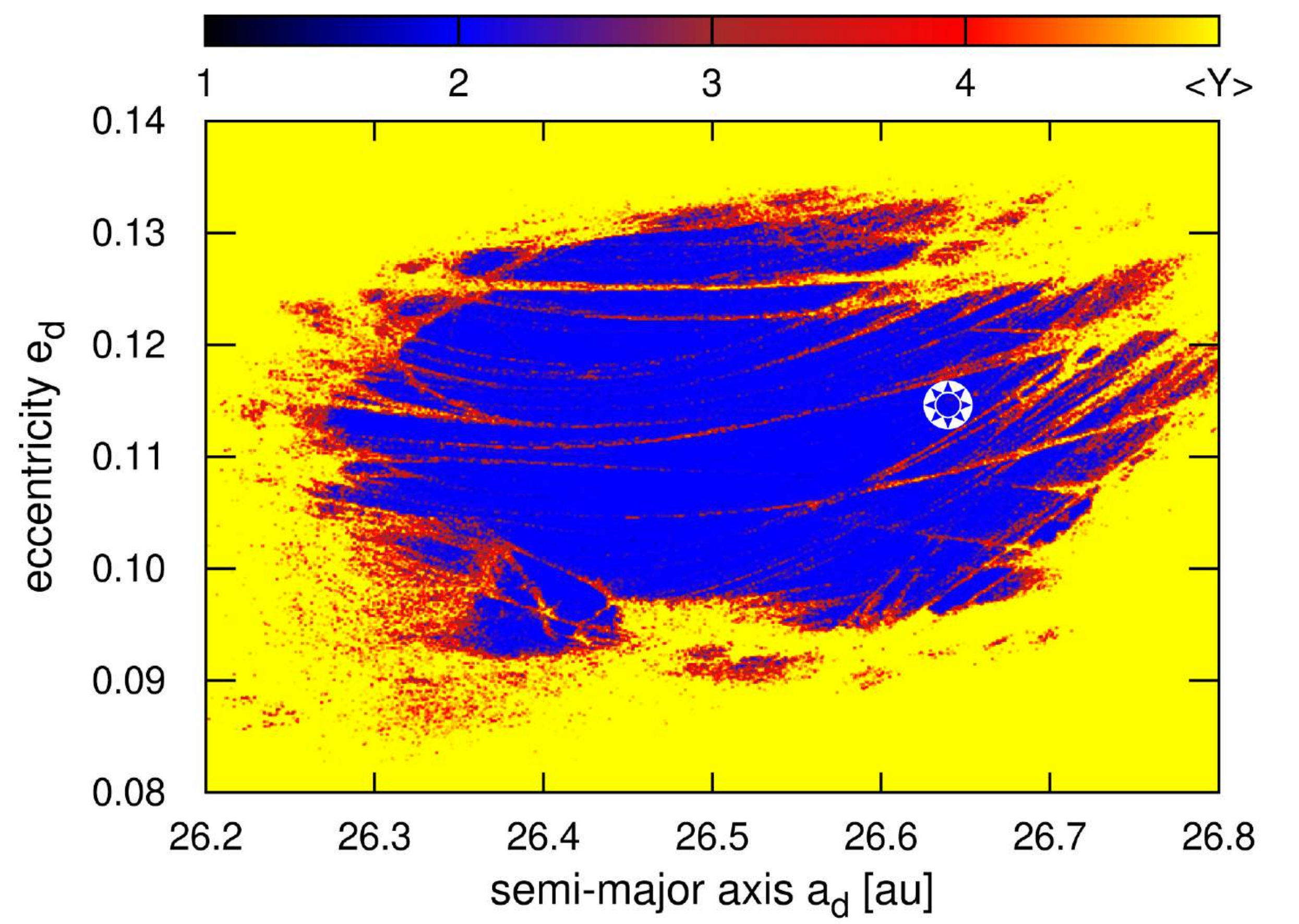} %hr8799dmapY.pdf}
\qquad
\includegraphics[width=0.4\textwidth]{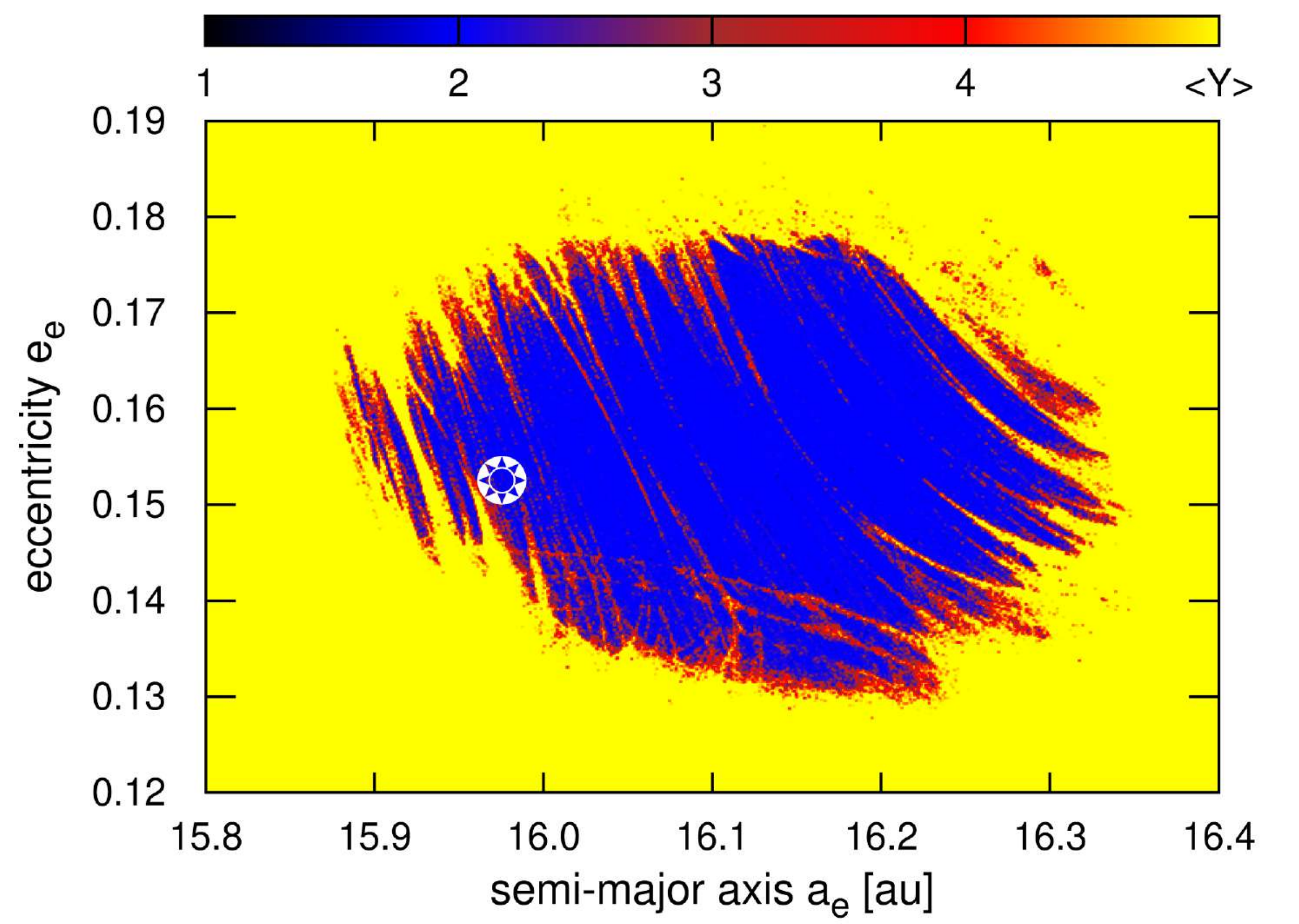} %hr8799emapY.pdf}
}
}
}
\caption{
Dynamical maps in the Keplerian, osculating astrocentric semi-major-axis--eccentricity plane for the quasi-periodic, near-resonant configuration described by Model~2. All { masses and elements} but the map coordinates are fixed at their best-fitting values in Table~\ref{tab:tab3}. The osculating epoch is \epk{}. Stable solutions are determined with $|\Ym-2| \simeq 0$ { and marked with blue color}.}
\label{fig:fig9}
\end{figure*}

The middle column (Model~2) in Fig.~\ref{fig:fig7} is for a model slightly displaced from the exact resonance, as measured by $|\Delta n| \simeq 10^{-7}$\drad{}, yet it yields {slightly} better fit quality. The critical arguments evolve in the same manner, as for the exact resonance, but their amplitudes and eccentricities become noticeably larger. This solution also yields a fully stable configuration of the planets. It is regular in the sense of the Lyapunov exponent, as it can be seen in Fig.~\ref{fig:fig9} showing dynamical maps in terms of the \megno{} fast indicator \citep{Cincotta2003}. The mean exponential growth factor of nearby orbits (MEGNO; also known as $\Ym$) is a numerical technique designed to efficiently characterize the stability of the $N$-body solutions in terms of the maximal Lyapunov exponent (MLE). We implemented MEGNO for the planetary problem in \cite{Gozdziewski2001} and in our CPU-parallelized $\mu${\sc Farm} numerical package. It is also clear that this solution lies close to the edge of the resonance island (dark blue color).

Finally, the right column (Model 3) in Fig.~\ref{fig:fig7} is for a marginally stable solution, in the sense of nonzero MLE, with $|\Delta n| \simeq 10^{-5}$\drad{}. Although it survived for the 1~Gyr integration interval, one of the critical angles in the outermost 2:1~MMR rotates. This solution is peculiar in the sense that, despite rotations of one of the 2:1~MMR critical arguments, the semi-amplitudes of other critical angles are similar to those for quasi-periodic Model~2. 

Model~3 in Fig.~\ref{fig:fig7} is also remarkable when we compare it with a sequence of solutions in Fig.~\ref{fig:fig8} illustrating chaotic, yet still long-term stable models characterised by $|\Delta n| \simeq 10^{-6}$\drad. All these models self-destruct between 800~Myr (Model~4, the left column in Fig.~\ref{fig:fig8}), 600~Myr (Model~5, the middle column), and just 200~Myr (Model~6, the right column). Despite this, for the initial few hundred million years, extending safely beyond most approximations of the stellar age, which range between 30--160~Myr \citep{2012ApJ...761...57B,2022AJ....163...52S}, the systems are bounded and locked in the resonance. Simultaneously, Models~3 and~4 yield masses of the inner planet in the $\simeq 10\,$\MJup range, and Model~4 is especially interesting given its low $\Chi$ compared to the initial starting PO configuration. These models yield a declining mass hierarchy that resembles that of the outer Solar System, and a mass of \hr8799{}e is consistent with the dynamical estimate in \cite{2021ApJ...915L..16B}. 

These examples are to justify that the system may be dynamically long-term stable in the planet mass range of $\simeq 10\,$\MJup, even if detuned from the exact resonance and mildly chaotic.
%, as well as with the observational evidence that planet spectra are matched very closely by some red field dwarfs \citep{2021A&A...648A..26W}.
{ In all unstable cases, one of the critical arguments of the 2:1 MMR of the two outermost planets progressively increases its libration amplitude and eventually begins to rotate.}
%In all cases of instability, it emerges through increasing librations of one of the critical 2:1~MMR arguments in the outermost pair \hr8799{}b--c which eventually begins to rotate. 
In this sense, the outermost pair \hr8799{}b--c is the weakest link in the resonance chain, provoking instability of the whole system displaced from the resonance. {The time for the onset of instability can be {relatively} very long, as shown by the evolution of Models 4 and 5. }
%Yet the event time may be very long, as simulations for Models~4 and~5 demonstrate.  
In any case, the libration amplitudes of the critical angles seem {to be a weak indicator of instability}, because there is no clear relationship between these amplitudes and the time of instability. {Also, Model~5 illustrates the difficulty in predicting the system behavior based on the variation of critical angles, especially if the system stability is tested for a limited period of time. The apparently regular, quasi-periodic, and bounded evolution of the critical angles for $\simeq~200$~Myrs does not prevent the system from eventually { becoming unstable} after $\simeq 400$--{ 500}~Myr. A similar effect may be observed for Model~6, although in this case, the critical argument of the Laplace resonance varies irregularly at the beginning of the integration.}

It is worth noting that all of the example solutions yield astrometric fits that differ little in quality in terms of $\Chi$ and RMS. The models are difficult to distinguish statistically and visually from each other. This may mean that for a coplanar, near-resonance, or resonance model we can hardly differentiate between perfectly regular and chaotic evolution of the system, as long as the instability time is sufficiently long relative to the age of the star. However, these two types of configurations occur near the exact Laplace resonance.

\subsection{Resonant structure of the inner debris disk}
\label{sec:debris}

Based on the updated orbital solutions in Table~\ref{tab:tab1}, we revised and extended simulations of the dynamical structure of the inner debris disk in \cite{GM2018}. These experiments rely on the concept of the so-called $\Ym$-model. We assume that the planets form a system of primaries in safely stable orbits robust to small perturbations. We then inject bodies with masses significantly smaller than those of the primaries and on orbits with different semi-major axes and eccentricities spanning the interesting region, and randomly selected orbital angles. Next, we integrate the synthetic configuration and determine its stability with the MEGNO ($\Ym$) fast indicator.  Calibration experiments spanning orbital evolution of debris disks in \hr8799{} for up to $70$~Myr are described in detail in \cite{GM2018}. A comparison of the results of direct numerical integrations with the outcomes of the $\Ym$-model confirms that these two approaches are consistent with one another, yet the $\Ym$-based method is much more CPU efficient and therefore makes it possible to obtain a clear representation of the structure of stable solutions. This algorithm is especially effective for strongly interacting systems. Also, these simulations revealed that the close proximity of the four primaries to the exact Laplace resonance makes the system stability robust even to apparently significant perturbations caused by probe masses as large as $\simeq 2$\MJup.
  
We considered two types of probe objects: Ceres-like asteroids with a mass of $10^{-6}$\MJup and Jupiter-like planets with a mass of $1$\MJup. To calculate the $\Ym$ values for the synthetic systems, we integrated the $N$-body equations of motion and their variational equations with the Gragg-Bulirsch-Stoer integrator \citep{Hairer} for $3$~Myr, which covers $6 \times 10^4$ orbital periods of \hr8799{}e and $7 \times 10^3$ orbital periods of \hr8799{}b, respectively. That integration time is consistent with the typical $10^4$ outermost orbits required to achieve $\Ym$ convergence. We also note that the most significant interactions are exerted by the inner planets. We chose the integration time that is optimal from the CPU overhead point of view.
 
The results for the less massive { Ceres-like asteroids} are illustrated in Fig.~\ref{fig:fig10} and Fig.~\ref{fig:fig11}. In this experiment, we sampled the semi-major axis $a_0 \in [4,18]$~au and eccentricity $e_0 \in [0,0.8]$ { of these objects}. We collected $3\times10^5$ $N$-body initial conditions with $|\Ym-2|<0.05$ that represent the structure of long-term-stable orbits in the inner debris disk. Subsequent panels in Figure~\ref{fig:fig10} illustrate snapshots of the disk at other epochs following the first observation ($t_0=$\epk{}) as seen on the sky plane.  To obtain the snapshots, we numerically integrated the whole set of initial conditions for planets and { the Ceres-like objects} defined at the epoch \epk{} up to the given final epoch $t$. For instance, $t=2009.575$ (middle-right panel) is for the first detection of the innermost planet in \cite{2010Natur.468.1080M}. 

We mark the model orbits of the two inner planets with red curves, and the astrometric measurements are over-plotted on them. Instant positions of the planets are marked with shaded large filled circles (for the first \hst{} epoch), and large filled circles for positions of the planets at the particular epoch labeled in the lower right corner of each panel. The time range of about 30~yr is more than half the orbital period of \hr8799{}e and roughly one orbital period of an asteroid involved in the 2:1e MMR with this planet.

Positions of the test particles are marked with different colors depending on their dynamical status: yellow and orange dots are for objects involved in 1:1e~MMR with the innermost planet \hr8799{}e; green dots are for the 3:2e MMR, blue dots are for the 2:1e~MMR, and gray dots are for the other stable asteroids. As can be seen in Fig.~\ref{fig:fig10}, the outer edge of the disk is highly nonsymmetric and quickly evolves in time. Its temporal structure strongly changes even for a relatively very short interval of 25~years spanned by the astrometric observations of the system.

The structure of the disk has a clear resonant structure that was also noted by \cite{GM2018}. This is demonstrated in Fig.~\ref{fig:fig11}, which shows the distribution of canonical osculating elements inferred in {the} Jacobi reference frame, in the $(a_0,e_0)$-plane (top panel), and on the plane of relative orbital phases, $(a_0,\lambda_{\idm{e}}-\lambda_0)$ (bottom panel). In both graphs, we mark the test particles with the same color scheme as in the previous plot, and the low-order resonances with \hr8799{}e are labeled. We also plot the geometrical collision curve (in gray) with planet \hr8799{}e constrained through the apocenter distance of the inner orbit equal to the pericenter distance of the planet, $a_0(1+e_0)=a_{\idm{e}}(1-e_{\idm{e}})$. Also, the curve depicted in red is an image of the collision curve shifted by $\Delta{}a_0={ 4}$~au towards the star and clearly marks a boundary of stable particles. 

The bottom graph shows objects involved in low-order MMRs with \hr8799{}e that are apparently widely spread on the sky plane, but they appear in narrow and well-localized islands on the planes of orbital elements; this is particularly clear on the $(a_0,\lambda_{\idm{e}}-\lambda_0)$-plane. Both graphs mark two families of asteroids in 1:1e MMR -- depicted in gold {colour} ($e_0<0.3$) and in orange ($e_0>0.3$). The low-eccentricity objects resemble classic Trojan asteroids in the Solar System, relative by $\pm 60^{\circ}$ to the planet. The second unusual family of highly eccentric particles in 1:1e MMR can be found on the sky plane far beyond the orbit of \hr8799{}e.  At some epochs (e.g., 2011.83), these appear closer to the orbit of \hr8799{}d, which may be counter-intuitive. Given the 1:1e MMR dynamical classification, they would be expected to share the same orbit as the planet.

{ This simulation reveals the dynamical structure composed of long-term-stable but mutually noninteracting asteroids in the system, on the same orbit as the innermost planet,
whose orbital evolution is governed by the gravitational interactions with the massive Jovian companions. Such objects should not be significantly affected by nongravitational
forces, such as Poynting-Robertson drag or the Yarkovsky effect.
However, such forces may be crucial for the investigation and modeling of
the actual emission profile of dust produced by collisional dynamics. The results indicate that the asteroid breakup events
may be frequent and violent. Given the large eccentricities and
the instant mixing of different resonant fractions of these objects in
the wide inner region between 4 au (and below) and 18 au,
including wide Lagrange clumps of planet HR 8799e, their
orbital velocity dispersion may be significant. The interplay
of all these factors ---and especially the clearing rate of the dust due to
radiation of the young and bright host star--- determines the
emission profile. The present observational evidence is limited, and the position of the warm disk is estimated down
to 610 au \citep[e.g.,][]{2009ApJ...705..314S,2014ApJ...780...97M,Chen}, which coincides with the 2:1e MMR (Fig.\ref{fig:fig11});
see also Sect 5. Given the topic is complex and somewhat out
of the scope of the present work, we postpone in-depth analysis
of the dynamical structure of the warm dust disk to a future
paper. }

\subsection{Putative fifth Jupiter-like planet}
\begin{figure*}
\centerline{ 
\vbox{
\hbox{
\includegraphics[width=0.33\textwidth]{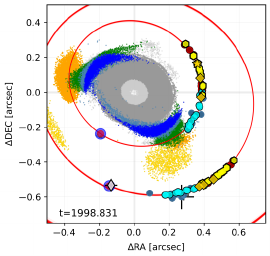}
\includegraphics[width=0.33\textwidth]{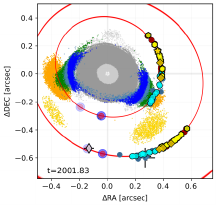}
\includegraphics[width=0.33\textwidth]{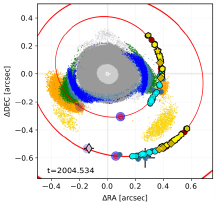}
}
\hbox{
\includegraphics[width=0.33\textwidth]{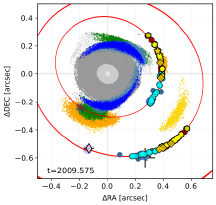}
\includegraphics[width=0.33\textwidth]{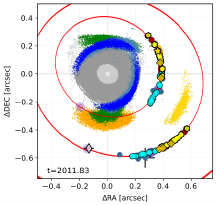}
\includegraphics[width=0.33\textwidth]{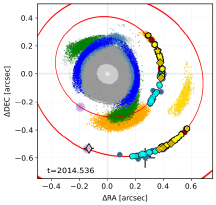}
}
\hbox{
\includegraphics[width=0.33\textwidth]{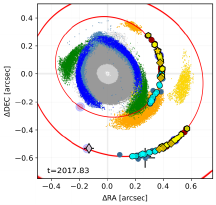}
\includegraphics[width=0.33\textwidth]{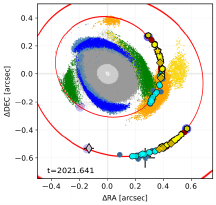}
\includegraphics[width=0.33\textwidth]{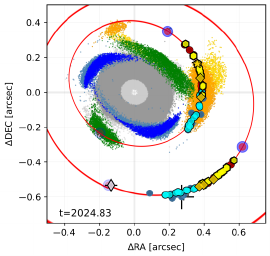}
}
}
}
\caption{
Evolution of the inner debris disk as seen on the plane of the sky.  The subsequent panels show the instantaneous positions of two inner planets and small asteroids in stable orbits (small colored dots) with a mass of $10^{-6}\,$\MJup at the epoch labeled in the lower-left corner of each graph.  We collected $3.3 \times 10^{5}$ particles. Their distribution in the $(a_0,e_0)$-plane of osculating, canonical elements at the initial epoch \epk{} is shown in Fig.~\ref{fig:fig11}.  The largest filled circles correspond to the positions of the planets \hr8799{}e and { \hr8799{}d} at the snapshot epoch (large filled circles) and the initial osculating epoch (shaded filled circles), respectively.  The asteroid colors are for the lowest order MMR types, as indicated in the bottom two plots, or those belonging to the innermost quasi-homogeneous disk (gray points). The 2024.83 epoch covers roughly one orbital period in a 2:1e~MMR resonance (dark blue circles) with the innermost planet.
}
\label{fig:fig10}
\end{figure*}

\begin{figure*}
\centerline{
\vbox{
\hbox{
\includegraphics[width=1.0\textwidth]{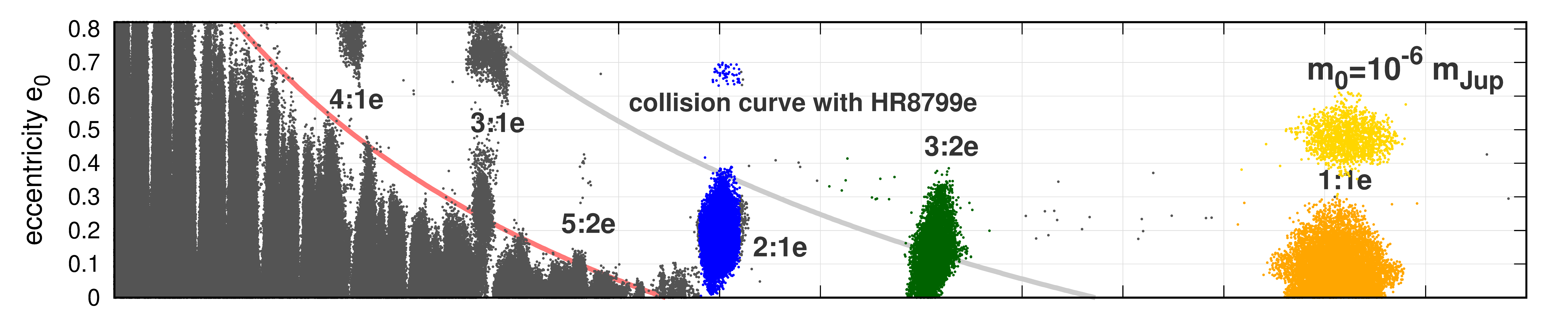} %fig12ae.pdf}
}
\vspace*{-2mm}
\hbox{
\includegraphics[width=1.0\textwidth]{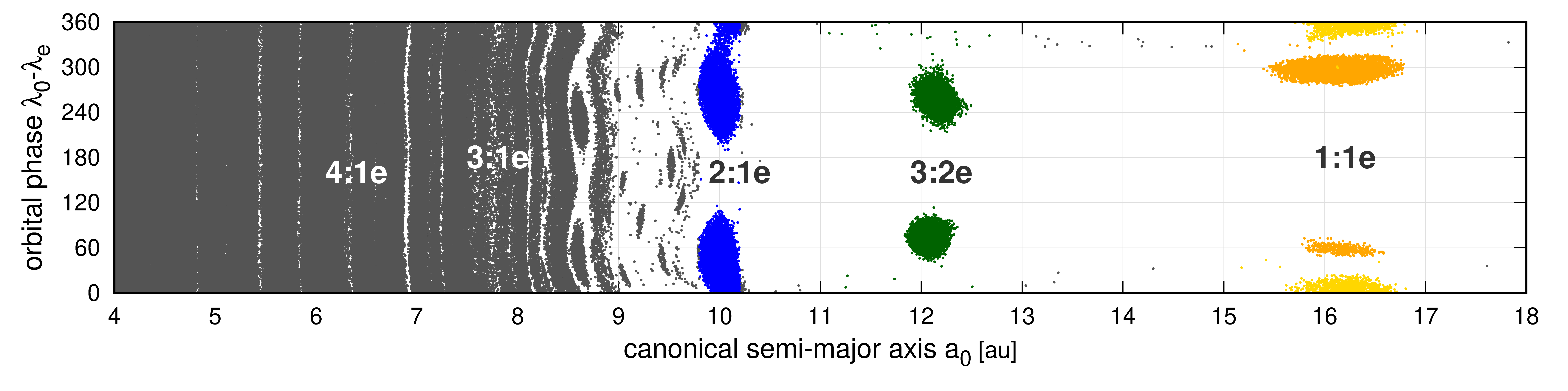} %fig12af.pdf}
}
}
}
\caption{
Structure of the inner debris disk composed of $3\times 10^5$ Ceres-like asteroids of mass $10^{-6}\,$\MJup on stable orbits, seen in the $(a_0,e_0)$-plane of Jacobian (canonical) osculating elements at the initial \epk{} epoch.  The colors of the asteroids indicate the lowest order MMR in which they are involved with \hr8799{}e, as indicated in the diagrams, or belong to the innermost, quasi-homogeneous disk (gray points) according to Fig.~\ref{fig:fig10}.  The light gray curve indicates the collision of the orbits with \hr8799{}e and the red curve is an image of the collision curve shifted by $\Delta{}a_0 \simeq { 4}$\,au toward the star. We note that loose points above the collision curve in the top panel represent immediately scattered asteroids on regular but hyperbolic (open) orbits.
}
\label{fig:fig11}
\end{figure*}

\begin{figure*}
\centerline{ 
\vbox{
\hbox{
\includegraphics[width=0.33\textwidth]{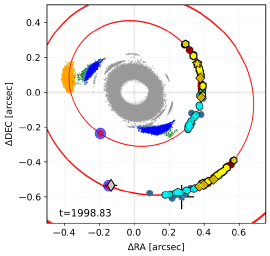}
\includegraphics[width=0.33\textwidth]{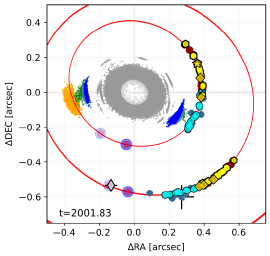}
\includegraphics[width=0.33\textwidth]{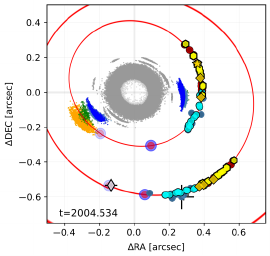}
}
\hbox{
\includegraphics[width=0.33\textwidth]{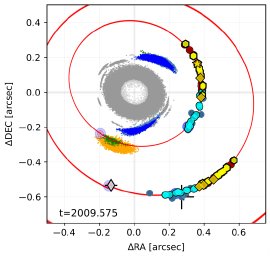}
\includegraphics[width=0.33\textwidth]{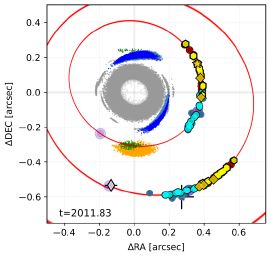}
\includegraphics[width=0.33\textwidth]{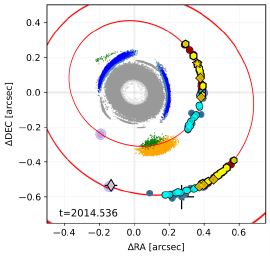}
}
\hbox{
\includegraphics[width=0.33\textwidth]{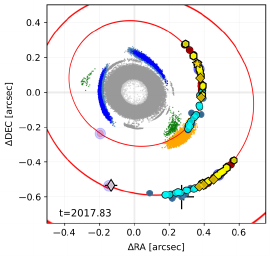}
\includegraphics[width=0.33\textwidth]{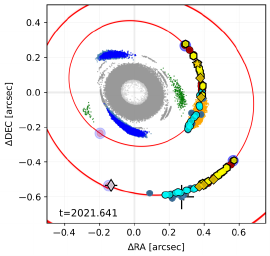}
\includegraphics[width=0.33\textwidth]{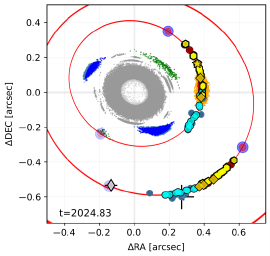}
}
\hbox{
\includegraphics[width=1.0\textwidth]{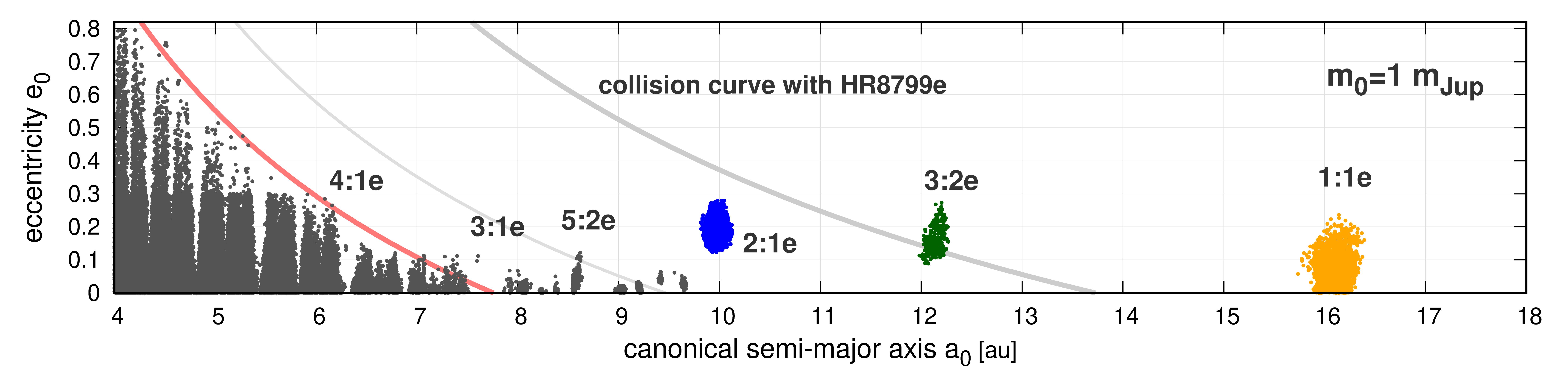}
}
}
}
\caption{
Possible instant positions (small gray and colored dots) of a hypothetical and still undetected planet of mass $1\,$\MJup on the sky plane, at the epoch labeled in the lower-left corner of each graph.  We collected $10^{5}$ initial conditions.  The bottom plot shows their distribution in the $(a_0,e_0)$--plane of canonical osculating elements at the initial epoch of \epk{}, similarly to Fig.~\ref{fig:fig11}. As in Fig.~\ref{fig:fig10}, the largest filled circles correspond to the positions of the planets \hr8799{}e and \hr8799{}d at the snapshot epoch (large filled circles) and the initial osculating epoch (shaded filled circles), respectively. 
Colors mark the lowest order MMR types, as labeled in the bottom plot.  The light-gray curve marks the collision zone with \hr8799{}e, the thinner curve marks the stability zone for $m_0 = 10^{-6}\,$\MJup (see Fig.~\ref{fig:fig10}), and the red curve is the image of the geometric collision curve shifted by $\Delta{}a_0 \simeq { 6}$\,au toward the star.
}
\label{fig:fig12}
\end{figure*}

We performed a very similar experiment for test particles of mass $1\,$\MJup, a hypothetical planet below the current detection level that may exist in the inner part of the \hr8799{} system.  The results are illustrated in Fig.~\ref{fig:fig12} in a similar way to in Figs.~\ref{fig:fig10}--\ref{fig:fig11}. We collected $10^5$ stable solutions. Again, the positions of the hypothetical planet are marked at characteristic epochs; for example, near the first epoch of the \hst{} detection, up to the last epoch of measurements in this work, and one epoch ahead of that time. Compared to the previous case, the distribution of stable objects on the plane of the sky is much narrower and more restricted. As in the case of low-mass asteroids, the positions of the putative planets can be seen to change rapidly relative to the background model orbits and astrometric measurements. 

We note that the stability limit (red curve) in the bottom plot for the $(a_0,e_0)$- plane is offset by $\Delta{}a_0=6$~au with respect to the collision curve with planet \hr8799{}e shown in gray. In addition to the clear dependence of the structure of stable solutions on the probe mass, this simulation indicates that the hypothetical planet may be located only on very narrow islands in the orbital parameter space, limited to low-order resonances with the innermost planet. The temporal evolution of these islands in the sky can be useful for analyzing AO images and provides clues as to where an additional, Jovian planet might be expected. If such objects exist beyond $\simeq 7$~au, they should be involved in a 2:1e, 3:1e, or 5:2e MMR with the inner planet, respectively. These low-order resonances may be preferred if the system has undergone a migration in the past. We discuss this further in  Sect.~\ref{Sec:his}.

Attempts to find the hypothetical innermost fifth planet have so far been unsuccessful. Very recently, \cite{2021A&A...648A..26W}, using 
\sphere{} measurements also reported here, did not detect planet~\hr8799{}f at the most plausible locations, namely 7.5 and 9.7~au, down to mass limits of 3.6 and $2.8\,$\MJup, respectively. Neither did these authors detect any new candidate companions at the smallest observable separation, of namely 0.1$"$ or $\simeq 4.1$ au, overlapping with the semi-major axes range in Fig.~\ref{fig:fig10}--\ref{fig:fig12}.  \cite{2021A&A...648A..26W} conclude that the planet may still exist with a mass of $2$-$3.6\,$\MJup at 7.5~au (3:1e~MMR) or $1.5$-$2.8\,$\MJup at 10~au, which is in the region we closely investigated in this section. The contrast curve in the interesting zone between roughly 7 and 16~au lies above roughly three~Jupiter masses, as also concluded here; see Fig.~\ref{diskplanets} and discussion in Sect.~\ref{s:planetdisk}. Therefore, we believe that steadily improved imaging { techniques}, gaining better contrast and lower detection limits, combined with dynamical simulations similar to those conducted in this section may eventually reveal the fifth planet. However, if a relatively massive planet of mass $\simeq 1\,$\MJup exists in the inner part of the system, the outer edge of the debris disk carved out by this planet would have a much smaller radius than predicted at $\simeq 10$~au under the current observational configuration of four planets (Fig.~\ref{fig:fig11}), and related to  the above-mentioned detections of warm dust at $6$--$10$~au \cite[e.g.,][]{2009ApJ...705..314S,2011ApJ...740...38H,2014ApJ...780...97M,Chen}. We defer the analysis of this situation to a future work as well, because of its complexity arising from different MMR scenarios.

\begin{table*}
\caption{
Osculating, heliocentric elements of the best-fitting solutions at the epoch of~\epk{}. The stellar mass is $m_{\star} = 1.52\,\msun$ or $m_{\star} = 1.47\,\msun$. 
Uncertainties for parameters in models IV$_{\rm PO}$ and IV$_{\rm \Delta n}$ are determined as the 16th and 86th percentiles of the samples. Parameter values with { 8--9}~significant digits are provided in order to reproduce the results of particular integrations.
Parameter $\nu=N_{\rm obs} - \dim{\vec{p}}$ is for the degrees of freedom, where $N_{\rm obs}$ denotes the number of $(\alpha,\delta)$ measurements, and $\dim{\vec{p}}$ is the number of free parameters.
%\ednote{This table must be checked numerically, its a draft}.
}  
\label{tab:tab1}
\centering
\begin{tabular}{l c c c c c c c}
%\hline
%\hline
%& $m\,[\mJ]$ & $a\,$[au] & $e$ & $I\,$[deg] & $\Omega\,$[deg] & $\varpi\,$[deg] & $\Mmean\,$[deg] \\
%\hline
%\hline
%\multicolumn{8}{c}{}\\
\multicolumn{8}{c}{ 
{ Model IV$_{\rm PO}$}: \quad  
$m_{\star} = 1.52\,\msun$, \quad
$\Pi = 24.2585686$\,{}mas $(24.26\pm 0.05)$\,{}mas, 
\quad  $\Chinu=4.32$, { $\nu=441$}, $\dim{\vec{p}}=11$
}\\
& $m\,[\mJ]$ & $a\,$[au] & $e$ & $I\,$[deg] & $\Omega\,$[deg] & $\varpi\,$[deg] & $\Mmean\,$[deg] \\
\hline
%\hline
\hr8799{}e & $8.1 \pm 0.7$ & $16.28 \pm 0.03$ & $0.147 \pm 0.001$  & & & $110.8 \pm 0.5$ & $336.8 \pm 0.5$\\
 &  8.24465258   &  16.2817072  & 0.147508399 & & &  110.765660  & 336.793683 \\
%8.24465258088969e+00 1.62817071967698e+01 1.47508399523400e-01 2.68923597483684e+01 6.22416411783964e+01 1.10765660337342e+02 -2.32063170342185e+01 
\hr8799{}d & $9.5 \pm 0.1$ & $26.78 \pm 0.08$ & $0.115 \pm 0.001$  & & & $29.4 \pm 0.9$ & $59.8 \pm 0.8$ \\
& 9.48438158 &  26.7888442  & 0.01154150 & & &  29.4676779   & 59.6448288 \\
% 9.48438158130628e+00 2.67888442412372e+01 1.15415020996667e-01 2.68923597483684e+01 6.22416411783964e+01 2.94676779296498e+01 5.96448287848556e+01
\hr8799{}c & $7.7 \pm 0.5$ & $ 41.40 \pm 0.10$ & $0.055 \pm 0.002$  & $26.9 \pm 0.2$ & $62.2 \pm 0.4$ & $92.1 \pm 0.5$ & $ 145.6 \pm 0.8$\\
& 7.64047381 &  41.3945527    &  0.05485344 &  26.8923597 & 62.2416412 & 92.1126122   & 145.7203489 \\
% 7.64047380981731e+00 4.13945526663305e+01 5.48534408400912e-02 2.68923597483684e+01 6.22416411783964e+01 9.21126121974773e+01 1.45720348992691e+02 
\hr8799{}b & $ 6.0\pm 0.4$ & $71.95 \pm 0.13$ & $0.017 \pm 0.001$ & & & $
44.4 \pm 3.0$ & $309.2 \pm 2.6$\\
& 6.01476193 & 71.9477733 &  0.017597398  & & &  44.6384383 &  309.291519 \\
%6.01476192530805e+00 7.19477732572401e+01 1.75973980490390e-02 %2.68923597483684e+01 6.22416411783964e+01 4.46384383455767e+01 -5.07084815358445e+01=309.2915184641555
%2.42585685737366e-02 2.68923597483684e+01 6.22416411783964e+01 1.57243044842289e+02 62.241641 44.638438  -50.708482
%
\hline
\hline
\multicolumn{8}{c}{}\\
%\hline
%\hline
\multicolumn{8}{c}{ 
{ Model IV$_{\rm PO}$}: \quad 
$m_{\star} = 1.47\,\msun$, \quad
$\Pi = 24.5256617$\,{}mas $(24.53\pm 0.04)$\,{}mas, 
\quad $\Chinu=4.32$, { $\nu=441$},
$\dim{\vec{p}}=11$
}\\
& $m\,[\mJ]$ & $a\,$[au] & $e$ & $I\,$[deg] & $\Omega\,$[deg] & $\varpi\,$[deg] & $\Mmean\,$[deg] \\
\hline
%\hline
\hr8799{}e & $7.6 \pm 0.9$ & $16.10 \pm 0.03$ & $0.148 \pm 0.001$  & & & $110.8 \pm 0.5$ & $336.8 \pm 0.3$\\
 &  7.4060918  &  16.1038912  &  0.147675953 & & & 110.816707   & 336.828050 \\
% 7.40609179465660e+00 1.61038911951884e+01 1.47675953591131e-01 2.68734270906510e+01 6.21852189006772e+01 1.10816706677204e+02 -2.31719499808413e+01=336.8280500191587
\hr8799{}d & $9.2 \pm 0.1$ & $26.46 \pm 0.09$ & $0.115 \pm 0.002$  & & & $29.1 \pm 1.0$ & $60.0 \pm 1.0$ \\
& 9.18999484 & 26.45172667   & 0.114600334 & & &   29.0594738   & 60.1752106 \\
% 9.18999484340388e+00 2.64517266650185e+01 1.14600334897462e-01 2.68734270906510e+01 6.21852189006772e+01 2.90594738297464e+01 6.01752105961338e+01
\hr8799{}c & $7.7 \pm 0.7$ & $ 40.97 \pm 0.10$ & $0.054 \pm 0.002$  & $26.9 \pm 0.2$ & $62.2 \pm 0.4$ & $92.3 \pm 0.6$ & $ 145.5 \pm 0.7$\\
& 7.79611845 &  40.9861801    &  0.05380902 & 26.8734271  & 62.1852189 & 92.3879462  & 145.5347082 \\
% 7.79611845337560e+00 4.09861801411293e+01 5.38090156109975e-02 2.68734270906510e+01 6.21852189006772e+01 9.23879461914203e+01 1.45534708232992e+02
\hr8799{}b & $ 5.8\pm 0.4$ & $71.14 \pm 0.15$ & $0.017 \pm 0.002$ & & & $
43.50 \pm 3.0$ & $310.35 \pm 3.0$\\
& 5.80235941 & 71.1427851  &  0.01667073  & & &  42.9704380 & 310.895065  \\
% 5.80235940619091e+00 7.11427850957351e+01 1.66707344313247e-02 2.68734270906510e+01 6.21852189006772e+01 4.29704380301362e+01 -4.91049353500623e+01=-4.91049353500623e+01
% 2.45256616913822e-02 2.68734270906510e+01 6.21852189006772e+01 1.57386105286261e+02 62.185219 42.970438  -49.104935
\hline
\hline
\multicolumn{8}{c}{}\\
%\hline
%\hline
\multicolumn{8}{c}{ 
% fit 11
{ Model IV$_{\rm \Delta n}$}: \quad
$m_{\star} = 1.47\,\msun$, \quad
$\Pi = 24.5256617$\,{}mas $(24.49 \pm 0.07)$\,{}mas, 
\quad $\Chinu=3.9$, {$\nu=429$}, $\dim{\vec{p}}=23$
} \\
& $m\,[\mJ]$ & $a\,$[au] & $e$ & $I\,$[deg] & $\Omega\,$[deg] & $\varpi\,$[deg] & $\Mmean\,$[deg] \\
\hline
%\hline
%\hline
\hr8799{}e & $7.5 \pm 0.6$ & $16.00 \pm 0.10$ & $0.150 \pm 0.004$  & & & $108.8 \pm 1.3$ & $336.00 \pm 1.5$\\
 &  7.40609179  &  16.1038912  &  0.147675954 & & & 110.816707   & 336.828050 \\
% 7.40609179465660e+00 1.61038911951884e+01 1.47675953591131e-01 2.68734270906510e+01 6.21852189006772e+01 1.10816706677204e+02 -2.31719499808413e+01=336.8280500191587
\hr8799{}d & $9.3 \pm 0.5$ & $26.55 \pm 0.12$ & $0.114 \pm 0.004$  & & & $28.66 \pm 1.2$ & $60.0 \pm 1.0$ \\
& 9.18999484 & 26.45172667   & 0.11460033 & & &   29.05947383   & 60.1752106 \\
% 9.18999484340388e+00 2.64517266650185e+01 1.14600334897462e-01 2.68734270906510e+01 6.21852189006772e+01 2.90594738297464e+01 6.01752105961338e+01
\hr8799{}c & $7.8 \pm 0.6$ & $ 41.00 \pm 0.15$ & $0.050 \pm 0.003$  & $26.5 \pm 0.5$ & $63.1 \pm 0.5$ & $91.8 \pm 1.0$ & $ 145.0 \pm 1.0$\\
& 7.79611845 &  40.9861801    &  0.05380902 & 26.8734271  & 62.1852189 & 92.3879462  & 145.5347082 \\
% 7.79611845337560e+00 4.09861801411293e+01 5.38090156109975e-02 2.68734270906510e+01 6.21852189006772e+01 9.23879461914203e+01 1.45534708232992e+02
\hr8799{}b & $ 5.8\pm 0.8$ & $71.3 \pm 0.2$ & $0.017 \pm 0.002$ & & & $
42.3 \pm 1.6$ & $310.8 \pm 2.0$\\
& 5.80235941 & 71.1427851  &  0.016670734  & & &  42.9704380 & 310.895065  \\
% 5.80235940619091e+00 7.11427850957351e+01 1.66707344313247e-02 2.68734270906510e+01 6.21852189006772e+01 4.29704380301362e+01 -4.91049353500623e+01=310.8950646499377
% 2.45256616913822e-02 2.68734270906510e+01 6.21852189006772e+01 1.57386105286261e+02 62.185219 42.970438  -49.104935
\hline
\hline
\end{tabular}
\end{table*}

\begin{table*}
\caption{
Osculating, heliocentric elements of the best-fitting solutions at the epoch
of $1998.83$ for models illustrated in Figure~\ref{fig:fig4} and in Figs.~\ref{fig:fig7}--\ref{fig:fig8}. 
}  
\label{tab:tab3}
\centering
\begin{tabular}{rrrrrrrr}
\multicolumn{8}{c}{{ Model 1}: \quad $m_{\star}=$1.47 $M_{\odot}$,   \quad $\Pi=$   24.525662 mas,   \quad $\chi^2$=1593.22,   \quad RMS$=$7.41~mas, \quad $\log |\Delta{}n[\mbox{rad}\,\mbox{d}^{-1}]|$=-12.0}\\
& $m\,[\mJ]$ & $a\,$[au] & $e$ & $I\,$[deg] & $\Omega\,$[deg] & $\omega\,$[deg] & $\Mmean\,$[deg] \\
\hline
HR8799e &7.4060918 & 16.1038912 & 0.1476760 & 26.8734271 & 62.1852189 & 110.8167067 & -23.1719500 \\
HR8799d &9.1899948 & 26.4517267 & 0.1146003 & 26.8734271 & 62.1852189 & 29.0594738 & 60.1752106 \\
HR8799c &7.7961185 & 40.9861801 & 0.0538090 & 26.8734271 & 62.1852189 & 92.3879462 & 145.5347082 \\
HR8799b &5.8023594 & 71.1427851 & 0.0166707 & 26.8734271 & 62.1852189 & 42.9704380 & -49.1049354 \\
\hline
\hline
\\
\multicolumn{8}{c}{{ Model 2}: \quad $m_{\star}=$1.47 $M_{\odot}$,   \quad $\Pi=$   24.451807 mas,   \quad $\chi^2$=1568.19,   \quad RMS$=$7.66~mas, \quad $\log |\Delta{}n[\mbox{rad}\,\mbox{d}^{-1}]|$=-6.2}\\
& $m\,[\mJ]$ & $a\,$[au] & $e$ & $I\,$[deg] & $\Omega\,$[deg] & $\omega\,$[deg] & $\Mmean\,$[deg] \\
\hline
HR8799e &7.8680000 & 15.9753973 & 0.1524708 & 26.2394739 & 63.7770393 & 108.3445443 & 335.6210695 \\
HR8799d &9.6490000 & 26.6396632 & 0.1145654 & 26.2394739 & 63.7770393 & 28.3126706 & 59.7394429 \\
HR8799c &7.3081300 & 40.9406369 & 0.0517422 & 26.2394739 & 63.7770393 & 91.3551907 & 145.0637721 \\
HR8799b &5.6590700 & 71.3289844 & 0.0173411 & 26.2394739 & 63.7770393 & 41.6627885 & 310.5199311 \\
\hline
\hline
\\
\multicolumn{8}{c}{{ Model 3}: \quad $m_{\star}=$1.52 $M_{\odot}$,   \quad $\Pi=$   24.347839 mas,   \quad $\chi^2$=1624.73,   \quad RMS$=$7.66~mas, \quad $\log |\Delta{}n[\mbox{rad}\,\mbox{d}^{-1}]|$=-4.5}\\
& $m\,[\mJ]$ & $a\,$[au] & $e$ & $I\,$[deg] & $\Omega\,$[deg] & $\omega\,$[deg] & $\Mmean\,$[deg] \\
\hline
HR8799e &8.9817200 & 16.1163303 & 0.1548634 & 27.3470617 & 62.9005413 & 108.0955330 & 336.2997163 \\
HR8799d &9.0680400 & 26.8768634 & 0.1068892 & 27.3470617 & 62.9005413 & 29.2629241 & 60.0277429 \\
HR8799c &7.2212900 & 41.0832702 & 0.0627005 & 27.3470617 & 62.9005413 & 91.9708727 & 144.7587008 \\
HR8799b &6.1950900 & 71.7822044 & 0.0192741 & 27.3470617 & 62.9005413 & 44.4483815 & 308.9107410 \\
\hline
\hline
\\
\multicolumn{8}{c}{{ Model 4}: \quad $m_{\star}=$1.52 $M_{\odot}$,   \quad $\Pi=$   24.305464 mas,   \quad $\chi^2$=1564.70,   \quad RMS$=$7.64~mas, \quad $\log |\Delta{}n[\mbox{rad}\,\mbox{d}^{-1}]|$=-6.2}\\
& $m\,[\mJ]$ & $a\,$[au] & $e$ & $I\,$[deg] & $\Omega\,$[deg] & $\omega\,$[deg] & $\Mmean\,$[deg] \\
\hline
HR8799e &10.6577500 & 16.1274692 & 0.1496852 & 26.9747476 & 63.6149718 & 107.6133926 & 336.6560551 \\
HR8799d &9.5739000 & 26.8852716 & 0.1169394 & 26.9747476 & 63.6149718 & 27.5720349 & 60.0200173 \\
HR8799c &6.8413200 & 41.2618982 & 0.0555455 & 26.9747476 & 63.6149718 & 93.1349542 & 143.0235158 \\
HR8799b &5.7085500 & 71.9446080 & 0.0217920 & 26.9747476 & 63.6149718 & 45.1149621 & 307.6876430 \\
\hline
\hline
\\
\multicolumn{8}{c}{{ Model 5}: \quad $m_{\star}=$1.47 $M_{\odot}$,   \quad $\Pi=$   24.466345 mas,   \quad $\chi^2$=1591.33,   \quad RMS$=$7.63~mas, \quad $\log |\Delta{}n[\mbox{rad}\,\mbox{d}^{-1}]|$=-6.9}\\
& $m\,[\mJ]$ & $a\,$[au] & $e$ & $I\,$[deg] & $\Omega\,$[deg] & $\omega\,$[deg] & $\Mmean\,$[deg] \\
\hline
HR8799e &7.1130900 & 15.9976012 & 0.1545297 & 26.5969851 & 62.8989533 & 108.8687608 & 336.1919760 \\
HR8799d &9.8864900 & 26.5290674 & 0.1122973 & 26.5969851 & 62.8989533 & 27.8297522 & 61.2151462 \\
HR8799c &7.7520500 & 41.1188833 & 0.0497536 & 26.5969851 & 62.8989533 & 92.4070918 & 144.9763772 \\
HR8799b &5.5380200 & 71.3013150 & 0.0172797 & 26.5969851 & 62.8989533 & 44.6821435 & 308.5634968 \\
\hline
\hline
\\
\multicolumn{8}{c}{{ Model 6}: \quad $m_{\star}=$1.47 $M_{\odot}$,   \quad $\Pi=$   24.539234 mas,   \quad $\chi^2$=1568.59,   \quad RMS$=$7.63~mas, \quad $\log |\Delta{}n[\mbox{rad}\,\mbox{d}^{-1}]|$=-6.2}\\
& $m\,[\mJ]$ & $a\,$[au] & $e$ & $I\,$[deg] & $\Omega\,$[deg] & $\omega\,$[deg] & $\Mmean\,$[deg] \\
\hline
HR8799e &7.3387300 & 15.9246991 & 0.1553777 & 26.5990408 & 62.7197878 & 108.3820177 & 336.0139713 \\
HR8799d &10.0516400 & 26.5092120 & 0.1117405 & 26.5990408 & 62.7197878 & 29.2671129 & 59.8339479 \\
HR8799c &8.0973100 & 40.9565104 & 0.0509428 & 26.5990408 & 62.7197878 & 91.8212168 & 145.6257159 \\
HR8799b &5.8932800 & 71.1073028 & 0.0167537 & 26.5990408 & 62.7197878 & 41.9648749 & 311.2692413 \\
\hline
\hline
\\
\end{tabular}

\end{table*}

\section{Possible history of the system}
\label{Sec:his}
 \label{history}
 
 Convergent migration due to tidal interaction with the nesting circumstellar disk is thought to be the most reliable mechanism for producing resonance trapping in  multiple-planet systems \citep{masset2001,lee2002,adams2005,thommes2005,beauge2006,crida2008,dangelo2012}. The different masses of the planets and progressive inside-out depletion of gas in the disk due to the presence of the planet itself and photo-evaporation lead to different migration speeds for the planets, which may end up in resonance. Numerical modeling by  \cite{hands2014} and \cite{szuskie2012} shows that during inward migration, planets are often trapped in resonances like the 2:1 and 3:2 (the most frequent). 

While in most resonant cases, such as   Gliese 876, HD 82943, and HD 37124 \citep{wright2011}, the planets are close to the star, in the case of HR\,8799 the planets are significantly far away.  This implies that the primordial disk from which they formed should have extended beyond 100 au which is compatible with the mass of the star being $\sim$1.5~\MSun. The most robust scenario is that in which the planets  were fully formed further out in the disk and then migrated inwards until they became progressively trapped in the multiple resonances.  This scenario raises the question of whether these planets were formed by core accretion \citep{1980PThPh..64..544M, 1996Icar..124...62P} or by gravitational instability \citep{1978M&P....18....5C, 1997Sci...276.1836B}. 

\begin{figure}
\centering
\includegraphics[width=0.9\columnwidth]{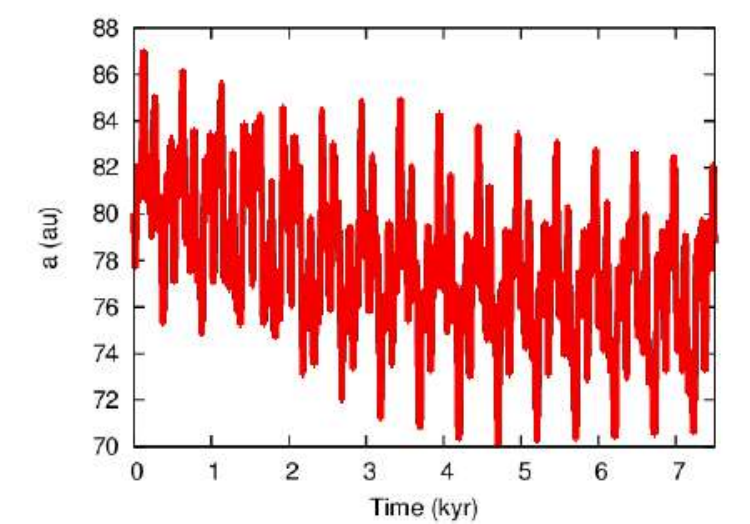}
\includegraphics[width=0.9\columnwidth]{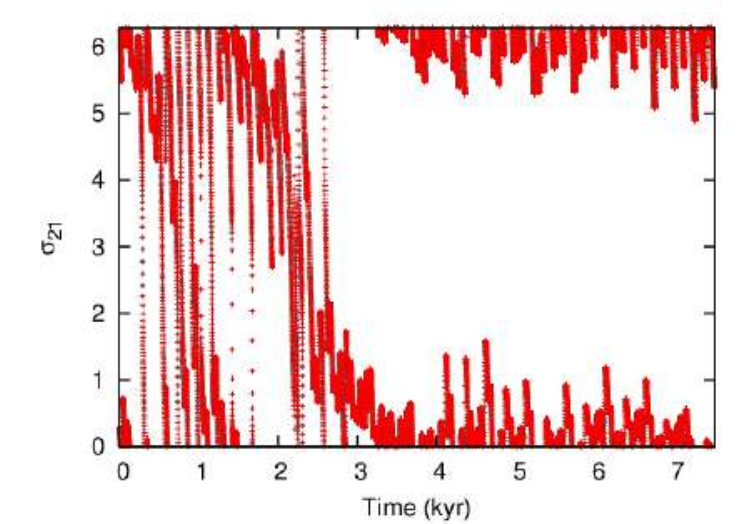}
\includegraphics[width=0.9\columnwidth]{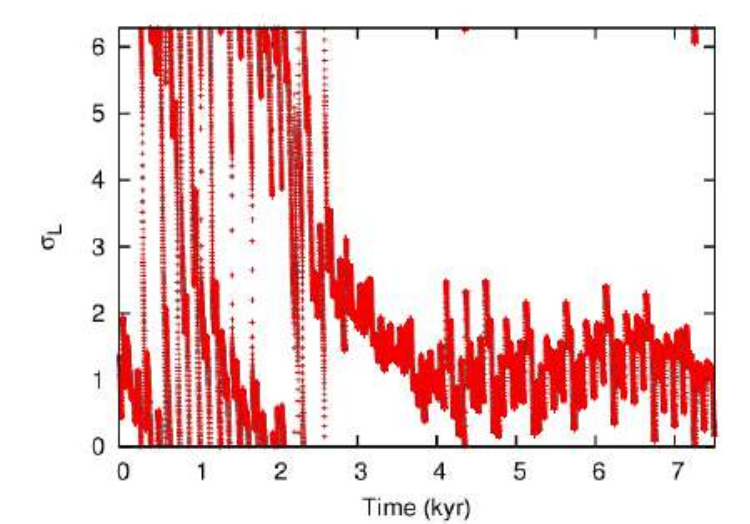}
\caption{{\it Resonance trapping of the \hr8799{} system during the migration phase.}
The upper panel shows the evolution of the semi-major axis of the outer planet at the moment of capture in resonance with the inner three bodies.   After the resonance trapping, the semi-major axis remains constant. The middle panel shows the critical argument of the resonance between the two outer planets. After the capture in resonance, the critical argument librates.  The same occurs for the Laplace critical argument shown in the bottom panel. 
}
\label{fig:mig1}
\end{figure}

The resonance trapping in this system at that distance from the star is confirmed by hydrodynamical simulations performed with the FARGO code \citep{2000A&AS..141..165M}. Fully radiative models have been adopted where the energy equation contains viscous heating and radiative cooling through the disk surface.  A polar grid with $682\times 512$ elements is used to cover the disk, extending from 1 to 120 au. The initial surface gas density is given by $\Sigma = \Sigma_0 r^{-1/2}$ with $\Sigma_0 = 50  $ g$/$cm$^2$. This low density is motivated by the evolved state of the disk when the planets are fully formed. To test the sequential resonance trapping, we start the inner planets on already resonant orbits; they are not affected by the disk,  and so do not migrate. The outer planet instead feels the disk perturbations and migrates inward until it is trapped in the 2:1 resonance and stops migrating. This behavior is illustrated in Fig.~\ref{fig:mig1} where in the top panel the semi--major axis of the fourth planet is shown while in the middle and bottom panels the critical argument of the 2:1 resonance between the third and fourth planet and the Laplace resonance argument is illustrated.  

\begin{figure}
%\centering
\hspace{-1.cm}
\includegraphics[width=1.2\columnwidth]{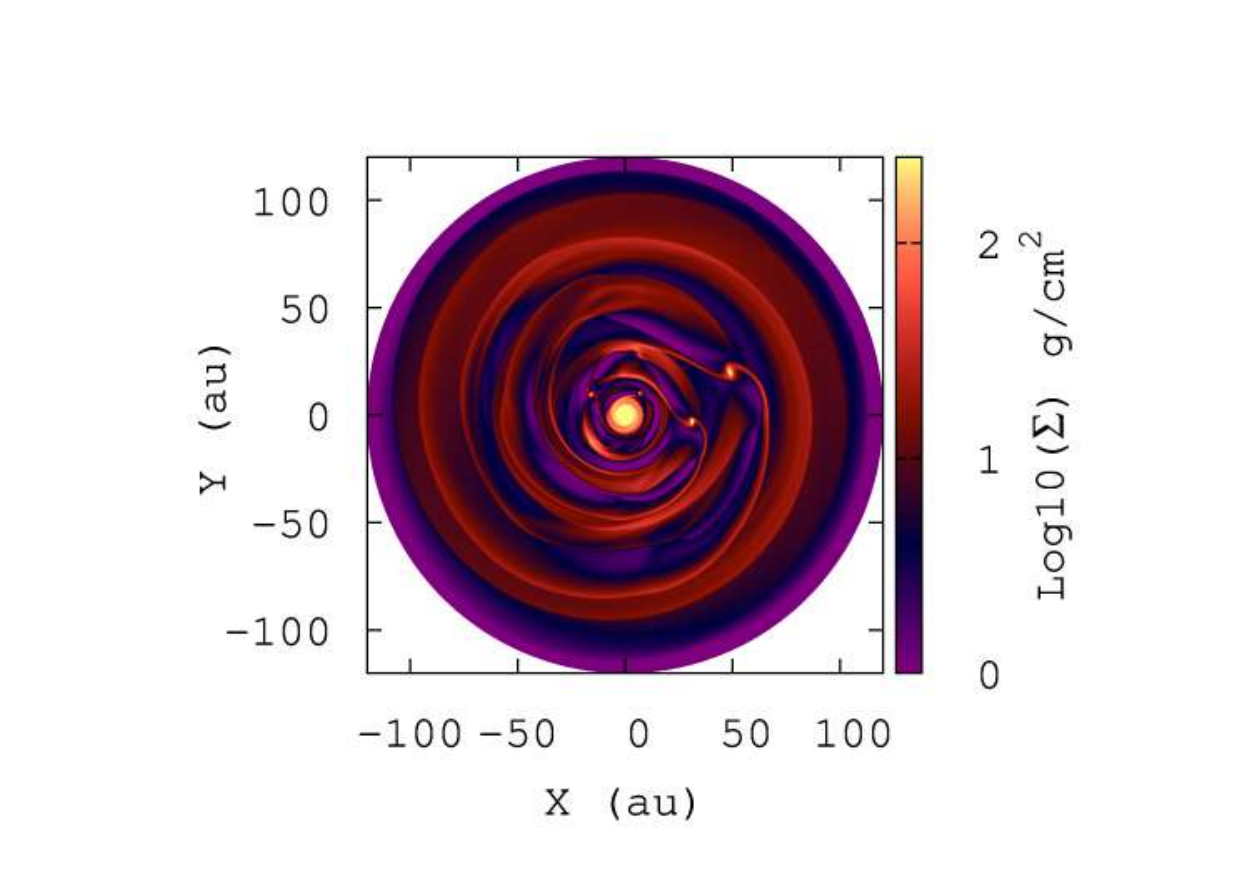}
\includegraphics[width=1.0\columnwidth]{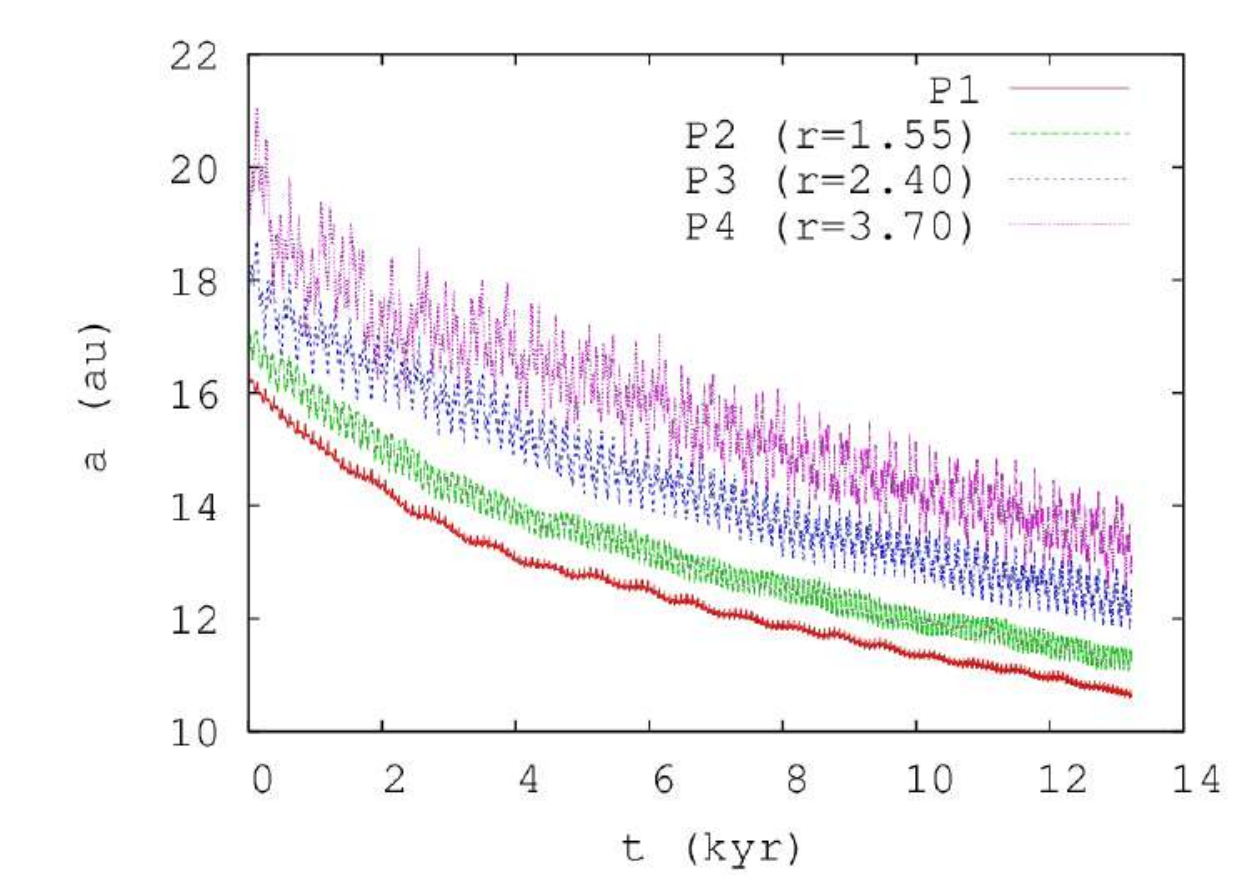}
\caption{
{\it Orbital evolution of the \hr8799{} system during the migration phase.}
The upper panel shows the gas density of the disk. The bottom panel shows the semi-major axes of the planets rescaled with constant values for ease of visualization in the same plot. 
}
\label{fig:mig2}
\end{figure}

It is noteworthy that, once all the planets are trapped in resonance, they share a common gap and migrate inwards. We performed an additional simulation where all four planets are affected by the disk perturbations and can freely migrate. In this simulation, we increased $\Sigma_0$ to 100 g$/$cm$^2$ in order to speed up the migration.   Figure~\ref{fig:mig2}  shows the gas density distribution after 8 Kyr from the beginning of the simulation. The planets create a common gap where some overdensities are still present in the corotation regions of each planet. The semi-major axes of the planets are illustrated in the bottom panel, rescaled to be shown in a single plot. They migrate inwards while trapped in resonance, suggesting that the present position of the planets is not necessarily the one where they became trapped in resonance. The capture in the mutual resonance may have occurred farther out, followed by an inward migration until the disk dissipates. This may push the location where the planet initially grew  even farther out. It is noteworthy that the wider oscillations observed in the critical arguments of the resonances and the semi-major axis of the planets compared to a pure $N$--body problem are related to the presence of the perturbations of the disk on the planets. 

\section{Planets--disk interaction}
\label{s:planetdisk}
Along with the four giant planets detected so far, the architecture of \hr8799{} is enriched by the presence of an extended debris disk with two components. The cold Kuiper-like component was  resolved in the FIR \citep{2014ApJ...780...97M} and at millimeter wavelengths \citep{Booth,Wilner} and was detected up to 450 au from the star, with a halo extending to thousands of astronomical units (au). The warm component was instead never {fully} resolved and its inferred position of $9.3$~au is given by modeling the IR excess at 155~K \citep{Chen} in the SED of the star under the assumption of dust particles behaving like black bodies. The two components are separated by a huge dust-free gap that extends from 109 au \citep{Wilner} to regions interior to the orbit of \hr8799{}e.

As the presence of the gap is tightly correlated with the planets in it, we can assess the planets--disk interactions and infer whether or not the planets detected are responsible for carving the entire gap or whether or not there is further dynamical space for hosting other undetected planets. For this analytical study, the only planets taken into account are the two closest to the edges of the gap, HR\,8799 b and e. The orbital parameters and masses adopted are the ones listed in Tables~\ref{tab:tab1} {and~\ref{tab:tab3}} for Model~1. To calculate the extension of the region from which dust particles are scattered away from the orbit of the planet (chaotic zone), we used Eqs.~9 and 10 of \cite{Lazzoni} for \hr8799{}b and \hr8799{}e, respectively. As a result, we obtained that the chaotic zone of the inner planet extends down to 9.8~au, { which is consistent with the direct $N$-body simulations in Sect.~\ref{sec:debris}}. Given the uncertainties on the position of the inner belt and on its width, we can safely state that \hr8799{}e is likely responsible for shaping the inner edge of the gap. On the other hand, the chaotic zone for \hr8799{}b can clear the orbit of the planet up to 91.4~au. Given the position of the outer edge of the gap at 109~au, we can try to infer the characteristics of a fifth planet able to extend the chaotic zone up to the detected position of the edge.  Following a very simple approach, we can use the same analytical tools to estimate the mass and semi-major axis of a planet able to carve a gap extending from 91.4 au to 109 au, that is, the free dynamical region left beyond the orbit of \hr8799{}b. As a result, we obtain a $0.12$\,\MJup planet orbiting at 100~au.

\begin{figure}
\centering
\includegraphics[width=\columnwidth]{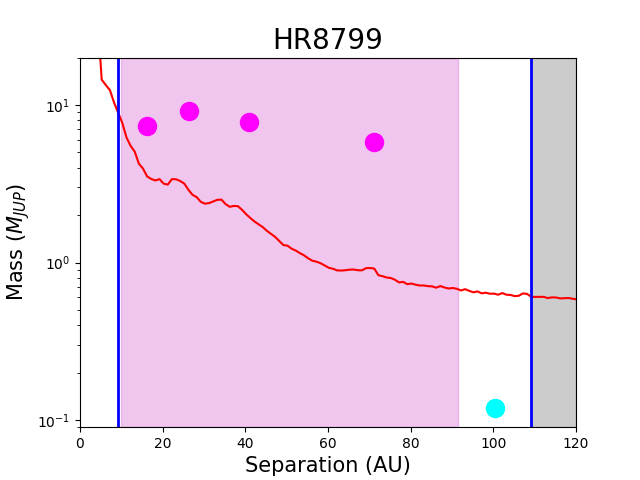}
\caption{
\hr8799{} architecture: two belts (blue lines), four detected planets (pink circles), and a putative fifth planet (light-blue circle). The pink shaded area represents the extension of the chaotic zone whereas the gray shaded zone represents the extension of the outer disk. The contrast curve for the system is shown in red.
}
\label{diskplanets}
\end{figure}

Figure \ref{diskplanets} shows a summary of the results discussed above. The four detected planets are represented as pink circles together with the extension of their chaotic zone (pink shaded area). As it emerges from the image, the inner extension of the clearing zone gets very close to the inner belt (blue vertical line) whereas it stops roughly $20$~au from the outer belt. Even a very low mass planet could carve the remaining part of the gap and, consistently with our observations, it would be too small to be detected with an instrument such as SPHERE, as proven by the detection limits (red curve; IRDIS data taken on 2017 October 12) shown in the figure.
{Simulations of the inner debris disk in the framework of the $N$-body resonant model of the system in Sect.
\ref{sec:debris} indicate possible localizations of such a planet.} We want to stress that this is only a preliminary analysis of the planet--disk interaction. More detailed studies will have to consider the dynamical interaction of the fifth planet with all four of the detected ones as well as the possibility of having a larger object locked in MMR with at least one of the other known planets, as discussed in the following.

{
Indeed, the results of numerical $N$-body simulations in \cite{GM2018} based on the $\Ym$-model described in Sect.~\ref{sec:debris} partially confirm the analytical predictions and address the likely real resonant configuration of the planets. These latter authors simulated the inner edge of the outer debris disk, also accounting for the presence of a hypothetical fifth planet with a~mass between $0.1$ and $1.66$~\MJup. These simulations are based on a quasi-periodic, near-resonant orbital model of the four known planets derived through migration simulations. However, given that \cite{GM2018} adopted the larger mass for the star of $m_{\star}=1.52$\MSun and a larger parallax $\Pi=25.4$~mas, the whole system appears more compact than predicted at present --- these authors found the osculating astrocentric semi-major axis of HR~8799b $\simeq 67.1$~au, compared to $\simeq 71$~au in the present models. The updated values of the parallax and the stellar mass translate to orbits expanded by a~few au. According to the simulations, the inner edge of the outer disk is globally highly nonsymmetric, and planets with a mass of between $0.1$ and $1.66$~\MJup may be present beyond the clearing zone, with semi-major axis $>90$~au. These planets could be involved in low-order resonances, such as 3:2b, 5:3b, 2:1b, or 5:2b with the outermost planet \hr8799{}b and in low and moderate eccentricity orbits; see Fig.~9 in \cite{GM2018}. The fifth planet,   even if it had a small mass, would be strongly affecting the shape of the inner parts of the outer debris disk.
}

\section{Conclusions}
\label{sec:con}
The system around HR\,8799 is a unique laboratory with which to study the mutual gravitational interactions between the planets and their relation with the circumstellar disk in the early evolutionary stages of  the system. To better understand the dynamics of this system, we followed up \hr8799{} with SPHERE at the Very Large Telescope and with LUCI at the Large Binocular Telescope to refine the orbital parameters of the system. We reduced the new data with state-of-the-art algorithms that apply the ADI technique, and for consistency, we also repeated the reduction of published SPHERE data from open-time programs. Precise astrometry for the four planets was obtained for 21 epochs from SPHERE and one epoch from LUCI. 

%{The estimates of the planetary masses were calculated using analytic and numerical dynamical models leading to} a value of $8$--$9$\MJup for planets \hr8799{}e,d,c, {while for the exterior planet \hr8799{}b we estimated a smaller mass} ($\simeq$ 6 \MJup),  consistent with other estimates in the literature.

We performed a detailed exploration of the orbital parameters for the four~planets with dynamical constraints imposed by the 8:4:2:1 Laplace resonance. This model was updated using 68 epochs from our reduction and the literature. As a result, we re-derived the dynamical masses of the planets and the parallax of the system with minimal prior information. The derived masses are consistent with the prediction from the evolutionary models and allow long-term stability of the orbits (over a~few hundred million years), even if they are $\sim$2 \MJup bigger than the masses proposed in the previous dynamical studies. We find masses of $8$--$9$\MJup for planets \hr8799 {}e, d, and c, while for the exterior planet \hr8799{}b we estimate a smaller mass of $\simeq$ 6 \MJup. Moreover, the dynamical parallax is consistent with $1\sigma$~uncertainty with the corrected, independently determined GAIA eDR3 value of $24.50 \pm 0.05$~mas, reinforcing both the adopted mass of the host star  $m_\star=1.47\,\msun$ and the assumed resonant or close-to-resonant configuration of the system.

Regarding the quality of the $N$-body astrometric models considered in this work, we did not find any significant or qualitative improvement of multi-parameter near-resonant configurations parametrized with 23 free orbital elements and masses compared to the  previously proposed,  strictly resonant periodic configuration described by 11~free parameters {by GM2020}. However, given that near-resonant, chaotic configurations may survive for hundreds of millions of years, it is still very difficult to differentiate such solutions from a strictly resonant configuration of the system. On the other hand, the well-defined PO $N$-body model with a minimal number of free parameters may be used as a~template solution in numerical experiments and simulations requiring sufficiently long stable orbital evolution of the planets. These may include simulations of debris disks and yet undetected objects.

As the system shows planets locked in MMRs, we ran numerical simulations to explore the possible migration histories which could have led to the present architecture of the system. Hydrodynamical modeling suggests that the planets migrated inwards and that a different migration speed or different formation times drove the system into the present quasi-exact resonance. 

We looked for the dynamical space where a putative fifth planet could sit using analytical formulae and estimates, and suggest that only a low-mass planet in the outer part of the system would agree with the currently  accepted scenario. Its location would be in between planet \hr8799{}b and the external debris belt, and its mass would be too small (<$0.1$ \MJup) to be detected by current instrumentation for high-contrast imaging. On the other hand, the interior belt is heavily shaped by planet~\hr8799{}e, and there is no need for an additional inner planet to explain it. However, we also determined the dynamical structure of this region with direct $N$-body simulations. The results may be useful for predicting the positions of small-mass objects below the present detection limits of $\simeq { 3}$\MJup.

\begin{acknowledgements}
  { We are thankful to the referee for the constructive report}. A.Z. acknowledges support from the FONDECYT Iniciaci\'on en investigaci\'on project number 11190837 and ANID -- Millennium Science Initiative Program -- Center Code NCN2021\_080. K.G. is very grateful to Dr Cezary Migaszewski for sharing a template $C$ code for computing periodic orbits in a few planet systems and  explanations regarding the PO approach.  K.G. thanks the staff of the Pozna\'n Supercomputer and Network Centre (PCSS, Poland) for the generous long-term support and computing resources (grant No.  529).  K.G.  also thanks the staff of the Tricity Supercomputer Centre, Gdańsk (Poland) for computing resources on the Tryton cluster that were used to conduct some numerical experiments and for their great and professional support. A.V. acknowledges funding from the European Research Council (ERC) under the European Union’s Horizon 2020 research and innovation programme, grant agreements No. 757561 (HiRISE). The LBTO AO group would like to acknowledge the assistance of R.T. Gatto with night observations. We would also like to thank J.Power for assisting in preparation of the observations using ADI mode. For the purpose of open access, the authors have applied a Creative Commons Attribution (CC BY) licence to any Author Accepted Manuscript version arising from this submission.
\end{acknowledgements}

\begin{tiny}
SPHERE is an instrument designed and built by a consortium consisting
of IPAG (Grenoble, France), MPIA (Heidelberg, Germany), LAM (Marseille,
France), LESIA (Paris, France), Laboratoire Lagrange (Nice, France), INAF
- Osservatorio di Padova (Italy), Observatoire de Gen\`eve (Switzerland), ETH
Zurich (Switzerland), NOVA (Netherlands), ONERA (France), and ASTRON
(Netherlands), in collaboration with ESO. SPHERE was funded by ESO, with
additional contributions from CNRS (France), MPIA (Germany), INAF (Italy),
FINES (Switzerland), and NOVA (Netherlands). SPHERE also received funding
from the European Commission Sixth and Seventh Framework Programmes as
part of the Optical Infrared Coordination Network for Astronomy (OPTICON)
under grant number RII3-Ct-2004-001566 for FP6 (2004-2008), grant number
226604 for FP7 (2009-2012) and grant number 312430 for FP7 (2013-2016).

\end{tiny}

%\end{acknowledgements}
 
\bibliographystyle{aa}
\bibliography{hr8799lit}

\clearpage
\onecolumn

\begin{appendix}

\section{Astrometric points used in the analysis}

\begin{landscape}
\small
\begin{longtable}{lccccccccccccccccc}  
%\begin{minipage}{1\textwidth}
\caption{List of the astrometric points.} \label{t:lite}\\ \hline
Date & \multicolumn{4}{c}{Planet b} & \multicolumn{4}{c}{Planet c} & \multicolumn{4}{c}{Planet d} & \multicolumn{4}{c}{Planet e} &  \\
 & $\Delta$RA & $\sigma_{\Delta{\rm RA}}$ & $\Delta$Dec & $\sigma_{\Delta{\rm Dec}}$ & $\Delta$RA & $\sigma_{\Delta{\rm RA}}$ & $\Delta$Dec & $\sigma_{\Delta{\rm Dec}}$ & $\Delta$RA & $\sigma_{\Delta{\rm RA}}$ & $\Delta$Dec & $\sigma_{\Delta{\rm Dec}}$ & $\Delta$RA & $\sigma_{\Delta{\rm RA}}$ & $\Delta$Dec & $\sigma_{\Delta{\rm Dec}}$ & Ref. \\  \hline\hline
1998.83 &       1411    &       9       &       986     &       9       &--&--&--&--&--&--&--&--&--&--&--&      -- & \cite{2009ApJ...694L.148L}    \\
 1998.83        &       1418    &       22      &       1004    &       20      &       -837    &       26      &       483     &       23      &       133     &       35      &       -533    &       34      &--&--&--&      -- & \cite{2011ApJ...741...55S}    \\
 2002.54        &       1481    &       23      &       919     &       17      &--&--&--&--&--&--&--&--&--&--&--&      --& \cite{2009ApJ...696L...1F}      \\
 2004.53        &       1471    &       6       &       884     &       6       &       -739    &       6       &       612     &       6       &--&--&--&--&--&--&--&  -- & \cite{Konopacky2016}  \\
 2005.54        &       1496    &       5       &       856     &       5       &       -713    &       5       &       630     &       5       &       -87     &       10      &       -578    &       10      &--&--&--&      -- & \cite{2012ApJ...755L..34C}    \\
 2007.58        &       1504    &       3       &       837     &       3       &       -683    &       4       &       671     &       4       &       -179    &       5       &       -588    &       5       &--&--&--&      --&\cite{Konopacky2016} \\
 2007.81        &       1500    &       7       &       836     &       7       &       -678    &       7       &       676     &       7       &       -175    &       10      &       -589    &       10      &--&--&--&      --&\cite{Konopacky2016} \\
 2008.52        &       1527    &       4       &       799     &       4       &       -658    &       4       &       701     &       4       &       -208    &       4       &       -582    &       4       &--&--&--&      -- &\cite{2008Sci...322.1348M}     \\
 2008.61        &       1527    &       2       &       801     &       2       &       -657    &       2       &       706     &       2       &       -216    &       2       &       -582    &       2       &--&--&--&-- &\cite{2008Sci...322.1348M}     \\
 2008.71        &       1516    &       4       &       818     &       4       &       -663    &       3       &       693     &       3       &       -202    &       4       &       -588    &       4       &--&--&--&      --&\cite{Konopacky2016} \\
 2008.89        &       1532    &       20      &       796     &       20      &       -654    &       20      &       700     &       20      &       -217    &       20      &       -608    &       20      &--&--&--&      -- &\cite{2010ApJ...716..417H}     \\
 2009.02        &--&--&--&--&   -612    &       30      &       665     &       30      &--&--&--&--&--&--&--&  -- &\cite{2010ApJ...716..417H}     \\
 2009.58        &       1526    &       4       &       797     &       4       &       -639    &       4       &       712     &       4       &       -237    &       3       &       -577    &       3       &       -306    &       7       &       -211    &       7&\cite{Konopacky2016}  \\
 2009.62        &       1536    &       10      &       785     &       10      &--&--&--&--&--&--&--&--&--&--&--&      -- & \cite{2011ApJ...729..128C}    \\
 2009.70        &1538&30&777&   30      &-634   &       30      &       697     &       30      &-282& 30 & -590 &30&--&--&--&--& \cite{2010ApJ...716..417H}   \\
 2009.76        &1535   &20     &816    &20     &-636   &40     &692    &40     &-270   &70     &-600   &70     &--     &--&--& --&\cite{2011AA...528A.134B}\\
 2009.77        &       1532    &       7       &       783     &       7       &       -627    &       7       &       716     &       7       &       -241    &       7       &       -586    &       7       &       -306    &       7       &       -217    &       7  &\cite{2011ApJ...729..128C}    \\
 2009.83&1524&10&795&10&-636&9&720&9&-251&7&-573&7&-310&        9&-187& 9&\cite{Konopacky2016}  \\
 2009.84        &       1540    &       19      &       800     &       19      &       -630    &       13      &       720     &       13      &       -240    &       14      &       -580    &       14      &       -304    &       10      &       -196    &       10 & \cite{2011ApJ...739L..41G}    \\
 2010.53        &       1532    &       5       &       783     &       5       &       -619    &       4       &       728     &       4       &       -265    &       4       &       -576    &       4       &       -323    &       6       &       -166    &       6&\cite{Konopacky2016}  \\
 2010.55        &       1547    &       6       &       757     &       9       &       -606    &       6       &       725     &       6       &       -269    &       6       &       -580    &       6       &       -329    &       6       &       -178    &       6 &\cite{2014ApJ...795..133C}     \\
 2010.83        &       1535    &       15      &       766     &       15      &       -607    &       12      &       744     &       12      &       -296    &       13      &       -561    &       13      &       -341    &       16      &       -143    &       16&\cite{Konopacky2016} \\
 2011.55        &       1541    &       5       &       762     &       5       &       -595    &       4       &       747     &       4       &       -303    &       5       &       -562    &       5       &       -352    &       8       &       -130    &       8&\cite{Konopacky2016}  \\
 2011.79        &       1579    &       11      &       734     &       11      &       -561    &       13      &       752     &       13      &       -299    &       13      &       -563    &       13      &       -326    &       13      &       -119    &       13& \cite{2013AA...549A..52E}       \\
 2011.86        &       1546    &       11      &       725     &       11      &       -578    &       13      &       767     &       13      &       -320    &       13      &       -549    &       13      &       -382    &       16      &       -127    &       16& \cite{2013AA...549A..52E}       \\
 2012.55        &       1545    &       5       &       747     &       5       &       -578    &       5       &       761     &       5       &       -339    &       5       &       -555    &       5       &       -373    &       8       &       -84     &       8&\cite{Konopacky2016}  \\
 2012.82        &       1549    &       4       &       743     &       4       &       -572    &       3       &       768     &       3       &       -346    &       4       &       -548    &       4       &       -370    &       9       &       -76     &       9&\cite{Konopacky2016}  \\
 2012.83        &       1558    &       6       &       729     &       9       &       -557    &       6       &       763     &       6       &       -343    &       6       &       -555    &       6       &       -371    &       6       &       -80     &       6&\cite{2014ApJ...795..133C}    \\
 2013.79        &       1545    &       22      &       724     &       22      &       -542    &       22      &       784     &       22      &       -382    &       16      &       -522    &       16      &       -373    &       13      &       -17     &       13&\cite{Konopacky2016} \\
 2013.81        &       1562    &       8       &       713     &       13      &       -538    &       6       &       784     &       13      &       -377    &       7       &       -538    &       11      &       -394    &       11      &       -36     &       17 & \cite{2015AA...579C...2M}     \\
 2013.88        &--&--&--&--&   -537    &       1       &       782     &       2       &       -370    &       1       &       -539    &       1       &       -381    &       2       &       -30     &       0 &\cite{2018AJ....156..192W}\footnote{Revised following the updated astrometric calibration in \cite{2020SPIE11447E..5AD}}      \\
 2014.53        &       --      &       --      &       --      &       --      &       --      &       --      &       --      &       --      &       -400    &       4       &       -512    &       4       &       -389    &       1       &       -22     &       2       &       IFS, This work\\
 2014.53        &       1570    &       3       &       704     &       3       &       -521    &       3       &       790     &       9       &       -391    &       2       &       -530    &       2       &       -387    &       2       &       -10     &       3       &       IRDIS, This work\\
 2014.54        &       1560    &       13      &       725     &       13      &       -540    &       13      &       799     &       13      &       -400    &       11      &       -534    &       11      &       -387    &       11      &       3       &       11&\cite{Konopacky2016} \\
 2014.62        &       --      &       --      &       --      &       --      &       --      &       --      &       --      &       --      &       -396    &       1       &       -524    &       2       &       -389    &       1       &       -17     &       2       &       IFS, This work\\
 2014.70        &       1569    &       4       &       707     &       2       &       -519    &       1       &       794     &       2       &       -397    &       1       &       -530    &       2       &--&--&--&      --&\cite{2018AJ....156..192W}$\color{red}^1$    \\
 2014.93        &       1575    &       2       &       702     &       4       &       -511    &       2       &       799     &       2       &       -400    &       2       &       -523    &       2       &       -385    &       3       &       12      &       2 & \cite{2017AA...598A..83W}     \\
 2014.93        &       1574    &       3       &       701     &       2       &       -514    &       3       &       798     &       4       &       -399    &       4       &       -525    &       4       &       -389    &       8       &       11      &       4       &       IRDIS, This work\\
 2014.93        &       1574    &       4       &       701     &       3       &       -512    &       3       &       798     &       4       &       -400    &       4       &       -523    &       4       &       -390    &       7       &       12      &       4       &       IRDIS, This work\\
 2014.93        &       1573    &       3       &       701     &       3       &       -512    &       3       &       797     &       4       &       -403    &       4       &       -524    &       4       &       -383    &       8       &       11      &       4       &       IRDIS, This work\\
 2015.51        &       --      &       --      &       --      &       --      &       --      &       --      &       --      &       --      &       -424    &       4       &       -509    &       3       &       -391    &       1       &       33      &       2       &       IFS, This work\\
 2015.51        &       1579    &       1       &       694     &       1       &       -498    &       1       &       806     &       1       &       -417    &       1       &       -517    &       1       &       -383    &       9       &       33      &       5       &       IRDIS, This work\\
 2015.58        &       1580    &       5       &       689     &       3       &       -495    &       2       &       806     &       2       &       -419    &       2       &       -516    &       1       &       -386    &       1       &       36      &       1       &       IRDIS, This work\\
 2015.65 & 1569 & 11 &666 & 7 & -482 & 5 & 813 & 6 &-436 & 11 & -510 & 12 & --    &       --      &       --      & -- & \citet{2022AJ....163...52S} \\
 2015.74        &       --      &       --      &       --      &       --      &       --      &       --      &       --      &       --      &       -420    &       4       &       -513    &       4       &       -392    &       1       &       39      &       3       &       IFS, This work\\
 2015.74        &       1580    &       1       &       688     &       1       &       -494    &       1       &       811     &       1       &       -426    &       1       &       -512    &       1       &       -382    &       9       &       50      &       5       &       IRDIS, This work\\
 2016.72        &--&--&--&--&   -466    &       1       &       821     &       2       &       -453    &       1       &       -499    &       1       &       -376    &       2       &       81      &       0&\cite{2018AJ....156..192W}$\color{red}^1$     \\
 2016.88        &       --      &       --      &       --      &       --      &       --      &       --      &       --      &       --      &       -464    &       1       &       -486    &       2       &       -382    &       2       &       94      &       6       &       IFS, This work\\
 2016.88        &       1589    &       2       &       666     &       1       &       -464    &       2       &       824     &       2       &       -454    &       2       &       -489    &       2       &       -378    &       4       &       90      &       2       &       IRDIS, This work\\
 2017.45        &       --      &       --      &       --      &       --      &       --      &       --      &       --      &       --      &       -473    &       2       &       -476    &       2       &       -377    &       1       &       115     &       3       &       IFS, This work\\
 2017.45        &       1591    &       1       &       653     &       1       &       -449    &       1       &       835     &       1       &       -472    &       2       &       -482    &       2       &       -373    &       3       &       118     &       2       &       IRDIS, This work\\
 2017.78        &       --      &       --      &       --      &       --      &       --      &       --      &       --      &       --      &       -480    &       2       &       -478    &       2       &       -372    &       1       &       129     &       2       &       IFS, This work\\
 2017.78        &       1595    &       1       &       647     &       1       &       -441    &       1       &       839     &       1       &       -480    &       1       &       -477    &       1       &       -369    &       1       &       128     &       1       &       IRDIS, This work\\
 2017.78        &       --      &       --      &       --      &       --      &       --      &       --      &       --      &       --      &       -492    &       5       &       -463    &       6       &       -370    &       2       &       135     &       3       &       IFS, This work\\
 2017.78        &       1595    &       1       &       647     &       1       &       -441    &       1       &       839     &       1       &       -480    &       1       &       -477    &       1       &       -371    &       2       &       128     &       2       &       IRDIS, This work\\
 2018.46        &       --      &       --      &       --      &       --      &       --      &       --      &       --      &       --      &       -495    &       1       &       -460    &       2       &       -360    &       2       &       162     &       2       &       IFS, This work\\
 2018.46        &       1601    &       1       &       635     &       1       &       -424    &       1       &       848     &       1       &       -497    &       2       &       -463    &       2       &       -358    &       2       &       156     &       2       &       IRDIS, This work\\
 2018.63        &       --      &       --      &       --      &       --      &       --      &       --      &       --      &       --      &       -509    &       2       &       -452    &       3       &       -361    &       2       &       166     &       2       &       IFS, This work\\
 2018.63        &       1601    &       2       &       632     &       3       &       -421    &       1       &       850     &       1       &       -502    &       1       &       -461    &       1       &       -357    &       1       &       162     &       2       &       IRDIS, This work\\
 2018.63        &       --      &       --      &       --      &       --      &       --      &       --      &       --      &       --      &       -503    &       2       &       -456    &       2       &       -359    &       1       &       167     &       2       &       IFS, This work\\
 2018.63        &       1600    &       1       &       632     &       1       &       -421    &       1       &       851     &       1       &       -502    &       2       &       -458    &       1       &       -358    &       2       &       163     &       1       &       IRDIS, This work\\
 2018.66        &--&--&--&--&--&--&--&--&--&--&--&--&   -358    & 0 &   163     &       0&   \cite{2019AA...623L..11G}\\
 2019.83        &       --      &       -       &       --      &       --      &       --      &       --      &       --      &       --      &       -527    &       3       &       -432    &       3       &       -337    &       3       &       210     &       3       &       IFS, This work\\
 2019.83        &       1606    &       2       &       615     &       2       &       -392    &       2       &       875     &       2       &       -532    &       3       &       -425    &       2       &       -338    &       2       &       215     &       2       &       IRDIS, This work\\
 2019.84        &       --      &       --      &       --      &       --      &       --      &       --      &       --      &       --      &       -528    &       1       &       -435    &       2       &       -337    &       2       &       208     &       2       &       IFS, This work\\
 2019.84        &       1611    &       1       &       611     &       1       &       -388    &       2       &       870     &       2       &       -530    &       2       &       -430    &       1       &       -335    &       1       &       210     &       1       &       IRDIS, This work\\
 2020.75        &       1620    &       1       &       591     &       3       &       -364    &       2       &       883     &       1       &       -551    &       1       &       -415    &       4       &       -315    &       3       &       242     &       5         &       LUCI, This work \\
 2021.64        &       --      &       --      &       --      &       --      &       --      &       --      &       --      &       --      &       -569    &       1       &       -390    &       2       &       -292    &       2       &       276     &       2       &       IFS, This work\\
 2021.64        &       1626    &       1       &       578     &       2       &       -339    &       2       &       890     &       2       &       -563    &       2       &       -391    &       2       &       -287    &       4       &       272     &       2       &       IRDIS, This work\\
 \hline
\end{longtable}
\end{landscape}

\end{appendix}

\end{document}